%% file: ex_article_merged_figs.tex
\documentclass[review,hidelinks,onefignum,onetabnum]{siamart251216}

\usepackage{bm}
\usepackage{physics,algorithm,algorithmic}
\usepackage{graphicx}
\usepackage{adjustbox}
\usepackage{lineno,tikz}
\usepackage{comment}
\usepackage{subcaption}
\usepackage{amssymb}
\usepackage{setspace}
\usepackage{amsmath}

\DeclareMathOperator{\arcsinh}{arcsinh}

\input{ex_shared}

\ifpdf
\hypersetup{
  pdftitle={Singular and nearly singular quadrature in arbitrary dimensions},
  pdfauthor={Jing Hu, Luqiang He, Shuyu Sun}
}
\fi

\begin{document}
\nolinenumbers
\maketitle

% ======================================================================
\begin{abstract}
The accurate numerical evaluation of integrals with singular or nearly singular
kernels remains a fundamental challenge in scientific computing, particularly
as the spatial dimension increases. While existing approaches, including
singularity extraction, adaptive element subdivision, and analytical integration
schemes, have achieved considerable success in one and two dimensions, their
extension to higher dimensions is often hindered by severe near-singular
behaviour and the resulting numerical instabilities. In this work, we propose a
systematic approach for the high-accuracy evaluation of singular integrals and nearly singular integrals in arbitrary dimensions, based
on variable transformation techniques derived from the Gauss hypergeometric
function. We further analyze the structure
of singularities and near-singularities in both the Duffy domain and the
physical Cartesian domain, and establish a unified treatment via the sinh transformation.
A convergence analysis based on the Bernstein ellipse parameter reveals that the sinh transformation ($\beta=1$) consistently outperforms alternative
hypergeometric transformations ($\beta=2,3$), which introduce secondary
near-singularities despite achieving formal cancellation. Numerical experiments
in dimensions two through five, together with an analysis of quadrature-induced
consistency errors in variational discretisations, demonstrate that the proposed
approach achieves exponential convergence and substantially outperforms direct
Gauss--Legendre quadrature. The method offers a practical and theoretically
grounded framework for high-dimensional nearly singular integration with broad
applicability in scientific computing.
\end{abstract}



% ======================================================================
\section{Introduction}
% ======================================================================

The numerical evaluation of integrals with singular or nearly singular kernels
is a recurring computational bottleneck across many areas of scientific
computing. Such integrals form the mathematical foundation of boundary integral equation formulations for elliptic partial differential equations~\cite{ying2006high, kolm2001numerical}, potential theory, and fast multipole methods~\cite{rokhlin1985rapid, barnett2014evaluation}. In physical modelling, they are ubiquitous in acoustic and electromagnetic scattering~\cite{colton2013integral, liu1992weakly, tong2010novel, amari1995efficient, kleinman1988single, yang2000investigation}, wave propagation~\cite{darbas2015well}, and elastostatics and fracture mechanics~\cite{anderson2005fracture, williams1952stress, hutchinson1987crack, zheng2014new, guiggiani1992general}. Furthermore, the need for accurate singular quadrature extends to domains as diverse as computational fluid dynamics~\cite{tlupova2019regularized}, biomechanics~\cite{chao2021fully}, computer graphics and radiosity~\cite{schroder1993numerical, ouellette2001numerical}, and mathematical finance~\cite{ilhan2004singular, widdicks2005black}. When the source
point lies on or near the integration domain, the integrand develops a sharp
peak or an algebraic singularity that renders standard Gaussian quadrature
ineffective~\cite{schwab1994variable,atkinson2008introduction,davis2007methods,
burden1997numerical}. In the weakly singular case, the integral diverges in the
ordinary sense and requires specialised regularisation; in the nearly singular
case, the integrand remains bounded but develops steep gradients that cannot
be adequately resolved by a fixed quadrature rule~\cite{beale2001method,
huybrechs2009generalized}. These computational difficulties become substantially
more severe as the spatial dimension increases, since the singularity structure
couples across multiple coordinate directions and the cost of tensor-product
quadrature grows exponentially~\cite{novak1996high}.

Over the past decades, a variety of techniques have been developed to address
this challenge. Singularity subtraction and extraction methods~\cite
{graglia1993numerical,jarvenpaa2003singularity,guiggiani1992general} isolate
and integrate the singular part analytically, leaving a regular remainder for
numerical quadrature. Singularity swapping techniques~\cite{bao2024singularity,
af2021accurate} replace the singular kernel with a smoother one that shares
the same integral. Continuation approaches~\cite{rosen1995continuation,
lenoir2012evaluation,vijayakumar1989new} construct a smooth continuation of
the integrand away from the singularity. Adaptive subdivision methods~\cite
{chi2022adaptive} refine the integration mesh near the singular region, while
analytical and semi-analytical schemes~\cite{latypov2021evaluation,
padhi2001analytic,han2022semi,niu2005semi} exploit closed-form expressions
where available. Gauss--Jacobi quadrature frameworks for singular and nearly
singular integrals on domains with edges and corners have been developed by
Tsalamengas~\cite{tsalamengas2015quadrature,tsalamengas2016gauss}, and
Montanelli and coworkers~\cite{montanelli2022computing,montanelli2024computing}
have extended near-singular quadrature to curved boundary elements. Among all
these techniques, variable transformation methods, including the Duffy
transformation~\cite{duffy1982quadrature,mousavi2010generalized}, the sinh
transformation~\cite{johnston2005sinh}, exponential and polynomial
transformations~\cite{ye2008new,xie2014improved}, and sigmoidal
transformations~\cite{yun2003new}, have proven particularly effective, as
they regularise the integrand by clustering quadrature points in regions of
rapid variation via a nonlinear change of variables.

Despite the wealth of available transformations, two significant gaps remain.
First, there is no systematic framework for comparing the convergence
properties of different transformations applied to the same singularity
structure, making it difficult for practitioners to select an appropriate
method for a given problem. The theory of exponential convergence for analytic
functions provides a rigorous tool for such comparisons: for an analytic
integrand, the Gauss--Legendre quadrature error on $m$ points is bounded by
$C\rho^{-2m}$, where $\rho = |z^* + \sqrt{(z^*)^2-1}|$ is the Bernstein
ellipse parameter associated with the nearest complex singularity
$z^*$~\cite{trefethen2019approximation,zhao2013sharp,xiang2012convergence,
xie2013exponential}. Yet, this analysis has rarely been applied systematically to evaluate competing variable transformations. Consequently, the question of which methods are genuinely effective, alongside the theoretical justifications for their success, remains largely unresolved.
Second, most existing treatments are confined to one or two
dimensions~\cite{chernov2012exponential,johnston2005sinh,chernov2013numerical},
while higher-dimensional scenarios---which arise, for example, in boundary
integral formulations involving volumetric body forces or in high-dimensional
parametric problems---have received comparatively little attention. Notable
exceptions include the work of Mousavi and Sukumar~\cite{mousavi2010generalized} which generalised the Duffy transformation to high-dimensional vertex
singularities, Latypov~\cite{latypov2021evaluation} which derived closed-form
expressions for four-dimensional singular and nearly singular double surface
integrals, and Gu et al.~\cite{gu2017general} which developed an algorithm for
nearly strong-singular integrals in three-dimensional boundary element analysis.
Sparse grid methods~\cite{gerstner1998numerical} can mitigate the exponential
growth of tensor-product quadrature but are not directly applicable to singular
integrands. A general treatment of nearly singular integrals in arbitrary
dimensions, supported by rigorous convergence analysis, remains absent from
the literature. Motivated by these considerations, Hu et al.~\cite
{hu2026transformation} recently proposed a unified framework based on the
Duffy transformation for two-dimensional weakly singular and nearly singular
integrals, providing a comparative analysis of different variable
transformations. The present work extends this line of investigation to higher
dimensions.

In this work, we make the following contributions. First, working within the
isoparametric framework, we systematically analyse the structure of
singularities and near-singularities by examining the asymptotic behaviour of
the integrand in both the Duffy domain and the physical Cartesian domain. We
show that the near-singular behaviour associated with each integration variable
can be reduced to a canonical form, enabling a unified treatment via the Gauss
hypergeometric transformation. Second, we provide a convergence
analysis based on the Bernstein ellipse parameter in the complex plane, which
explains why the sinh transformation (the $\beta=1$ member of the hypergeometric
family) is uniquely effective; the transformations with $\beta=2$ and $\beta=3$
introduce secondary near-singularities through their Jacobians, thereby
degrading convergence despite achieving formal algebraic cancellation in the
denominator. Third, we present numerical experiments in dimensions two through
five that corroborate the theoretical analysis and demonstrate the substantial
improvement in convergence rate afforded by the sinh transformation. Finally,
we discuss the implications of accurate quadrature for the consistency error
in variational discretisations, illustrating how the quality of numerical
integration directly affects the total error of numerical solutions through Strang's first lemma~\cite{strang1974analysis}.

The remainder of the paper is organised as follows. Section~2 analyses the
singularity and near-singularity structure and presents the regularisation
strategies, with convergence results integrated into each subsection.
Section~3 provides numerical experiments in dimensions two through five.
Section~4 discusses the significance of accurate quadrature for variational
discretisations. Section~5 analyses the computational complexity, and
Section~6 concludes with a summary and outlook.

% ======================================================================
\section{Structure and regularisation of singular and nearly singular integrals}
\label{sec:structure}
% ======================================================================

\subsection{Problem formulation}
\label{sec:problem}

We consider integrals defined over an $n$-dimensional domain. In practical
computations, the physical domain is discretised
using curved elements with general geometries. To evaluate integrals over such
elements in a systematic manner, it is standard practice to map the physical
domain onto a reference isoparametric domain $\mathbf{D}=[-1,1]^n$, where
quadrature rules are canonically defined. In the isoparametric setting, the
same shape functions are used to interpolate both the geometry and the field
variables. An important consequence of this construction is that the mapping from the physical domain to the reference domain is a polynomial of fixed degree. As a
result, a Taylor expansion of the geometry truncated at the order of the shape
functions is exact; higher-order geometric terms, which would otherwise
complicate the analysis of the singularity structure, do not arise. This
property substantially simplifies the asymptotic analysis that follows. For
more complex integration domains, it suffices to first map the physical domain
onto the isoparametric reference domain; the subsequent analysis then applies
without modification.

With this motivation, we consider an integral on a $n$-dimensional isoparametric space:
\begin{equation}
    I = \int_{\mathbf{D}} \frac{f(\mathbf{x})}{\|\mathbf{x} - \mathbf{x}_{\text{s}}\|^{\alpha}} \, \mathrm{d}\mathbf{D},
    \qquad  \mathbf{D} \subset \mathbb{R}^n, \label{eq:Initial}
\end{equation}
where $f(\mathbf{x})$ is smooth, $\alpha>0$, and $\mathbf{x}_{\text{s}}$ is
the source point. Let $\mathbf{x}_0 \in \mathbf{D}$ be the point on $\mathbf{D}$
closest to $\mathbf{x}_{\text{s}}$, and define $\mathbf{r}_0 = \mathbf{x}_0 -
\mathbf{x}_{\text{s}}$ (Figure~\ref{schematic_pro}). Eq.~\eqref{eq:Initial} becomes
\begin{equation}
    I = \int_{\mathbf{D}} \frac{f(\mathbf{x})}{\|\mathbf{r}_0+\mathbf{x} - \mathbf{x}_0\|^{\alpha}} \, \mathrm{d}\mathbf{D},
    \qquad  \mathbf{D} \subset \mathbb{R}^n. \label{eq:integral_dD}
\end{equation}
Two regimes are distinguished:
\begin{itemize}
    \item Singular ($\mathbf{r}_0 = \mathbf{0}$): the source lies on
      the domain. The integral requires $\alpha < n$ for absolute convergence.
    \item Nearly singular ($0 < r_0 = \|\mathbf{r}_0\| \ll 1$): the
      source is close to the domain. The integrand is bounded but sharply peaked.
\end{itemize}

% ======================================================================
\subsection{Singular integrals: the Duffy transformation}
\label{sec:duffy}
% ======================================================================

\begin{figure}[!htbp]
    \centering
    \captionsetup[subfigure]{justification=centering}
    \subcaptionbox{Schematic of the problem\label{schematic_pro}}{%
        \includegraphics[width=4.2cm,height=3.2cm]{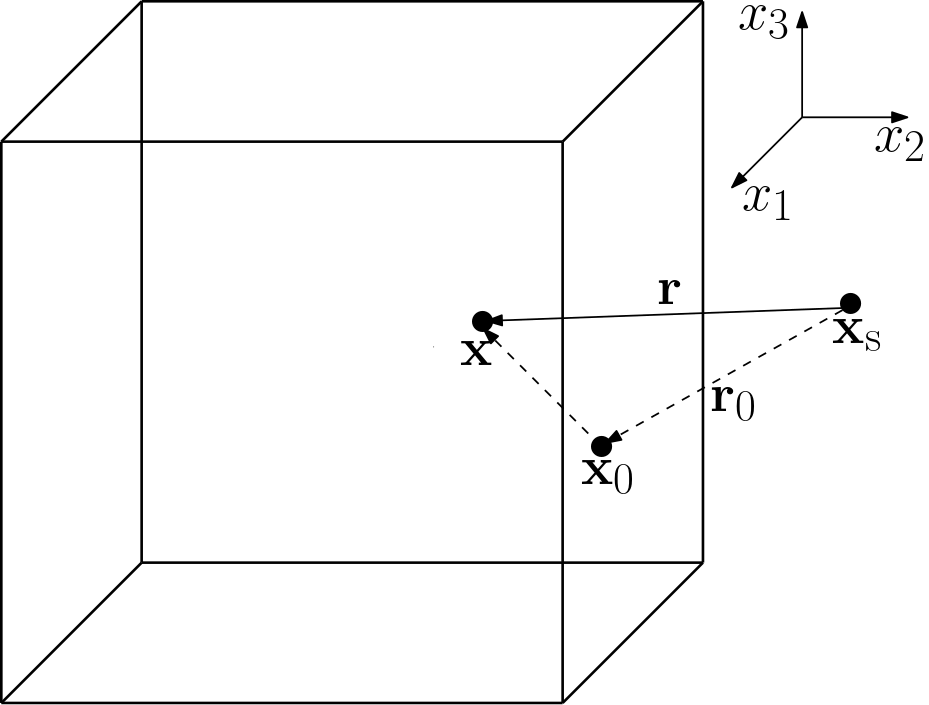}}\quad
    \subcaptionbox{Duffy domain\label{schematic_duffy}}{%
        \includegraphics[width=3.2cm,height=3.2cm]{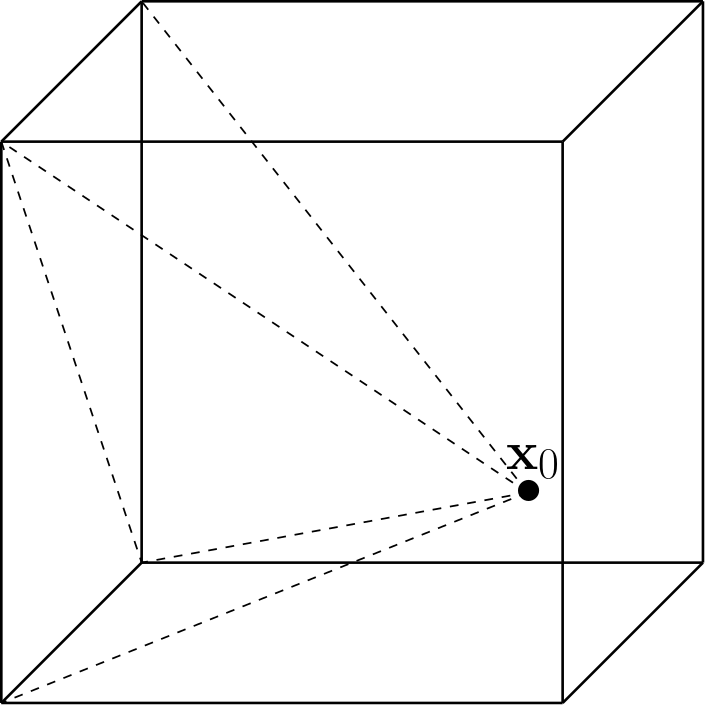}}\quad
    \subcaptionbox{Physical Cartesian domain\label{schematic_carte}}{%
        \includegraphics[width=3.2cm,height=3.2cm]{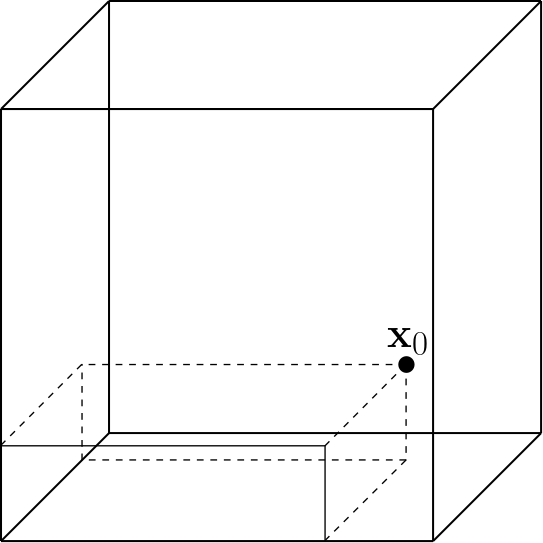}}
    \caption{Schematics of the domain partition for implementation in Duffy domain and physical Cartesian domain.}\label{fig:schematic}
\end{figure}

When $\mathbf{r}_0 = \mathbf{0}$, the integral reduces to
\begin{equation}
    I = \int_{\mathbf{D}} \frac{f(\mathbf{x})}{\|\mathbf{x} - \mathbf{x}_0\|^{\alpha}} \, \mathrm{d}\mathbf{D}.
\end{equation}
Let $\bar{\mathbf{x}}_0$ be the isoparametric coordinate of $\mathbf{x}_0$.
Introduce the local coordinate $\bm{\xi} = \bar{\mathbf{x}} - \bar{\mathbf{x}}_0$
and partition $\mathbf{D}=[-1,1]^n$ into $m_1 = 2\cdot n!$ $n$-simplices, each
sharing $\bar{\mathbf{x}}_0$ as a common vertex (Figure~\ref{schematic_duffy}).
Eq.~\eqref{eq:integral_dD} becomes
\begin{equation}
    I = \sum_{l=1}^{m_1} I_l = \sum_{l=1}^{m_1} \int_{\mathbf{D}_l}
    \frac{f(\mathbf{x})}{\|\mathbf{x}(\bm{\xi}) - \mathbf{x}_0\|^{\alpha}} \,
    \mathrm{d}\mathbf{D}_l. \label{eq:subintegral}
\end{equation}

\subsubsection{Generalised Duffy transformation and cancellation}

On each simplex $\mathbf{D}_l$ with vertices $\{\bar{\mathbf{x}}_0, \bar{\mathbf{x}}_1, \dots, \bar{\mathbf{x}}_n\}$,
the generalised $n$-dimensional Duffy transformation maps $[0,1]^n$ onto
$\mathbf{D}_l$:
\begin{equation}
    \bm{\xi}^{(k)}(\boldsymbol{\theta}) = \sum_{j=1}^{n} \left( \prod_{i=1}^{j} \theta_i \right)
    \bigl( \bm{\xi}^{(k)}_j - \bm{\xi}^{(k)}_{j-1} \bigr),
    \qquad k = 1, \dots, n, \label{eq:duffy}
\end{equation}
where $\bm{\xi}^{(k)}_j = \bar{\mathbf{x}}_j^{(k)} - \bar{\mathbf{x}}_0^{(k)}$
and $\bm{\xi}^{(k)}_0 \equiv 0$. The Jacobian is
\begin{equation}
    J_{\text{Duffy}} = n! \prod_{j=1}^{n-1} \theta_j^{\,n-j} |S_l|, \label{eq:jacobian}
\end{equation}
where $|S_l|$ is the $n$-dimensional Lebesgue measure of $\mathbf{D}_l$.

To analyse the denominator, define the geometric Jacobian
$\mathbf{S}^{(k)} = \partial \mathbf{x} / \partial \bm{\xi}^{(k)}|_{\bar{\mathbf{x}}_0}$
and the vector
\begin{equation}
    \mathbf{P}_1(\theta_2,\dots,\theta_n) = \sum_{k=1}^{n} \mathbf{S}^{(k)}
    \sum_{j=1}^{n} \Bigl( \prod_{i=2}^{j} \theta_i \Bigr)
    \bigl( \bm{\xi}^{(k)}_j - \bm{\xi}^{(k)}_{j-1} \bigr), \label{eq:P1}
\end{equation}
with the convention $\prod_{i=2}^{1}\theta_i \equiv 1$. In the isoparametric
setting, a Taylor expansion of $\mathbf{x}$ in $\theta_1$ about $\mathbf{x}_0$
is exact at linear order:
\begin{equation}
    \mathbf{x}(\bm{\xi}(\boldsymbol{\theta})) = \mathbf{x}_0 +
    \mathbf{P}_1(\theta_2,\dots,\theta_n)\,\theta_1. \label{eq:expansion}
\end{equation}
Hence the distance factorises as
\begin{equation}
    r = \|\mathbf{x} - \mathbf{x}_0\| = \sqrt{c_2}\; \theta_1,
    \qquad c_2 = \mathbf{P}_1\!\cdot\!\mathbf{P}_1. \label{eq:r_factor}
\end{equation}

Substituting the Jacobian~\eqref{eq:jacobian} and the distance~\eqref{eq:r_factor}
into the sub-integral~\eqref{eq:subintegral} yields the cancellation that is
the central mechanism of the Duffy method:
\begin{equation}
    I_l = \int_{[0,1]^n}
    \frac{f(\mathbf{x})\, n! \prod_{j=2}^{n-1} \theta_j^{\,n-j} |S_l|}
    {c_2^{\alpha/2}\, \theta_1^{\alpha-n+1}}
    \, \prod_{j=1}^{n} \mathrm{d}\theta_j. \label{eq:I_l_after_duffy}
\end{equation}
The Jacobian factor $\theta_1^{\,n-1}$ cancels against $\theta_1^{\alpha}$ from
$r^{\alpha}$. The net exponent $p = \alpha - n + 1$. The integral over
$\theta_1 \in [0,1]$ converges iff $p < 1$, i.e.\ $\alpha < n$.

\subsubsection{Convergence of the Duffy-transformed singular integral}

We now quantify the convergence rate of Gauss--Legendre quadrature applied to
Eq.~\eqref{eq:I_l_after_duffy}. The analysis separates the radial ($\theta_1$)
and angular ($\theta_2,\dots,\theta_n$) directions.

\textbf{Radial direction.} For $\theta_1 \in [0,1]$, the integrand behaves as
$\theta_1^{-p}$ with $p = \alpha - n + 1$. Three regimes arise:
\begin{enumerate}
    \item \textbf{$\alpha \le n-1$ ($p \le 0$):} The integrand is bounded or
      vanishes at $\theta_1 = 0$. Gauss--Legendre converges exponentially:
      $|E_1| \le C_1 \rho_1^{-2m_1}$ with $\rho_1 > 1$.
    \item \textbf{$n-1 < \alpha < n$ ($0 < p < 1$):} The integrand has an
      integrable algebraic singularity. Convergence is algebraic:
      $|E_1| \le C_1 m_1^{-2(n-\alpha)}$~\cite
      {trefethen2019approximation}.
    \item \textbf{$\alpha \ge n$ ($p \ge 1$):} The integral diverges.
\end{enumerate}
In boundary integral applications, the most common case is $\alpha = n-1$ (i.e.
weakly singular kernel), for which $p = 0$ and radial convergence is exponential.

\textbf{Angular directions.} For each $\theta_j$ ($j \ge 2$), the integrand is
analytic provided $c_2 > 0$. The convergence rate is governed by the nearest
complex singularity of $c_2(\theta_2,\dots,\theta_n) = 0$. The Bernstein ellipse
parameter is
\begin{equation}
    \rho_j = \bigl| z_j^* + \sqrt{(z_j^*)^2 - 1} \bigr|,
    \qquad z_j^* = 2\theta_j^* - 1,
\end{equation}
where $\theta_j^*$ is the nearest complex root of $c_2 = 0$. The per-dimension error is
$|E_j| \le C_j \rho_j^{-2m_j}$.

\textbf{The angular factor $c_2$ and the effect of dimension.}
To analyse $c_2$ uniformly across all $n$, define the geometric vectors
\begin{equation}
    \mathbf{v}_j = \sum_{k=1}^{n} \mathbf{S}^{(k)}
    \bigl( \bm{\xi}^{(k)}_{j+1} - \bm{\xi}^{(k)}_j \bigr), \qquad
    j = 0, 1, \dots, n-1,
\end{equation}
with $\bm{\xi}^{(k)}_0 \equiv 0$ so that $\mathbf{v}_0 = \sum_k \mathbf{S}^{(k)} \bm{\xi}^{(k)}_1$.
Substituting into Eq.~\eqref{eq:P1} gives the expansion
\begin{equation}
    \mathbf{P}_1(\theta_2,\dots,\theta_n)
    = \mathbf{v}_0
    + \theta_2\,\mathbf{v}_1
    + \theta_2\theta_3\,\mathbf{v}_2
    + \theta_2\theta_3\theta_4\,\mathbf{v}_3
    + \cdots
    + \theta_2\theta_3\cdots\theta_n\,\mathbf{v}_{n-1}. \label{eq:P1_termwise}
\end{equation}
For $n=2$, the sum terminates at $\theta_2\mathbf{v}_1$, giving
$\mathbf{P}_1 = \mathbf{v}_0 + \theta_2\mathbf{v}_1$., the sum contains two terms: $\mathbf{P}_1 = \mathbf{v}_0 + \theta_2\mathbf{v}_1$.
In this case $c_2$ reduces to the quadratic form
\begin{equation}
    c_2(\theta_2) = t_1\theta_2^2 + t_2\theta_2 + t_3, \qquad \theta_2 \in [0,1],
    \label{eq:c2_quadratic}
\end{equation}
with the coefficients
\begin{equation}
    \begin{split}
        t_1 &= \Bigl(\sum_{k=1}^{n}\mathbf{S}^{(k)}\bigl(\bm{\xi}_{2}^{(k)}-\bm{\xi}_{1}^{(k)}\bigr)\Bigr)\cdot
              \Bigl(\sum_{k=1}^{n}\mathbf{S}^{(k)}\bigl(\bm{\xi}_{2}^{(k)}-\bm{\xi}_{1}^{(k)}\bigr)\Bigr),\\[4pt]
        t_2 &= 2\Bigl(\sum_{k=1}^{n}\mathbf{S}^{(k)}\bigl(\bm{\xi}_{2}^{(k)}-\bm{\xi}_{1}^{(k)}\bigr)\Bigr)\cdot
              \Bigl(\sum_{k=1}^{n}\mathbf{S}^{(k)}\bigl(\bm{\xi}_{1}^{(k)}-\bm{\xi}_{0}^{(k)}\bigr)\Bigr),\\[4pt]
        t_3 &= \Bigl(\sum_{k=1}^{n}\mathbf{S}^{(k)}\bigl(\bm{\xi}_{1}^{(k)}-\bm{\xi}_{0}^{(k)}\bigr)\Bigr)\cdot
              \Bigl(\sum_{k=1}^{n}\mathbf{S}^{(k)}\bigl(\bm{\xi}_{1}^{(k)}-\bm{\xi}_{0}^{(k)}\bigr)\Bigr).
    \end{split}
\end{equation}
A key observation from~\eqref{eq:P1_termwise} is that for $n \ge 3$, the series
includes terms coupling at least two angular variables (i.e. $\theta_2\theta_3\mathbf{v}_2$
and beyond). For general $n$, the $j$-th term couples the product of the first
$j$ angular variables, with the final term $\theta_2\cdots\theta_n\mathbf{v}_{n-1}$
coupling all $n-1$ angular variables simultaneously. Squaring $\mathbf{P}_1$
gives the general expression
\begin{equation}
    c_2 = \sum_{j=0}^{n-1} \|\mathbf{v}_j\|^2 \Bigl(\prod_{i=2}^{j+1}\theta_i\Bigr)^{\!2}
    \;+\; 2\!\!\sum_{0 \le j < \ell \le n-1}\! (\mathbf{v}_j\!\cdot\!\mathbf{v}_\ell)
    \Bigl(\prod_{i=2}^{j+1}\theta_i\Bigr)\Bigl(\prod_{i=2}^{\ell+1}\theta_i\Bigr). \label{eq:c2_uniform}
\end{equation}
For $n=2$, the double sum reduces to the single term $(j,\ell)=(0,1)$, and
Eq.~\eqref{eq:c2_uniform} reproduces the quadratic~\eqref{eq:c2_quadratic}
with $t_1 = \|\mathbf{v}_1\|^2$, $t_2 = 2(\mathbf{v}_0\!\cdot\!\mathbf{v}_1)$,
and $t_3 = \|\mathbf{v}_0\|^2$. For $n \ge 3$, terms with $\ell-j \ge 2$ generate
monomials such as $\theta_2\theta_3$ (from $j=0,\ell=2$) that couple distinct
angular variables; the double sum cannot be expressed as a sum of univariate
functions.

\textbf{Geometric obstruction for distorted simplices.}
For a well-shaped simplex, $c_2$ is bounded away from zero; all $\rho_j$ are
bounded away from unity; and Gauss--Legendre converges exponentially in every
direction regardless of $n$. For a distorted simplex ($c_2 \approx 0$ for some
angular values), the near-singular set
\begin{equation}
    \mathcal{M} = \{(\theta_2,\dots,\theta_n) \in [0,1]^{n-1} : c_2 \approx 0\}
\end{equation}
is generically an $(n-2)$-dimensional manifold:
\begin{itemize}
    \item $n=2$: $\mathcal{M}$ is a point (dimension 0);
    \item $n=3$: $\mathcal{M}$ is a curve (dimension 1);
    \item $n=4$: $\mathcal{M}$ is a surface (dimension 2).
\end{itemize}
A tensor-product quadrature rule places nodes on a Cartesian grid in
$[0,1]^{n-1}$. Applying independent 1D nonlinear transformations to each angular
variable can cluster nodes near any axis-aligned point (a 0-dimensional locus),
but cannot cluster nodes around a manifold of dimension $\ge 1$ that is
oblique to the coordinate axes. No separable change of variables can alter this,
because separability preserves the Cartesian-product structure of the node set.
The angular convergence therefore degrades for distorted simplices when $n>2$.
In summary, the Duffy transformation provides a valid method for singular
integrals in any dimension $n$: the radial singularity is cancelled by the
Jacobian when $\alpha < n$, and for well-shaped simplices all variables converge
exponentially. For distorted simplices with $n>2$, the angular convergence
degrades because the $(n-2)$-dimensional near-singular manifold may not be well
resolved by a tensor-product rule. This is a robustness limitation of Duffy transformation; adaptive simplex subdivision can mitigate it in practice~\cite{mousavi2010generalized}.

% ======================================================================
\subsection{Nearly singular integrals: limitation of the Duffy domain}
\label{sec:duffy_nearly}
% ======================================================================

We now return to the nearly singular case ($0<r_0\ll1$). Applying the Duffy
transformation~\eqref{eq:duffy} to Eq.~\eqref{eq:integral_dD} and using the
orthogonality $\mathbf{r}_0\!\cdot\!\mathbf{P}_1 = 0$ (since $\bar{\mathbf{x}}_0$
is the projection of $\mathbf{x}_{\text{s}}$ onto $\mathbf{D}$), we obtain
\begin{equation}
    I_l = \int_{[0,1]^n}
    \frac{f(\mathbf{x})\, n! \prod_{j=1}^{n-1} \theta_j^{\,n-j} |S_l|}
    {(r_0^2 + c_2 \theta_1^2)^{\alpha/2}}
    \, \prod_{j=1}^{n} \mathrm{d}\theta_j. \label{eq:nearly_after_duffy}
\end{equation}

Two critical differences from the singular case emerge. First, the denominator
does not factorise: the Jacobian factor $\theta_1^{\,n-1}$ provides no
cancellation when $r_0 > 0$. Second, both $\theta_1$ and the angular
variables exhibit near-singular behaviour. The near-singularity in $\theta_1$
has the canonical form $(r_0^2 + c_2\theta_1^2)^{-\alpha/2}$, which is a 1D
structure treatable with a 1D nonlinear transformation in any dimension.
The angular near-singularity enters through $c_2^{-\alpha/2}$, exactly as in
the singular case.

For $n=2$, the angular variable is a single scalar $\theta_2$, and $c_2(\theta_2)$
is a quadratic. Duffy transformation followed by independent
1D sinh transformations on $\theta_1$ and $\theta_2$ completely regularises
the integral. For $n>2$, however, the angular variables $\theta_2,\dots,\theta_n$
are coupled through $c_2$ (Eq.~\ref{eq:c2_uniform}). The near-singular manifold
$\mathcal{M} = \{c_2 \approx 0\}$ has dimension $n-2 \ge 1$. As argued in
Section~\ref{sec:duffy}, a tensor-product rule cannot resolve such a manifold.
The Duffy-domain strategy therefore does not extend to $n > 2$ for nearly
singular integrals. This
geometric obstruction motivates the physical Cartesian domain treatment developed next.

% ======================================================================
\subsection{Nearly singular integrals: the physical-domain approach}
\label{sec:physical}
% ======================================================================

To overcome the geometric obstruction, we abandon the Duffy parametrisation and
work directly in the physical (Cartesian) domain. The integration domain is
subdivided into $2^n$ hyper-rectangles, each sharing a vertex at
$\bar{\mathbf{x}}_0$ (Figure~\ref{schematic_carte}). The integral becomes
\begin{equation}
    I = \sum_{l=1}^{m_2} I_l = \sum_{l=1}^{m_2}
    \int_{a_1}^{b_1}\!\cdots\!\int_{a_n}^{b_n}
    \frac{f(\mathbf{x})}{\|\mathbf{r}_0+\mathbf{x}(\bm{\xi}) - \mathbf{x}_0\|^{\alpha}}
    \,\mathrm{d}\bm{\xi}^{(1)}\!\cdots\mathrm{d}\bm{\xi}^{(n)}. \label{eq:cart_sum}
\end{equation}
Singular integrals cannot be treated in this domain because the Duffy Jacobian
is absent.

Expanding $\mathbf{x} = \mathbf{x}_0 + \sum_{k=1}^{n} \mathbf{S}^{(k)}\bm{\xi}^{(k)}$
and using $\mathbf{r}_0\!\cdot\!\mathbf{S}^{(k)} = 0$,
\begin{equation}
    I_l = \int_{a_1}^{b_1}\!\cdots\!\int_{a_n}^{b_n}
    \frac{f(\mathbf{x})}
    {\bigl(r_0^2 + \sum_{j=1}^{n}\sum_{k=1}^{n} \mathbf{S}^{(j)}\!\cdot\!\mathbf{S}^{(k)}\,
      \bm{\xi}^{(j)}\bm{\xi}^{(k)}\bigr)^{\alpha/2}}
    \,\mathrm{d}x_1\!\cdots\mathrm{d}x_n. \label{eq:cart_denom}
\end{equation}

Defining $c_k = \mathbf{S}^{(k)}\!\cdot\!\mathbf{S}^{(k)}$, the near-singular
structure in each coordinate direction takes form
\begin{equation}
	\frac{1}{d_0^2 + c_k (\bm{\xi}^{(k)})^2}. \label{eq:cart_canonical}
\end{equation}
where $d_0^2 = r_0^2 + \sum_{j \neq k} c_j (\bm{\xi}^{(j)})^2$ aggregates the squared distance $r_0^2$ and the diagonal contributions except for $c_k (\bm{\xi}^{(k)})^2$. The cross-coupling terms through $\mathbf{S}^{(j)}\!\cdot\!\mathbf{S}^{(k)}$ for $j \neq k$ are neglected under the assumption of an affine (or nearly affine) geometric mapping; see Remark~\ref{rem:cross_terms}. A direct transformation of this coupled structure compromises the separability of the integration variables, which prevents the use of tensor-product multidimensional Gaussian quadrature. However, it can be seen that the sharpest near-singularity occurs when the contribution of the other variables to $d_0^2$ is minimal, since each diagonal term is non-negative. As a result, the above structure degenerates into
\begin{equation}
	\frac{1}{r_0^2 + c_k (\bm{\xi}^{(k)})^2}. \label{eq:cart_canonical_degenerated}
\end{equation}
At this point, the transformation to be performed is independent of the other variables, and the nearly singularity is effectively decoupled.

In summary, the physical-domain approach resolves the geometric obstruction
identified in Section~\ref{sec:duffy_nearly}. By subdividing the domain into
hyper-rectangles and expanding the geometry in Cartesian coordinates, the
near-singularities align with the coordinate axes. Each direction then carries
an independent 1D near-singularity of the canonical form
$(r_0^2 + c_k\xi_k^2)^{-\alpha/2}$, which can be treated by a 1D sinh
transformation. This provides a robust method for nearly singular integrals in
arbitrary dimension $n$. The price is that singular integrals cannot be treated
in this domain, since the Duffy Jacobian which provides the cancellation of
the $1/r^\alpha$ singularity is absent.

\begin{remark}[Cross-terms in the physical domain]
\label{rem:cross_terms}
The derivation of~\eqref{eq:cart_canonical} neglects the cross-terms
$\mathbf{S}^{(j)}\!\cdot\!\mathbf{S}^{(k)}\bm{\xi}^{(j)}\bm{\xi}^{(k)}$ for
$j \neq k$. This is equivalent to assuming that the isoparametric Jacobian
matrix $[\mathbf{S}^{(k)}]$ is diagonal at $\bar{\mathbf{x}}_0$, i.e.\ the
element edges are orthogonal at the projection point. For general curved or
non-orthogonal elements, these cross-terms contribute additional coupling.
A rigorous treatment requires either a coordinate diagonalisation or an
estimate of the cross-term contribution, which we defer to future work.
\end{remark}

% ======================================================================
\subsection{The sinh transformation and its competitors}
\label{sec:sinh}
% ======================================================================

The canonical near-singular structure $(r_0^2 + c\,\xi^2)^{-\alpha/2}$ that
appears in both the Duffy domain (for $\theta_1$ and, when $n=2$, for $\theta_2$)
and the physical Cartesian domain (for each $\xi^{(k)}$) motivates a unified treatment.
We exploit the Jacobian of a nonlinear change of variables to eliminate or
weaken the near-singular behaviour.

\subsubsection{The Gauss hypergeometric family}

For an integrand of the prototypical form
\begin{equation}
    \frac{1}{\bigl(\sqrt{t_1(\theta+b)^2 + d^2}\,\bigr)^{\beta}},
    \qquad t_1, d > 0, \;\; b \le 0, \label{eq:proto_integrand}
\end{equation}
the unified transformation is
\begin{equation}
    \Theta(\theta) = d^{-\beta}\Bigl(
    -b\,{}_2F_1\!\Bigl(\frac12,\frac{\beta}{2};\frac32;-\frac{b^2 t_1}{d^2}\Bigr)
    + (b+\theta)\,{}_2F_1\!\Bigl(\frac12,\frac{\beta}{2};\frac32;-\frac{t_1(b+\theta)^2}{d^2}\Bigr)
    \Bigr), \label{eq:trans_hyper}
\end{equation}
where ${}_2F_1$ is the Gauss hypergeometric function. The transformation
requires an explicit inverse for use with Gauss--Legendre quadrature. Closed-form
inverses exist only for $\beta = 1,2,3$, corresponding to the sinh, tangent,
and algebraic transformations. For $b = 0$ (the case relevant to the canonical
form), the inverse functions and their Jacobians are
\begin{equation}
    \theta(\Theta) =
    \begin{cases}
        \dfrac{d}{\sqrt{t_1}}\sinh(\sqrt{t_1}\,\Theta), & \beta=1 \\[12pt]
        \dfrac{d}{\sqrt{t_1}}\tan(d\sqrt{t_1}\,\Theta), & \beta=2 \\[12pt]
        \dfrac{d^3\Theta}{\sqrt{1 - t_1 d^4\Theta^2}}, & \beta=3,
    \end{cases}
    \qquad
    J(\Theta) =
    \begin{cases}
        d\cosh(\sqrt{t_1}\,\Theta), & \beta=1 \\[6pt]
        d^2\sec^2(d\sqrt{t_1}\,\Theta), & \beta=2 \\[6pt]
        \dfrac{d^3}{(1 - t_1 d^4\Theta^2)^{3/2}}, & \beta=3.
    \end{cases}
\end{equation}

When $\alpha = \beta$, the transformation achieves exact cancellation of the
denominator, leaving only the smooth numerator and the Jacobian. It is
therefore expected that matching $\beta$ to $\alpha$ yields optimal performance, however the convergence analysis below shows this intuition to be incorrect for
$\beta = 2,3$.

\subsubsection{Convergence of the sinh transformation ($\beta=1$)}

For an analytic integrand, the $m$-point Gauss--Legendre error is bounded by
$C\rho^{-2m}$, where $\rho = |p_s + \sqrt{p_s^2-1}|$ is the Bernstein ellipse
parameter associated with the nearest complex singularity $p_s$
of the transformed integrand~\cite{trefethen2019approximation,zhao2013sharp,
xiang2012convergence,xie2013exponential}. The singularities of $(r_0^2 + c\xi^2)^{-\alpha/2}$ lie at
$\xi_s = \pm i r_0/\sqrt{c}$. Mapping to $p = 2\xi-1 \in [-1,1]$ gives
$p_s = -1 \pm 2i r_0/\sqrt{c}$, with $\rho_0 = 1 + 2r_0/\sqrt{c} + O(r_0^2)
\to 1$ as $r_0 \to 0$. Standard Gauss--Legendre therefore fails.

\textbf{Sinh transformation.} After applying the sinh transformation, the upper
limit is
\begin{equation}
    \Theta_1 = \frac{\arcsinh(\sqrt{c}/r_0)}{\sqrt{c}}
    \;\xrightarrow{r_0\to 0}\; \frac{\ln(2\sqrt{c}/r_0)}{\sqrt{c}}.
\end{equation}
The Jacobian $\cosh(\sqrt{c}\,\Theta)$ has its nearest singularity at
$\Theta_s = i\pi/(2\sqrt{c})$. Mapping to the $p$-domain:
\begin{equation}
    p_s^{(\beta=1)} = \frac{2\Theta_s}{\Theta_1} - 1
    = \frac{i\pi}{\arcsinh(\sqrt{c}/r_0)} - 1.
\end{equation}
The imaginary part approaches zero only at a logarithmic rate as
$r_0 \to 0$. The Bernstein parameter therefore departs from unity slowly,
preserving exponential convergence for practical $m$. The $\arcsinh$ function
acts as a regulariser: it pushes complex singularities away from the real axis.

For $\alpha > 1$, the denominator retains a factor $\cosh^{\alpha}(\sqrt{c}\,\Theta)$,
so cancellation is not exact. The logarithmic mechanism still ensures the
transformed integrand is substantially smoother than the original, and the
imaginary part of $p_s$ does not tend to zero.

\begin{figure}[!htbp]
    \centering
    \includegraphics[width=0.4\textwidth]{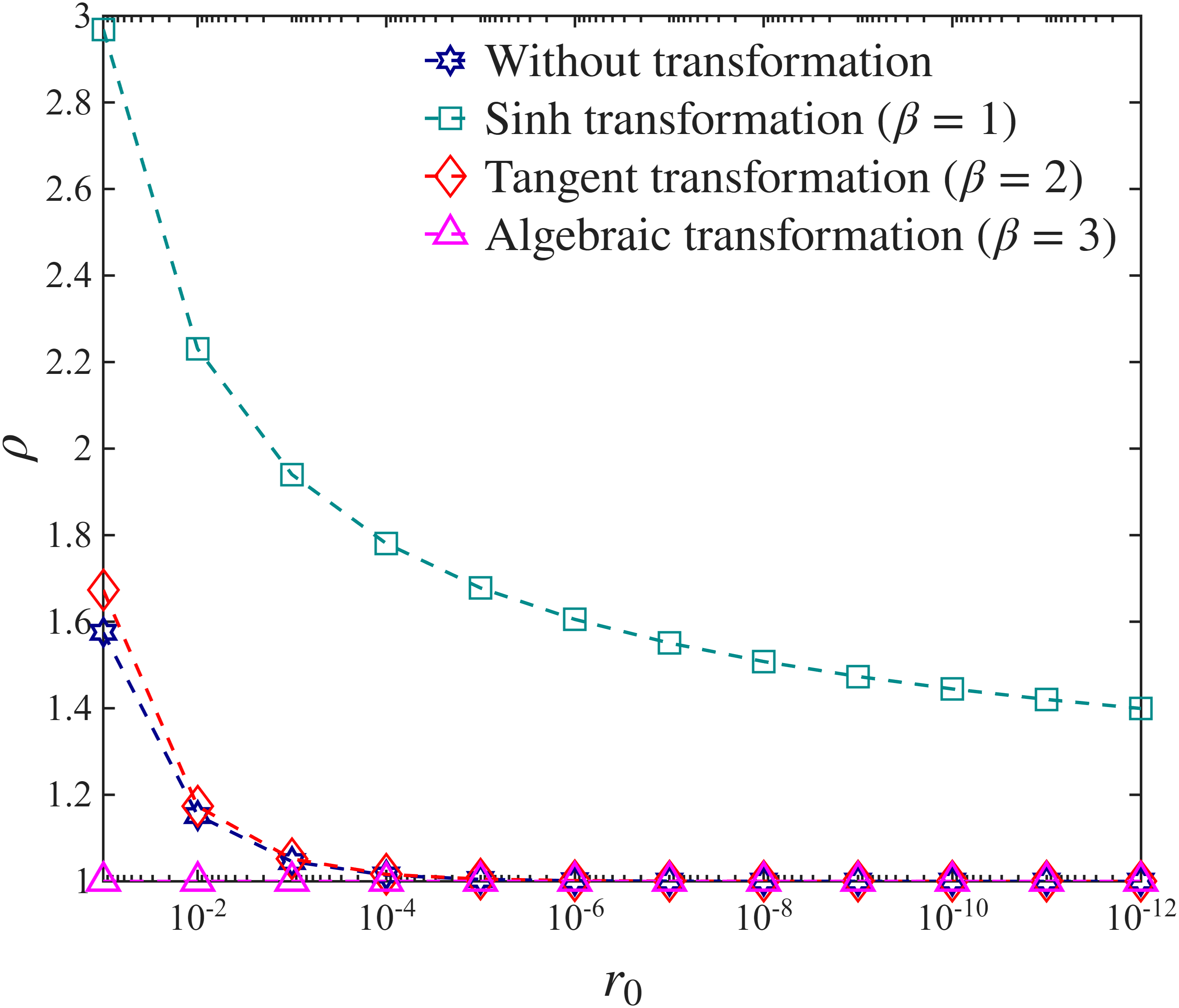}
    \caption{Bernstein ellipse parameter of the three hypergeometric
      transformations as a function of $r_0$ (with $c=1$). The sinh
      transformation ($\beta=1$) approaches unity at the slowest rate.}
    \label{fig:ellipse_para}
\end{figure}

\textbf{Tangent ($\beta=2$) and algebraic ($\beta=3$) transformations.}
The analysis (detailed in Appendix~\ref{apdx_beta23}) shows that both
transformations introduce new singularities through their Jacobians that map
to the real axis after mapping to $[-1,1]$, collapsing $\rho$ to
unity. The quadrature error therefore decays arbitrarily slowly. Figure~\ref{fig:ellipse_para} compares the Bernstein ellipse parameter as a
function of $r_0$ for all three transformations (with $c=1$).

In summary, among the three closed-form members of the Gauss hypergeometric
family, only the sinh transformation ($\beta=1$) consistently improves
convergence. The tangent ($\beta=2$) and algebraic ($\beta=3$) transformations
introduce secondary near-singularities through their Jacobians; after domain
mapping, the corresponding Bernstein ellipse parameters collapse to unity,
yielding arbitrarily slow convergence. The sinh transformation is therefore
the unique effective member of the family, and it is the transformation employed
throughout the remainder of this work.

% ======================================================================
\section{Numerical examples}
\label{sec:numerical}
% ======================================================================

To verify the proposed framework, we present computational examples in
dimensions two through five. In all cases, the sinh transformation ($\beta=1$)
is compared against direct Gauss--Legendre quadrature; for $n \ge 3$,  we further compare it with the tangent ($\beta=2$) and algebraic ($\beta=3$) transformations. Benchmark
values are obtained using a large number of Gauss points in the transformed
domain. The error is $E = |I - I_{\text{num}}|$.

\begin{figure}[!htbp]
    \centering
    \captionsetup[subfigure]{justification=centering}
    \subcaptionbox{$c=10^{-3}$, centred}{%
        \includegraphics[width=3.2cm,height=2.9cm]{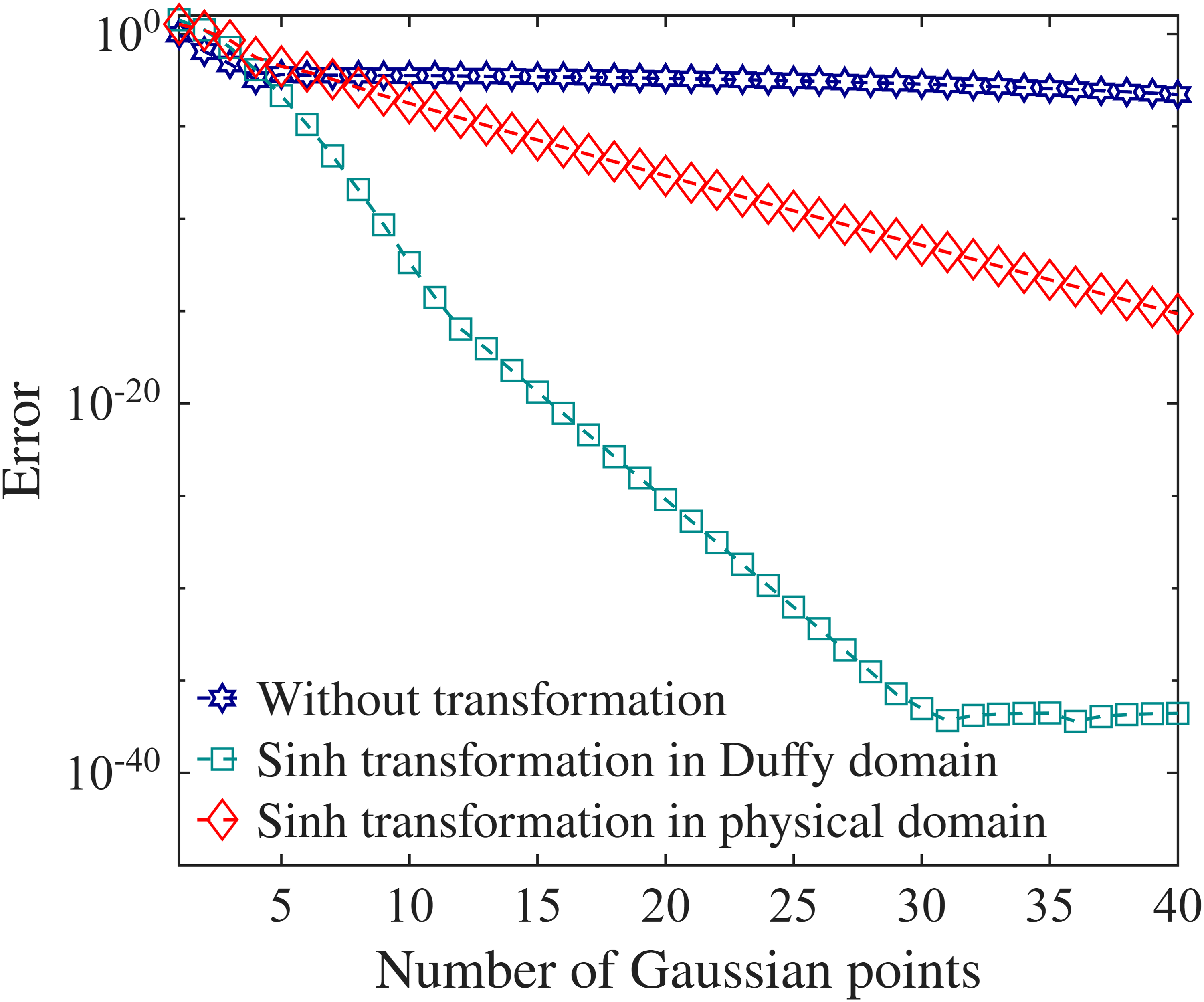}}\hfill
    \subcaptionbox{$c=10^{-4}$, centred}{%
        \includegraphics[width=3.2cm,height=2.9cm]{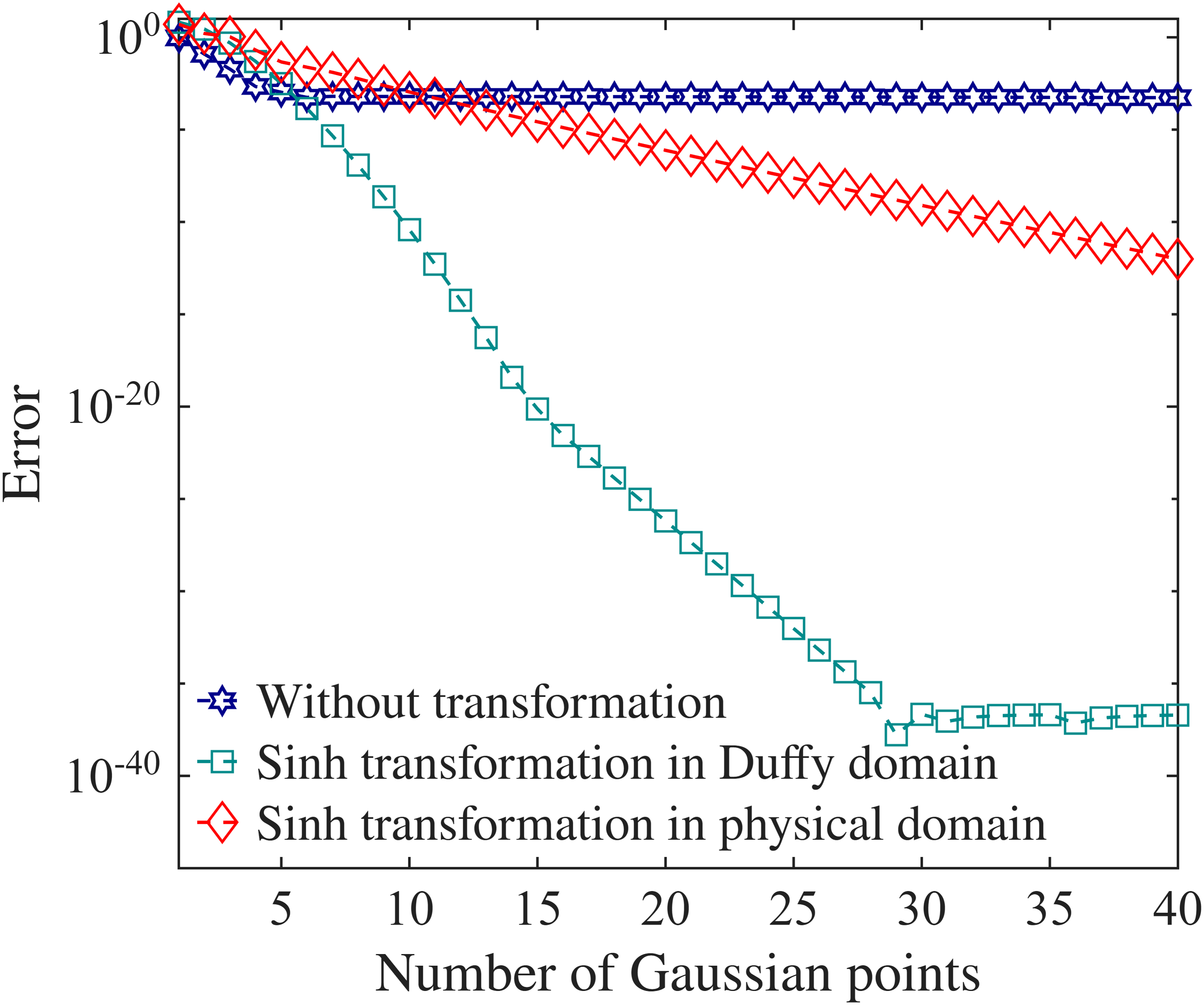}}\hfill
    \subcaptionbox{$c=10^{-5}$, centred}{%
        \includegraphics[width=3.2cm,height=2.9cm]{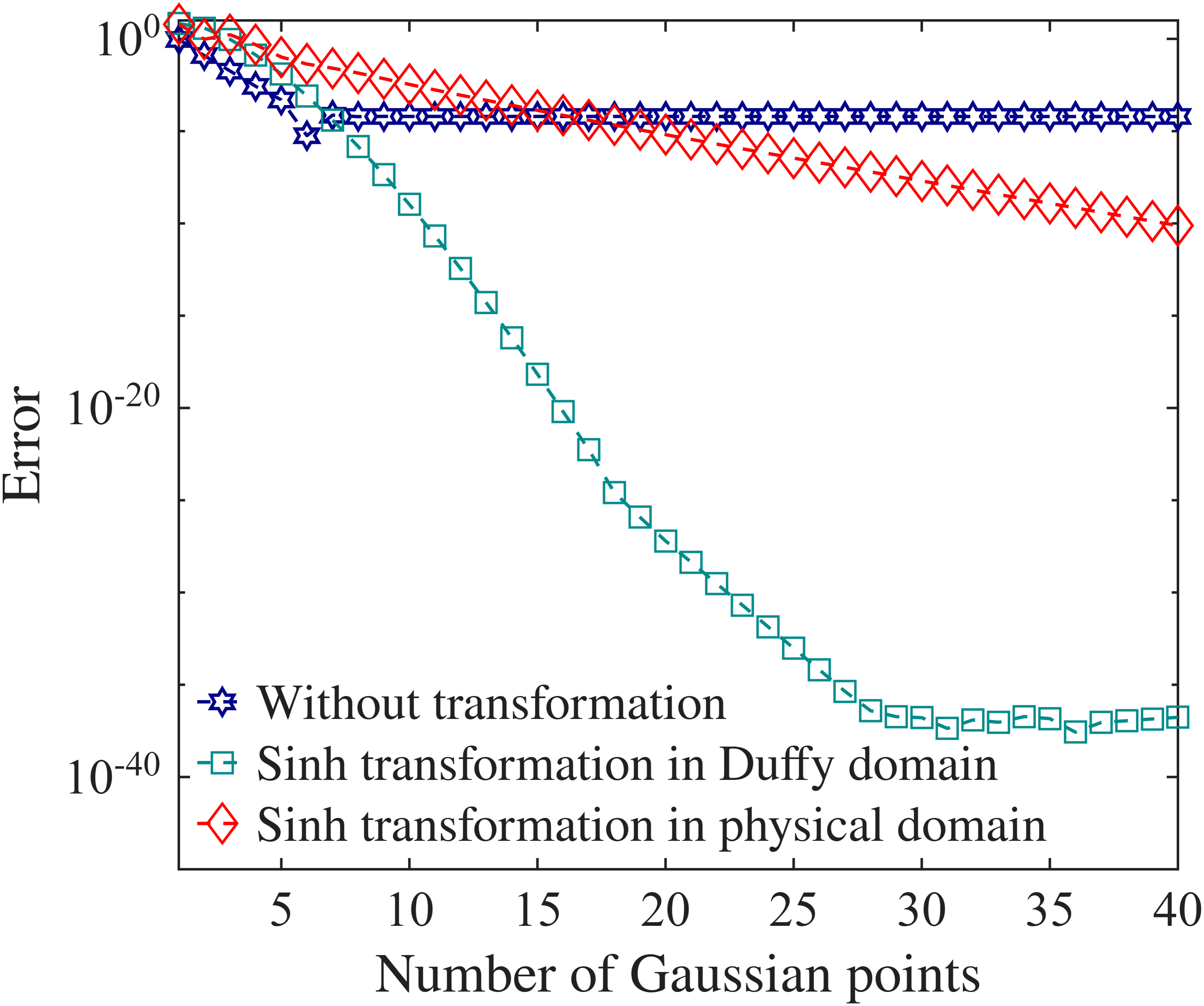}}\hfill
    \subcaptionbox{$c=10^{-6}$, centred}{%
        \includegraphics[width=3.2cm,height=2.9cm]{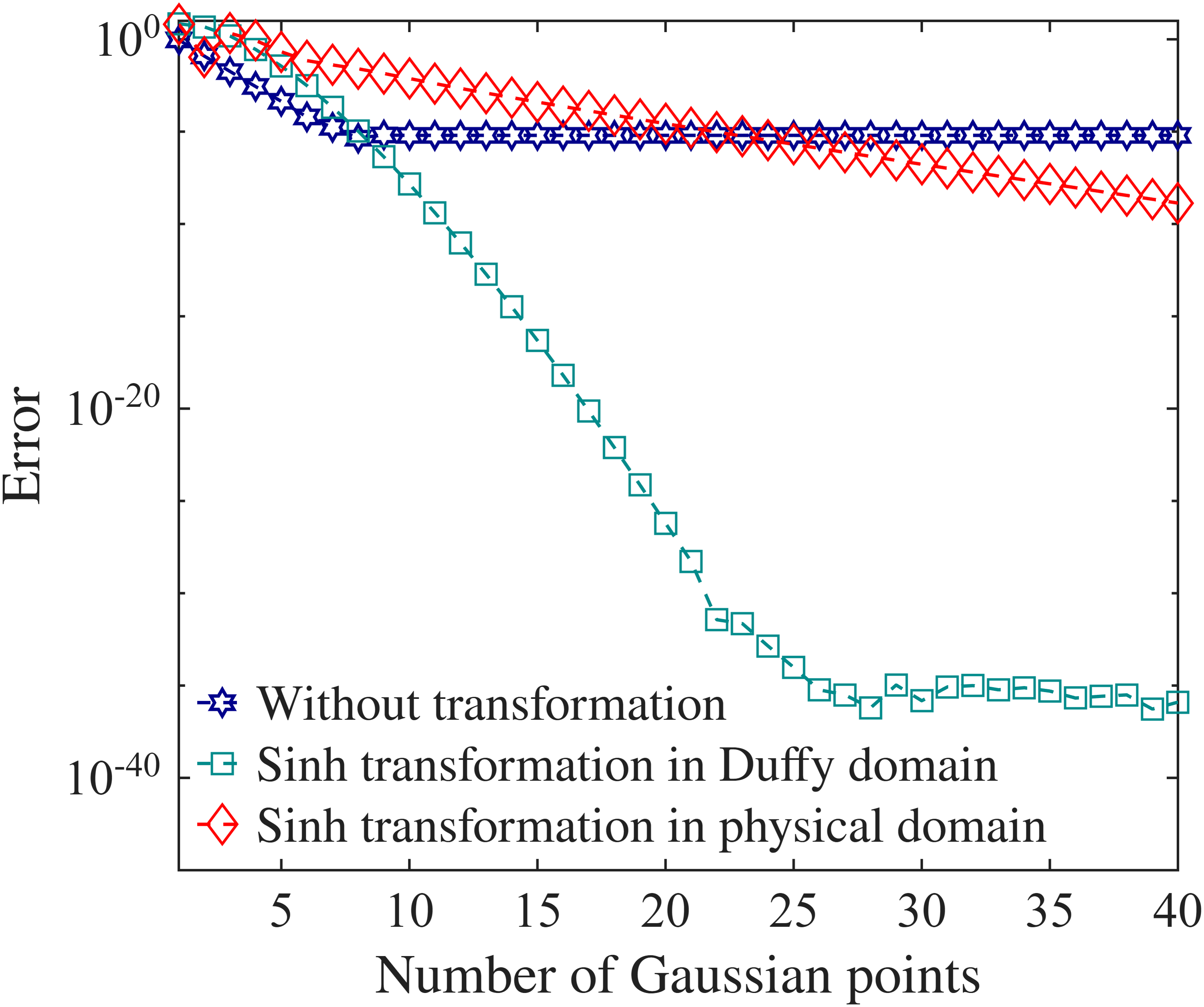}}\\
    \subcaptionbox{$c=10^{-3}$, near edge}{%
        \includegraphics[width=3.2cm,height=2.9cm]{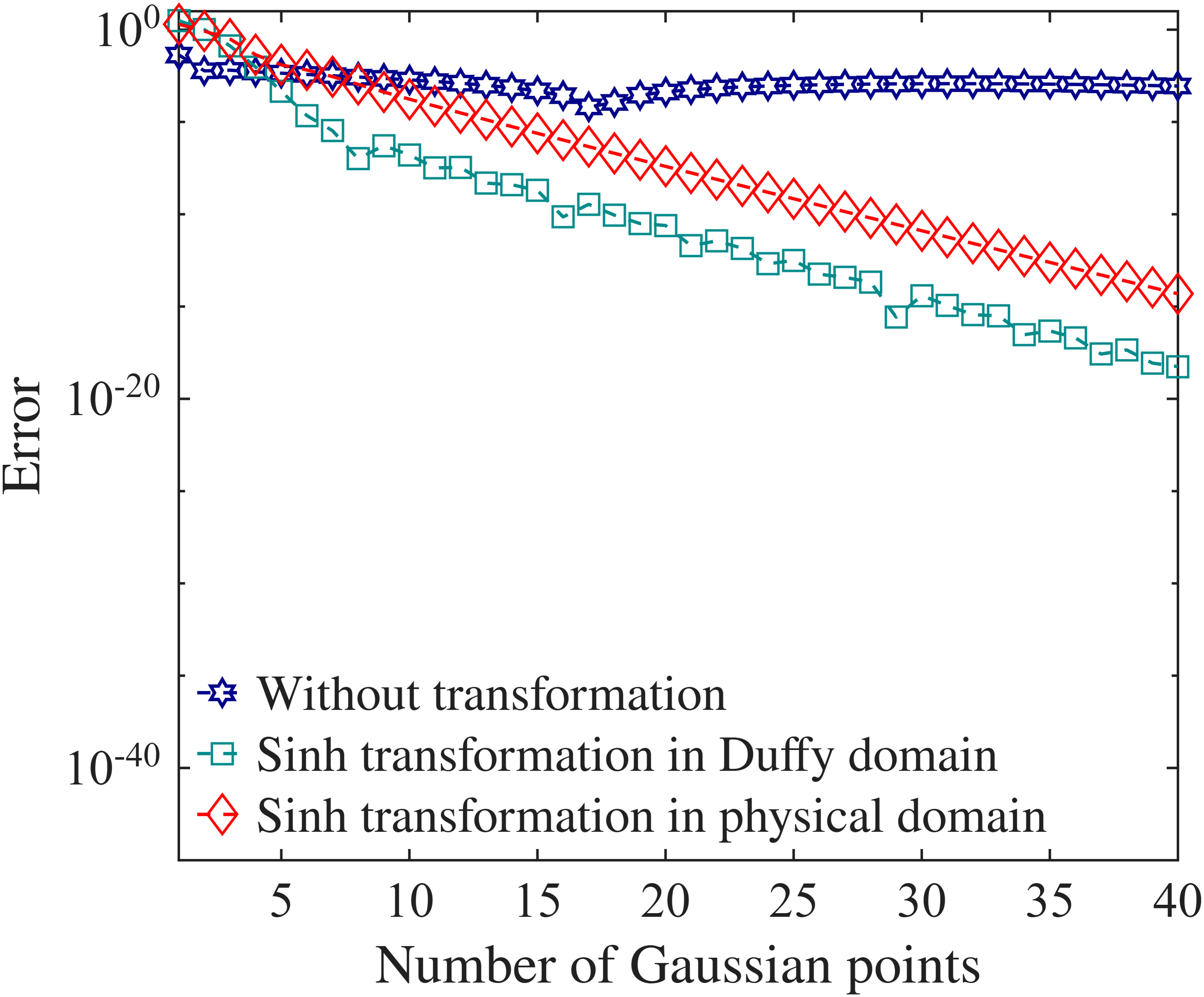}}\hfill
    \subcaptionbox{$c=10^{-4}$, near edge}{%
        \includegraphics[width=3.2cm,height=2.9cm]{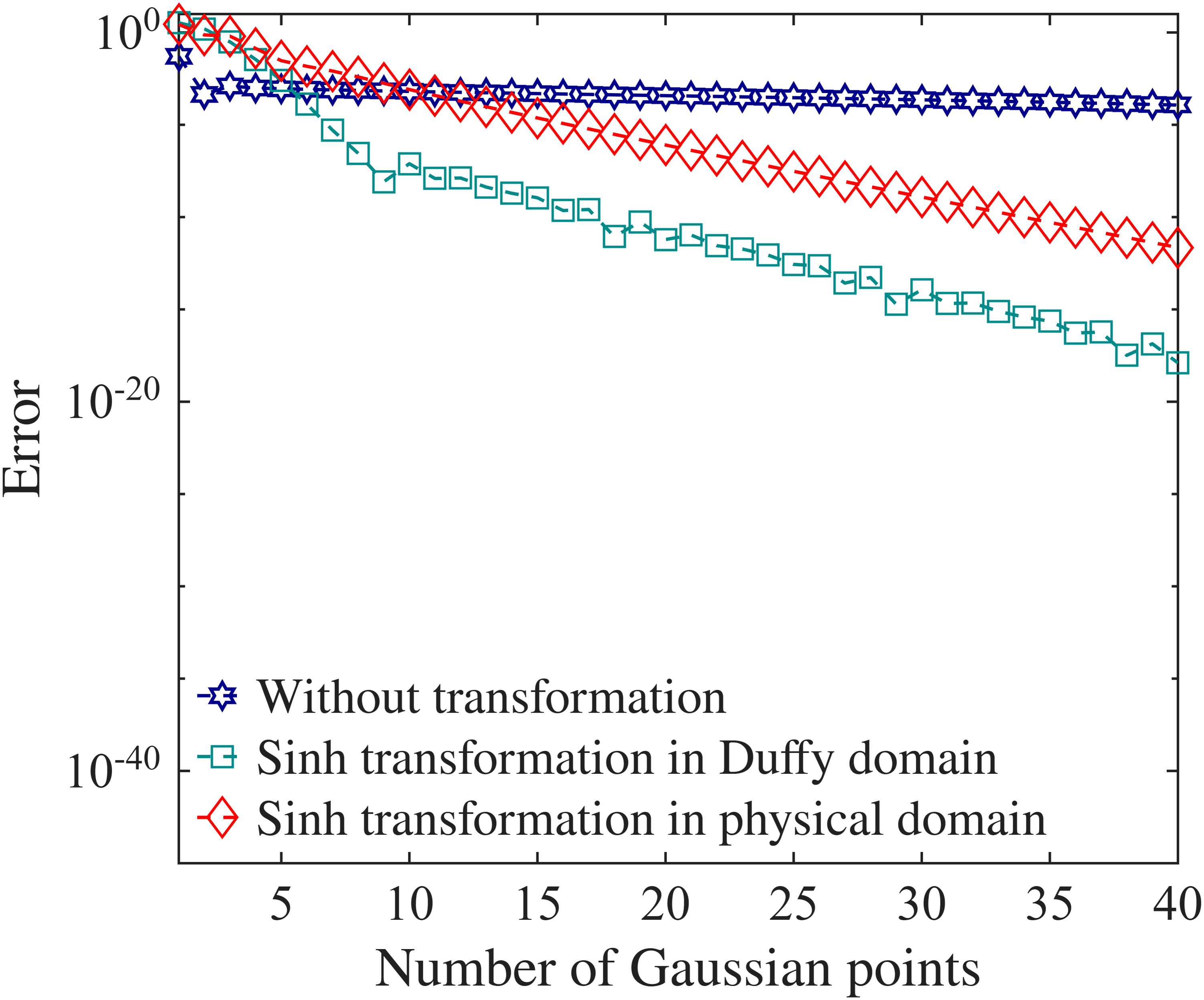}}\hfill
    \subcaptionbox{$c=10^{-5}$, near edge}{%
        \includegraphics[width=3.2cm,height=2.9cm]{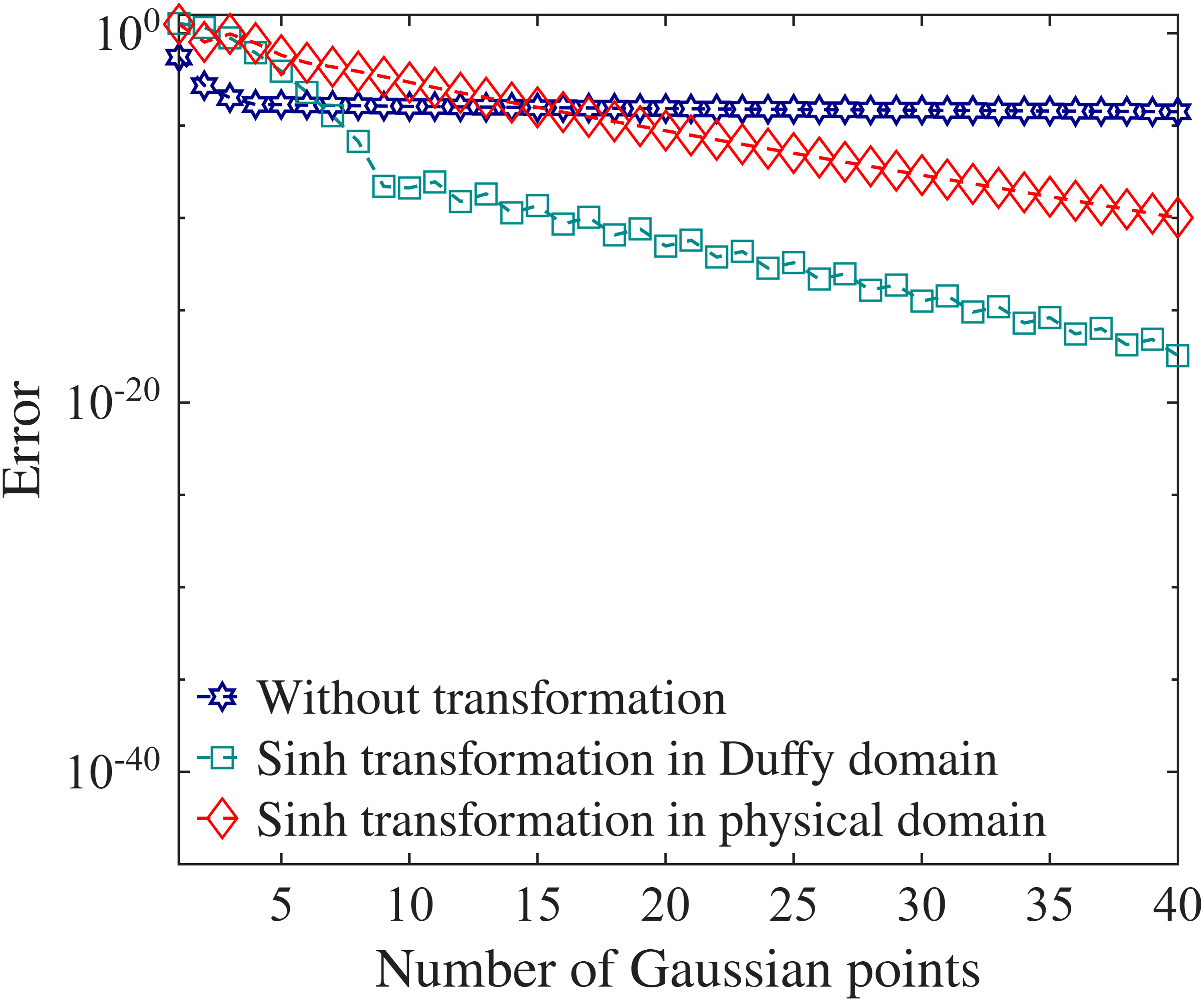}}\hfill
    \subcaptionbox{$c=10^{-6}$, near edge}{%
        \includegraphics[width=3.2cm,height=2.9cm]{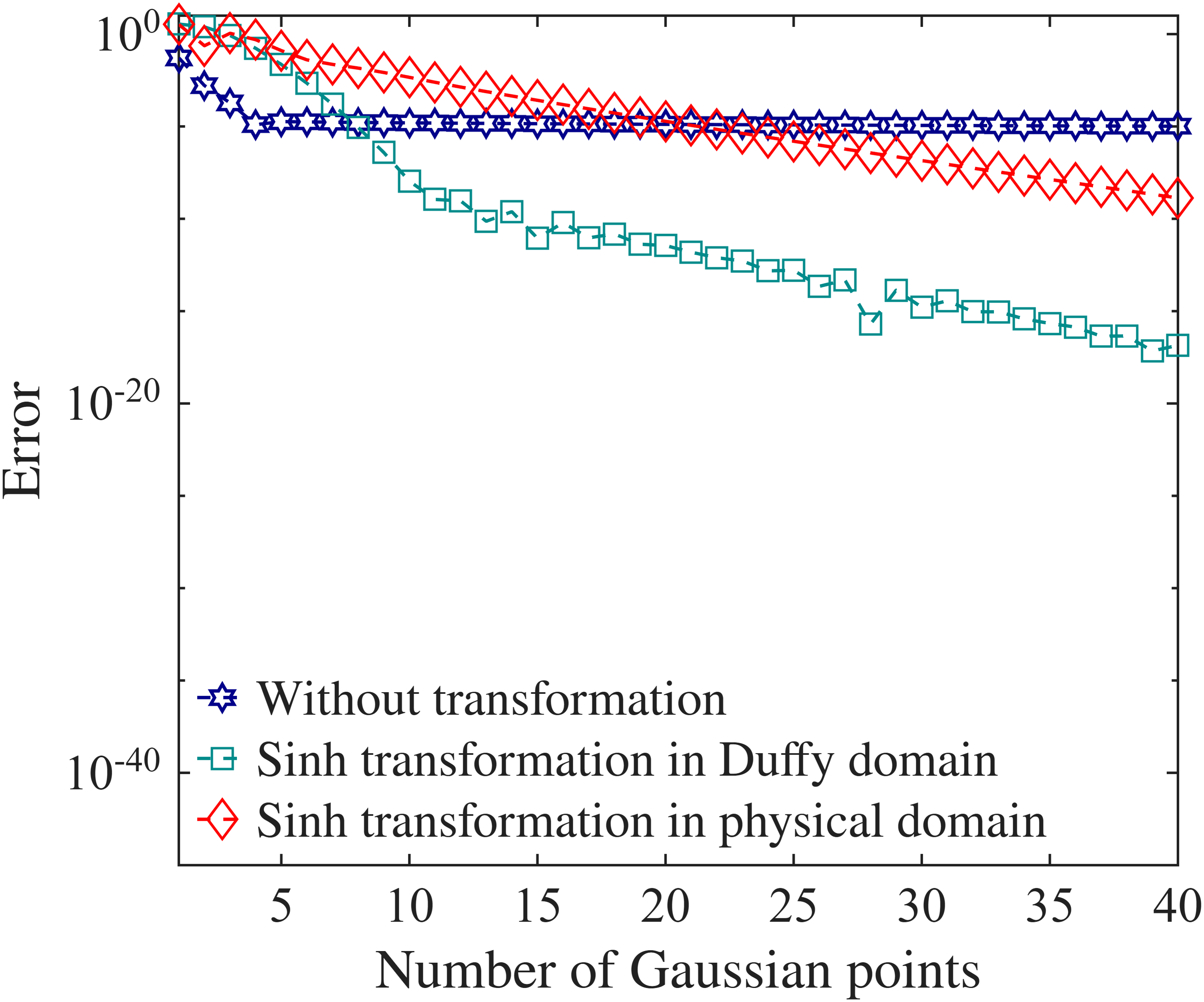}}
    \caption{Errors with and without sinh transformation, $n=2$. Top row: source
      centred ($e_1=e_2=0$). Bottom row: source near boundary
      ($e_1=e_2$ as indicated).}
    \label{fig:Double_integral}
\end{figure}

\paragraph{Double integral ($n=2$)}
Consider
\begin{equation}
    \iint_{[-1,1]^2} \frac{1}{\sqrt{(x-e_1)^2 + (y-e_2)^2 + c^2}} \,\mathrm{d}x\,\mathrm{d}y,
\end{equation}
the 2D weakly singular kernel arising in boundary element formulations of the
Laplace and Helmholtz equations~\cite{colton2013integral,sauter2011bem}, heat
conduction~\cite{carslaw1959heat}, Darcy flow~\cite{hu2022darcy},
and fracture mechanics~\cite{bonnet1999bie,hu2024hydraulic}.
Figure~\ref{fig:Double_integral} shows results for $e_1 = e_2 = 0$
(source centred, top row) and $e_1, e_2$ close to the boundary (bottom row),
both with and without the sinh transformation in the Duffy domain and physical (Cartesian) domain.
Applying the sinh transformation in the Duffy domain
yields better performance than in the physical domain, owing to additional
regularisation from the Duffy Jacobian.

\begin{figure}[!htbp]
    \centering
    \captionsetup[subfigure]{justification=centering}
    \subcaptionbox{$c=10^{-3}$, $n=3$}{%
        \includegraphics[width=3.0cm,height=2.7cm]{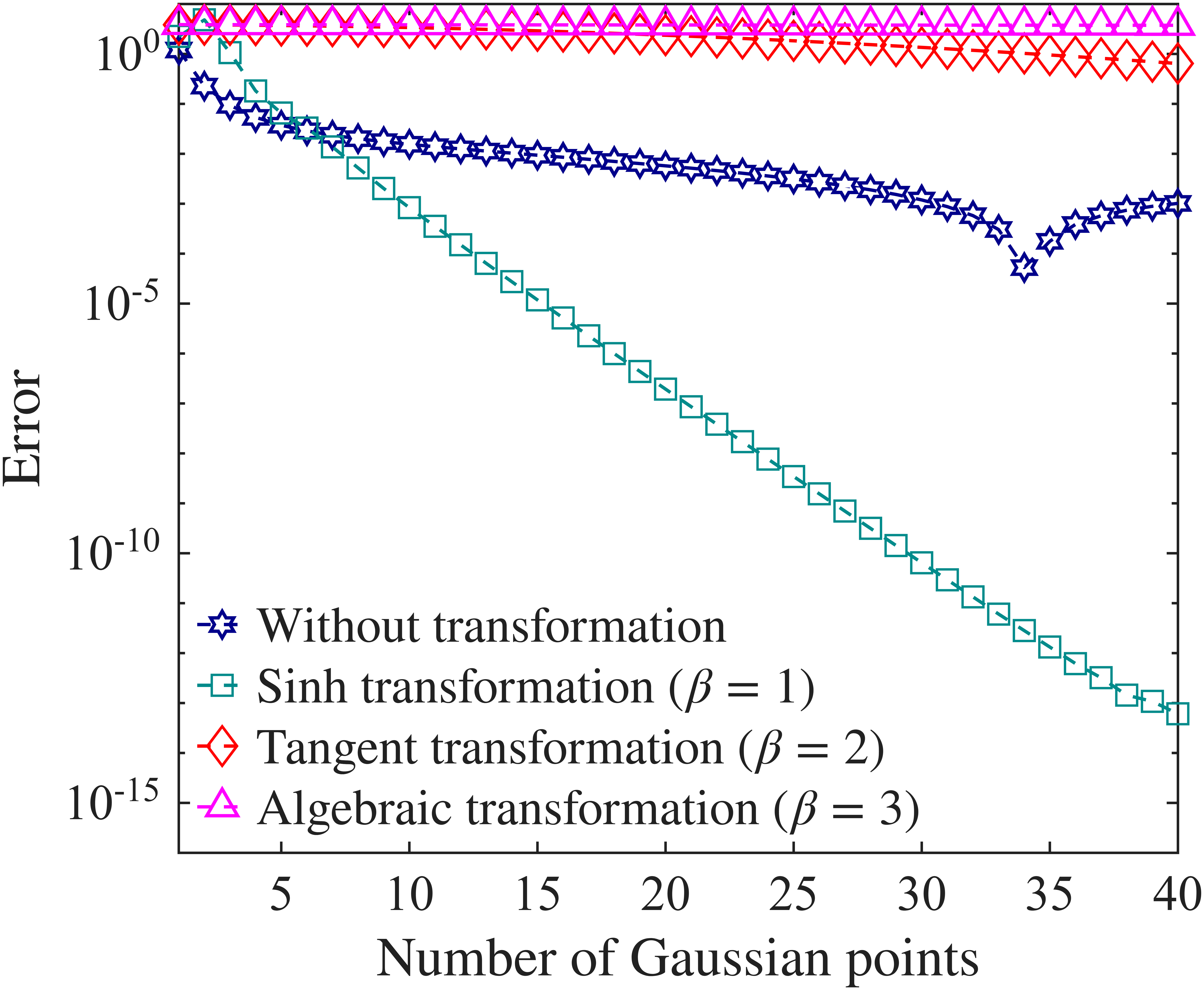}}\hfill
    \subcaptionbox{$c=10^{-4}$, $n=3$}{%
        \includegraphics[width=3.0cm,height=2.7cm]{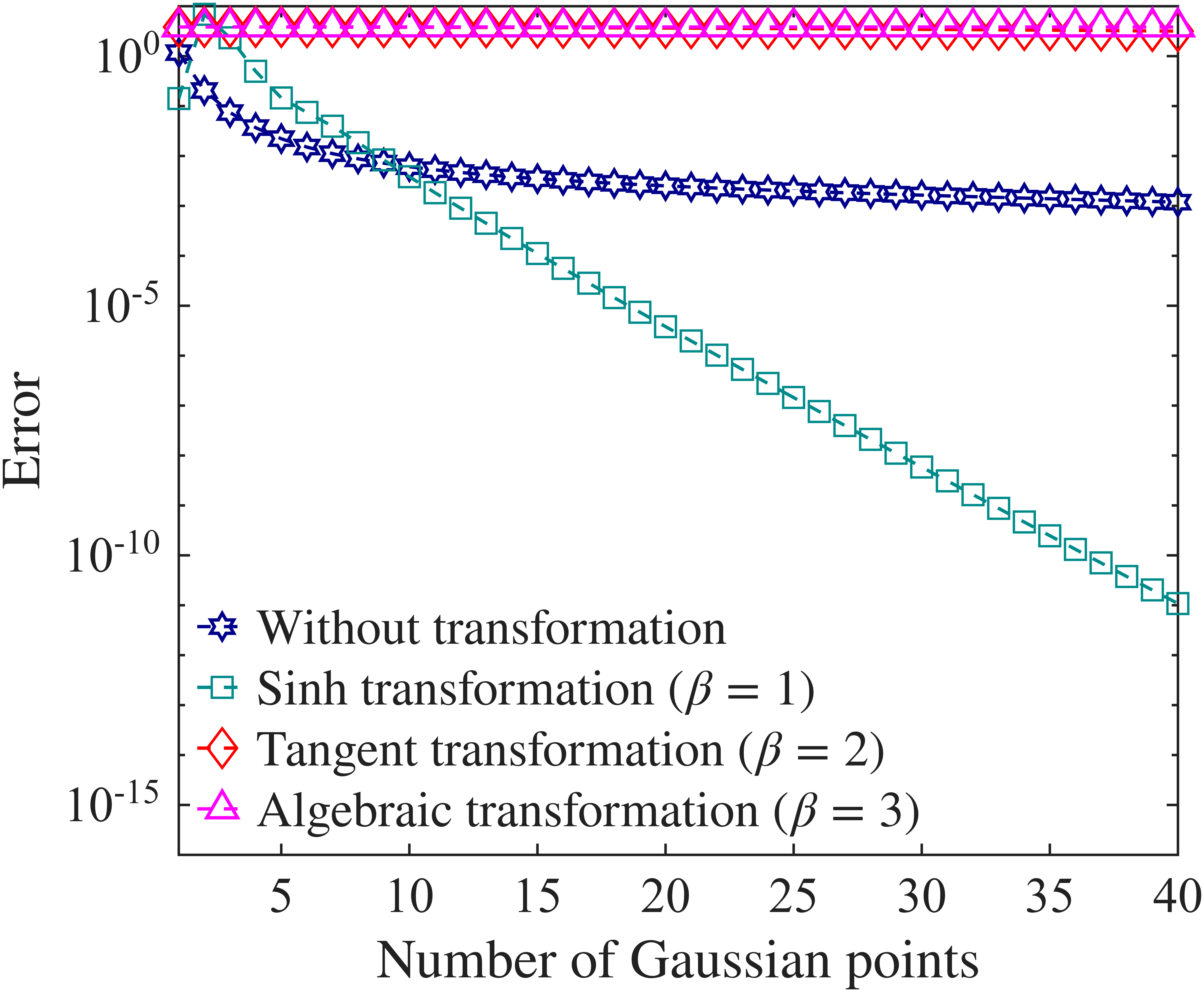}}\hfill
    \subcaptionbox{$c=10^{-5}$, $n=3$}{%
        \includegraphics[width=3.0cm,height=2.7cm]{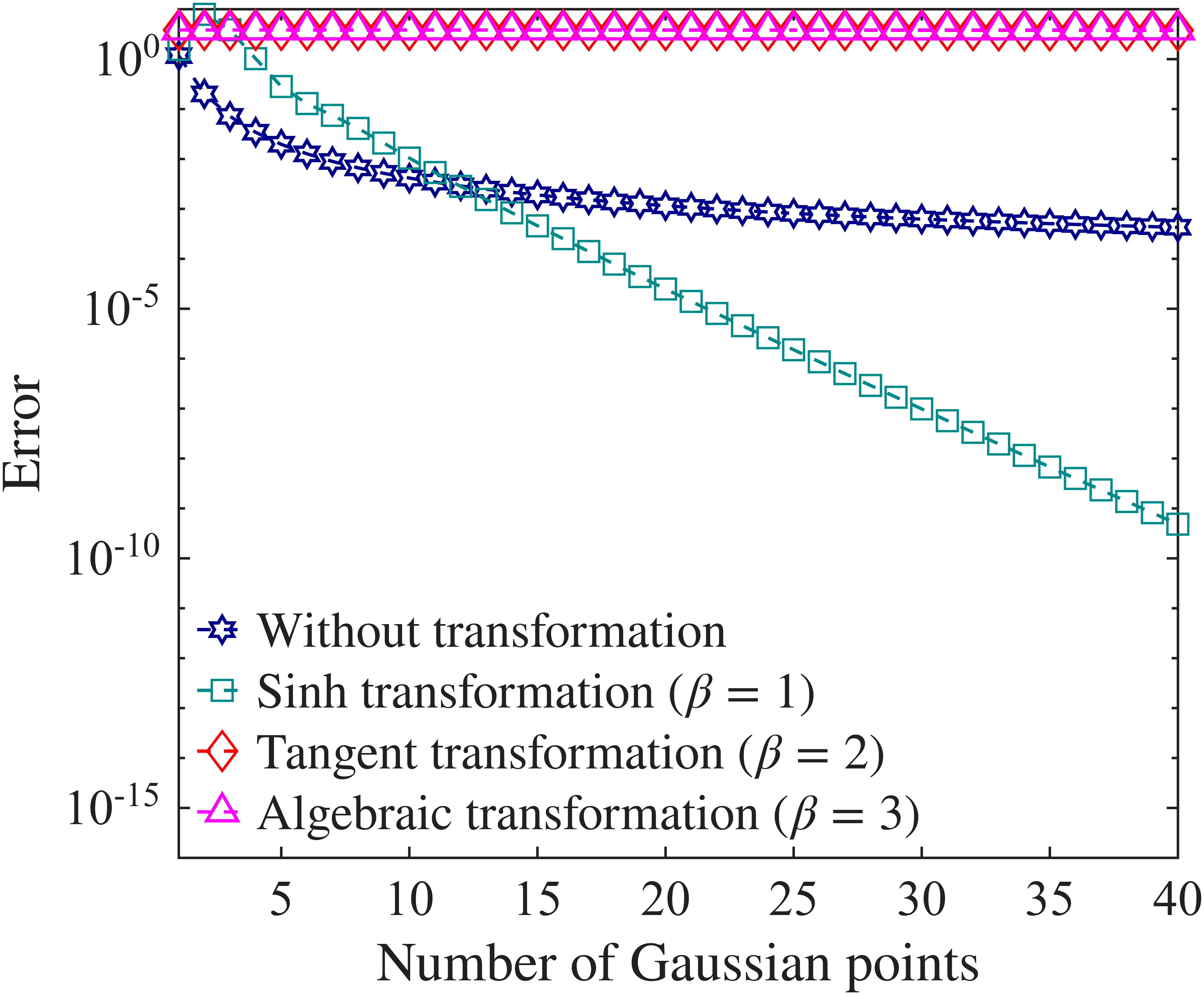}}\hfill
    \subcaptionbox{$c=10^{-6}$, $n=3$}{%
        \includegraphics[width=3.0cm,height=2.7cm]{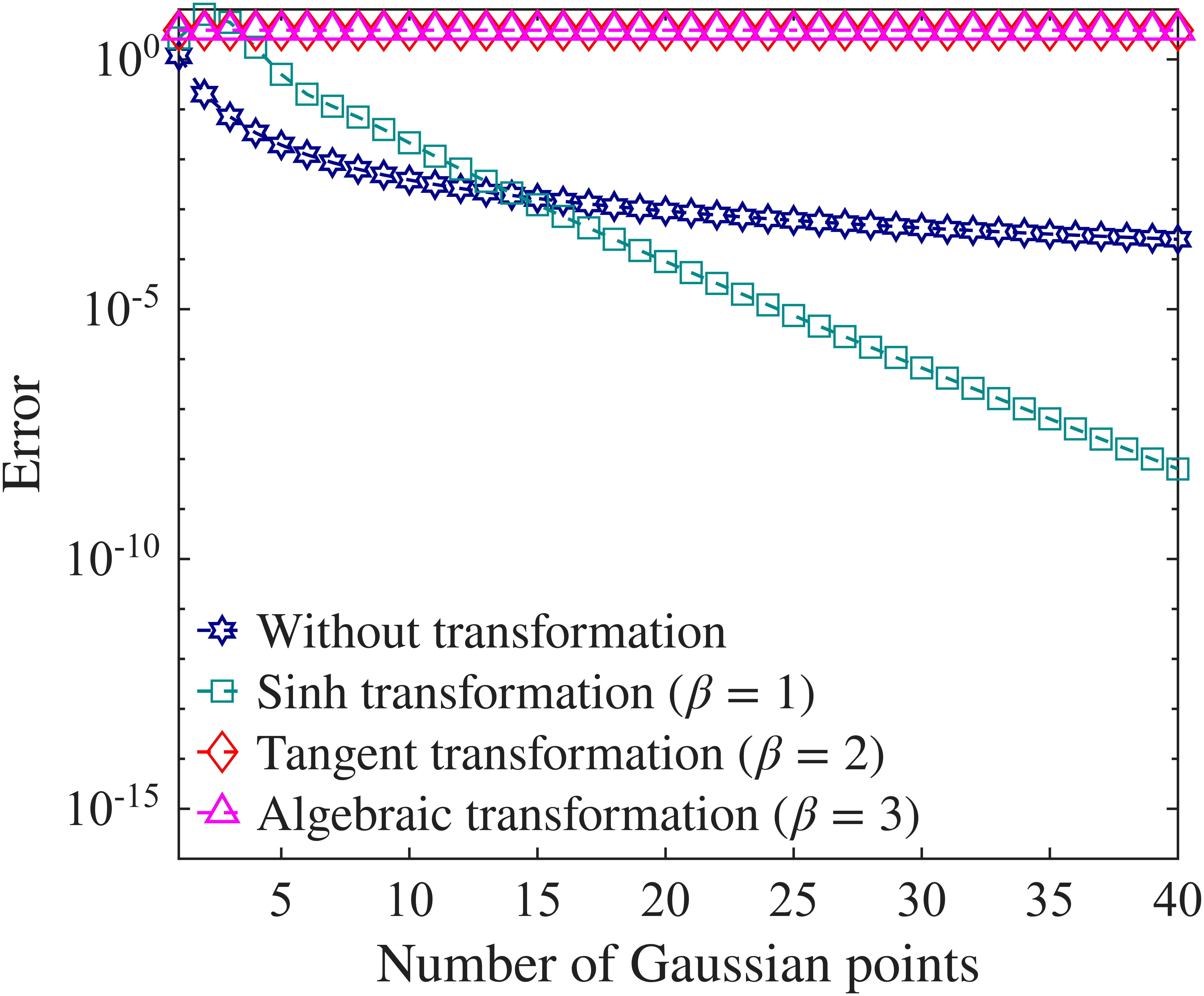}}\\
    \subcaptionbox{$c=10^{-3}$, $n=4$}{%
        \includegraphics[width=3.0cm,height=2.7cm]{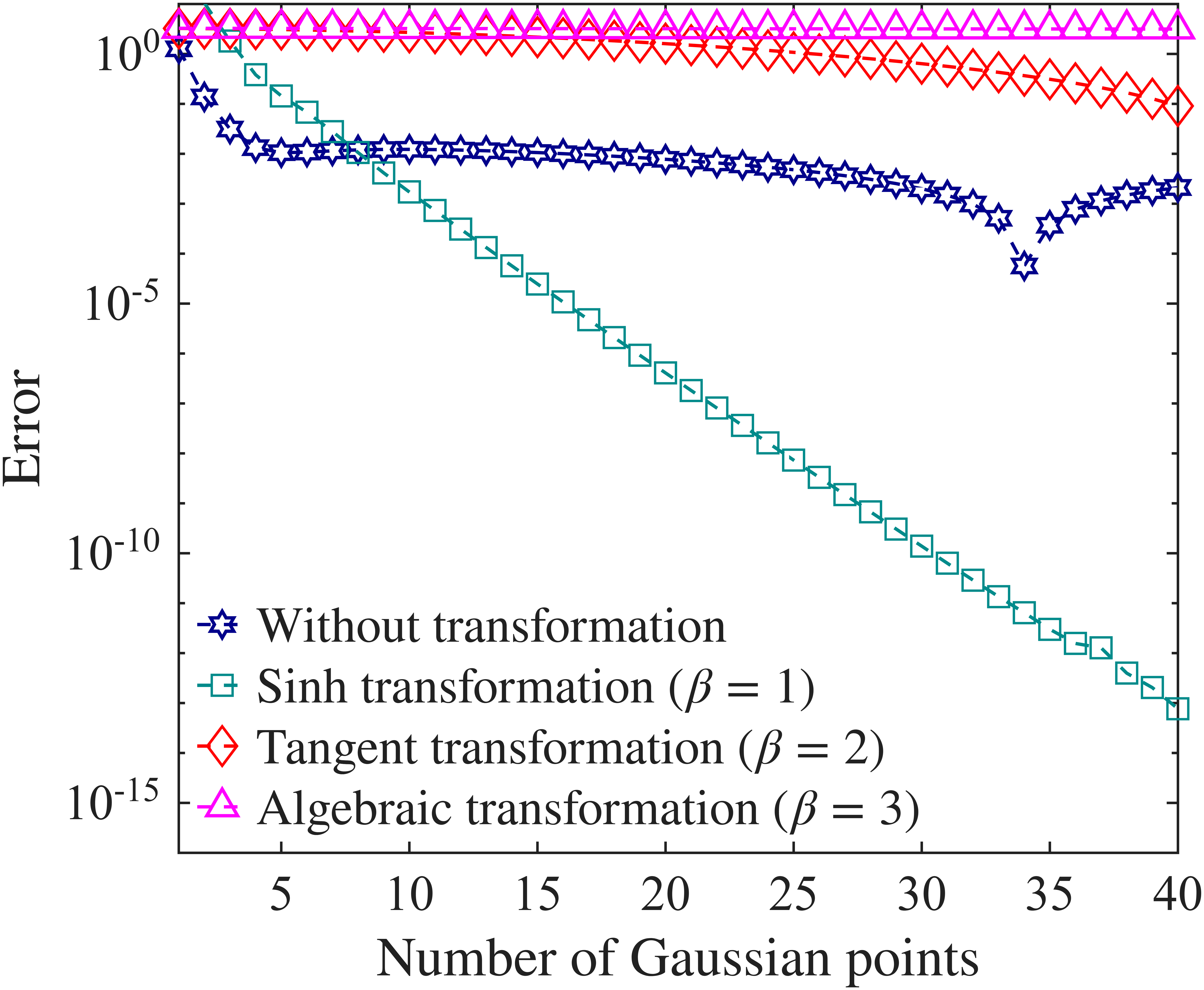}}\hfill
    \subcaptionbox{$c=10^{-4}$, $n=4$}{%
        \includegraphics[width=3.0cm,height=2.7cm]{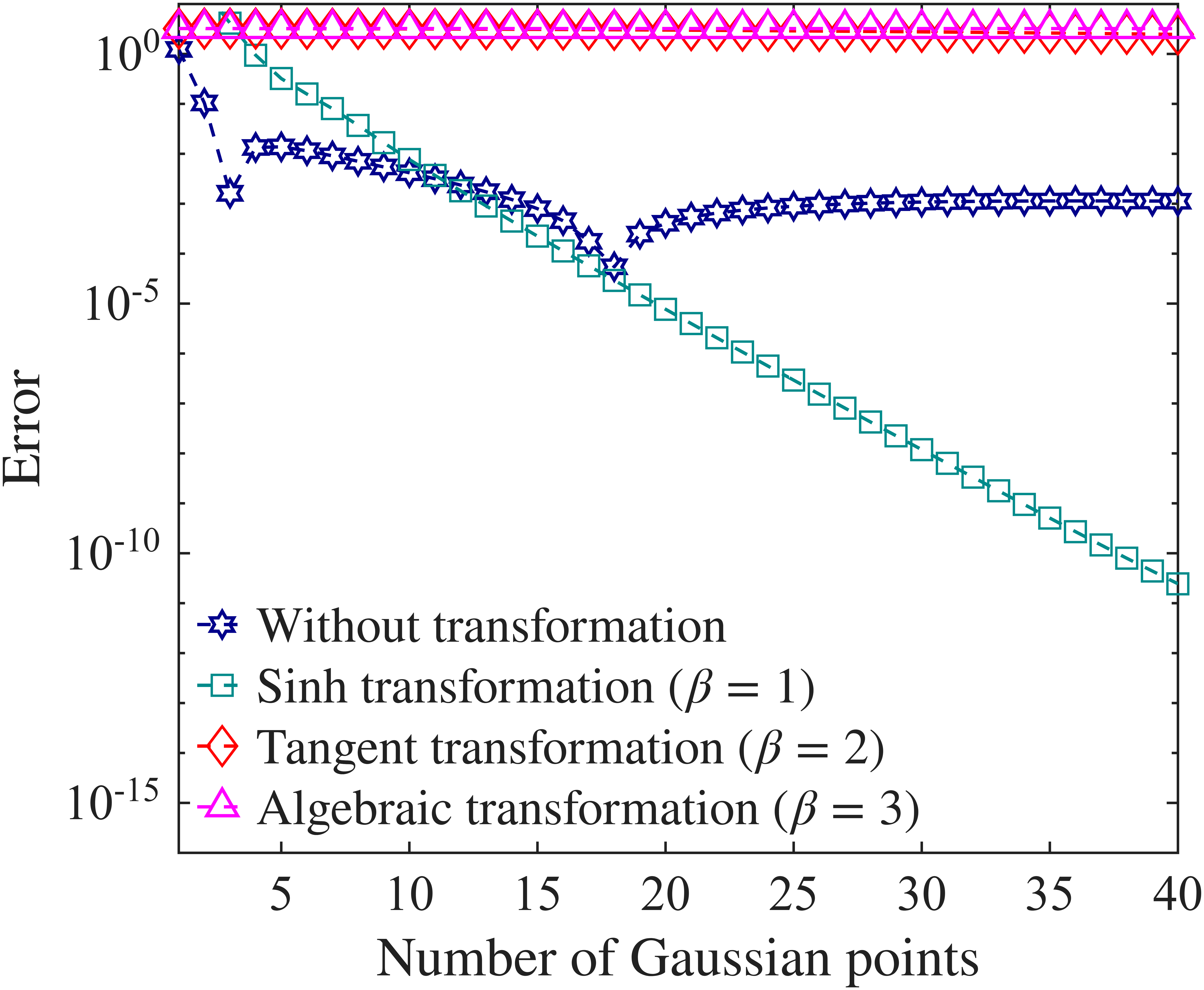}}\hfill
    \subcaptionbox{$c=10^{-5}$, $n=4$}{%
        \includegraphics[width=3.0cm,height=2.7cm]{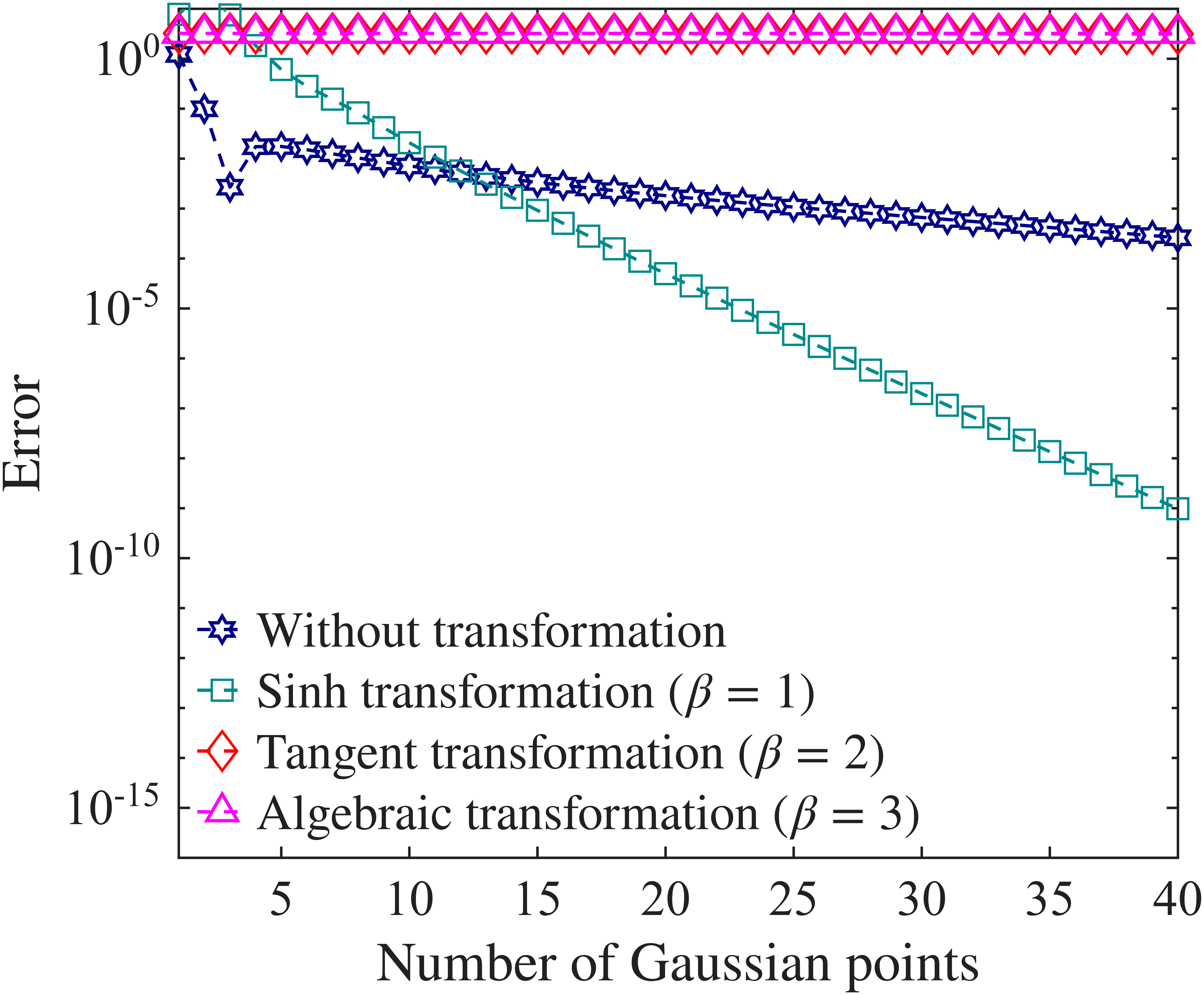}}\hfill
    \subcaptionbox{$c=10^{-6}$, $n=4$}{%
        \includegraphics[width=3.0cm,height=2.7cm]{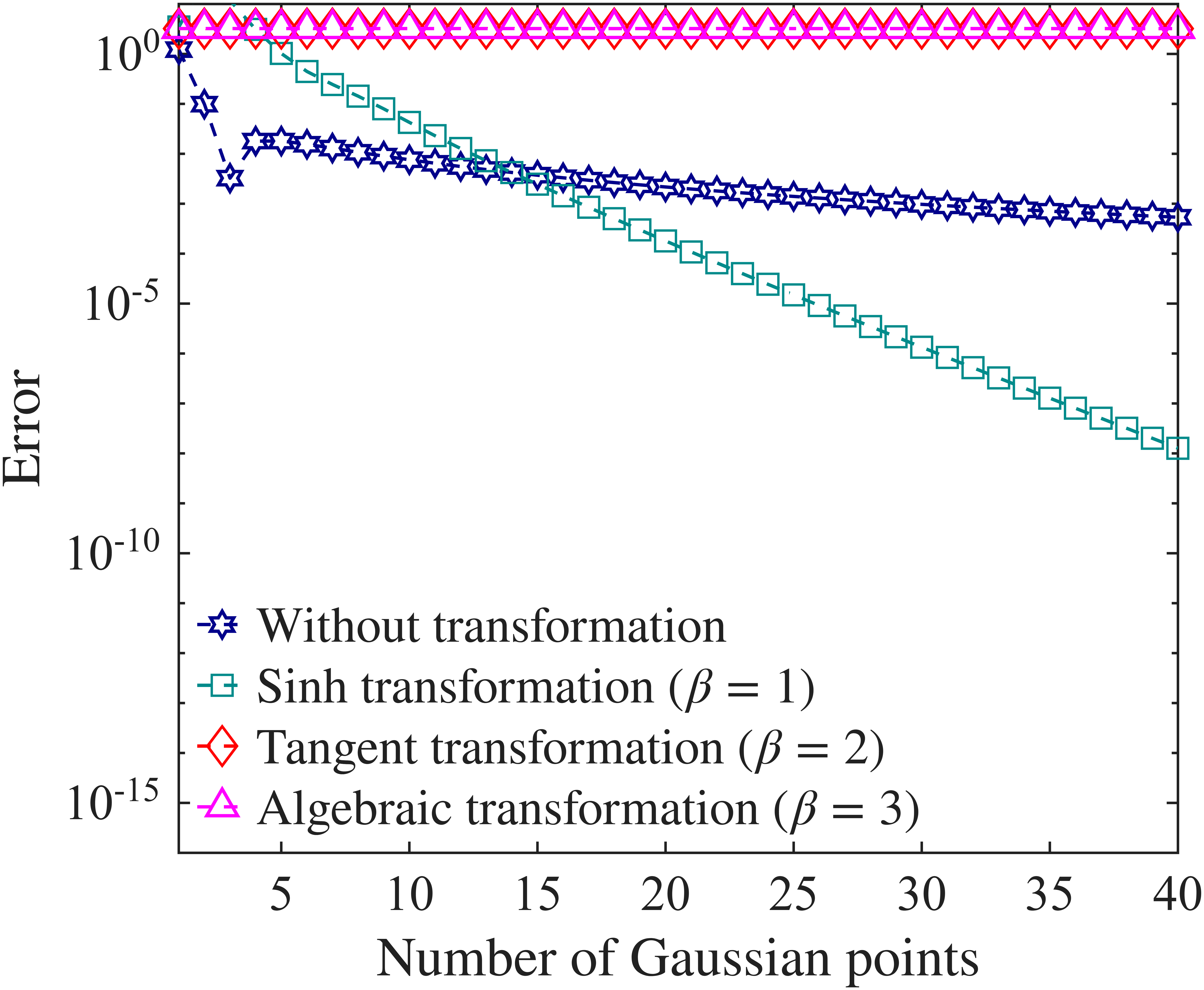}}\\
    \subcaptionbox{$c=10^{-3}$, $n=5$}{%
        \includegraphics[width=3.0cm,height=2.7cm]{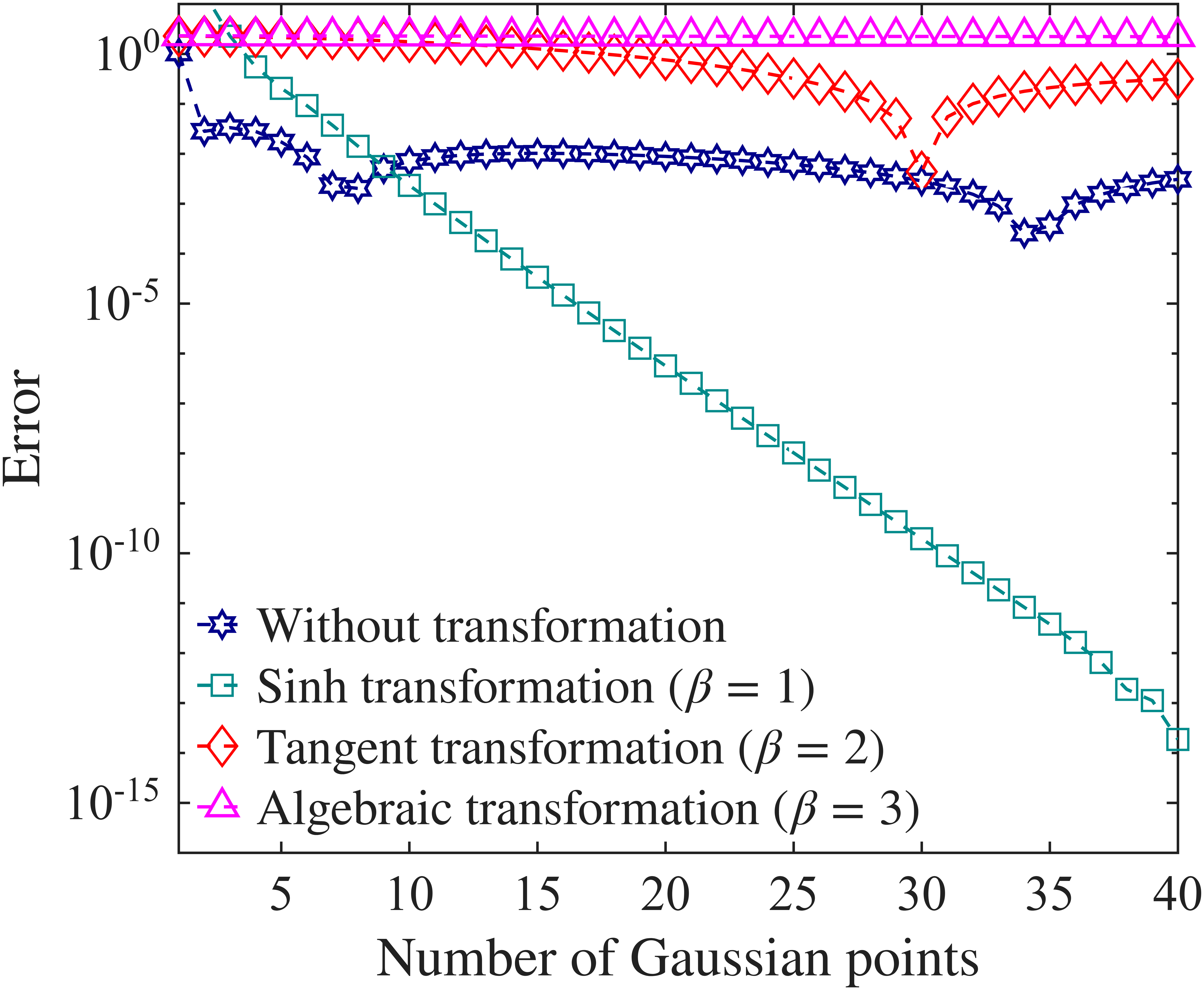}}\hfill
    \subcaptionbox{$c=10^{-4}$, $n=5$}{%
        \includegraphics[width=3.0cm,height=2.7cm]{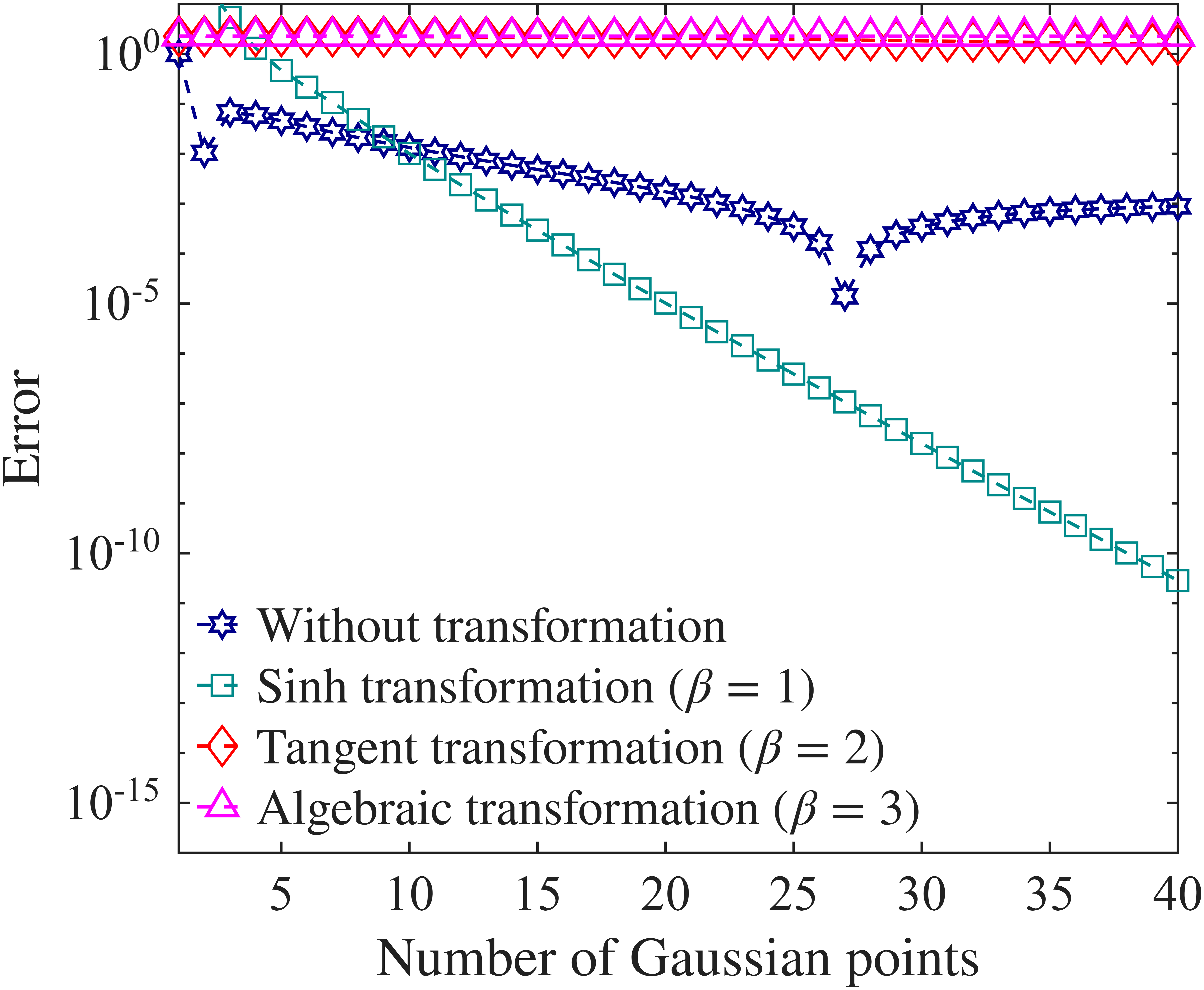}}\hfill
    \subcaptionbox{$c=10^{-5}$, $n=5$}{%
        \includegraphics[width=3.0cm,height=2.7cm]{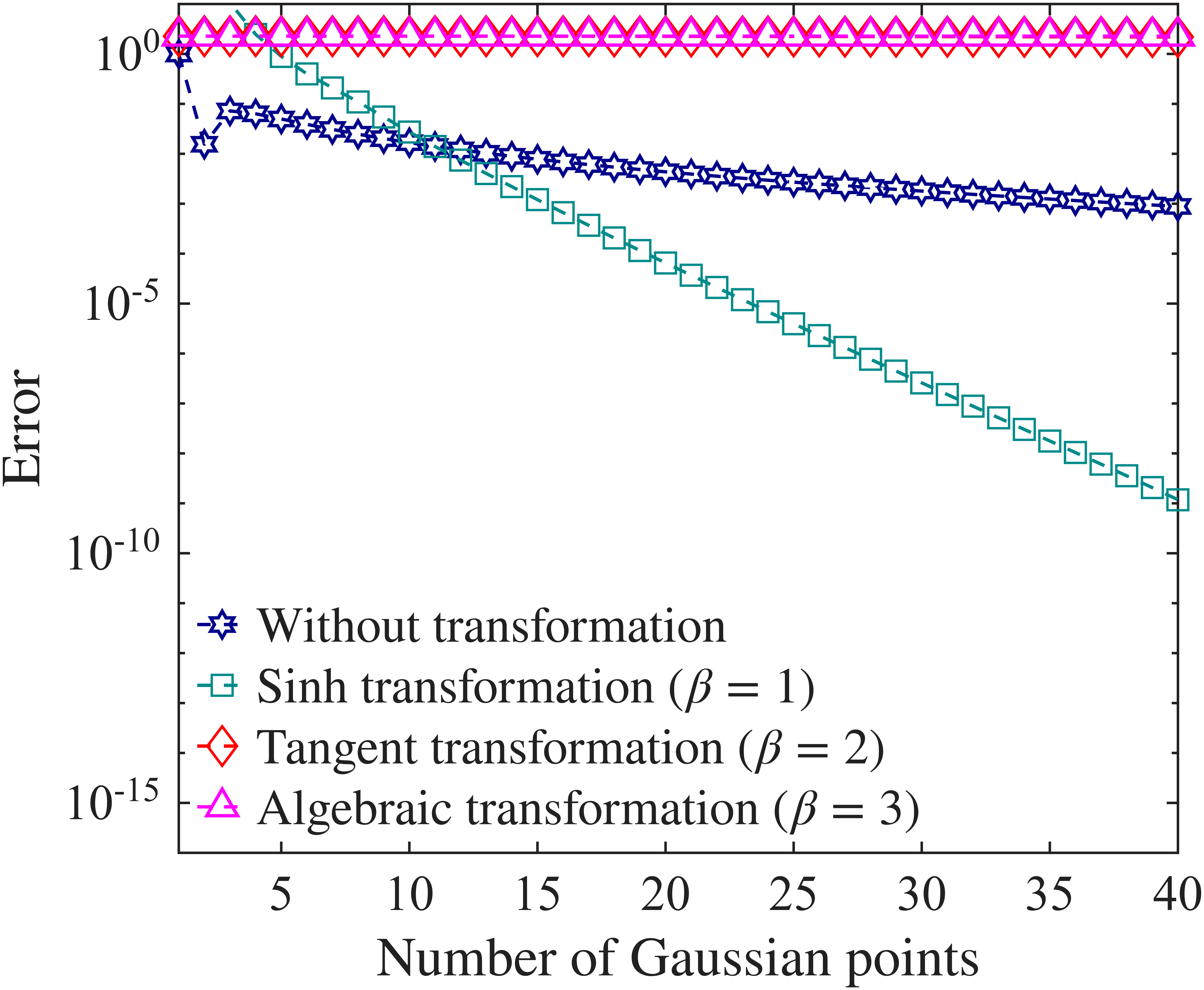}}\hfill
    \subcaptionbox{$c=10^{-6}$, $n=5$}{%
        \includegraphics[width=3.0cm,height=2.7cm]{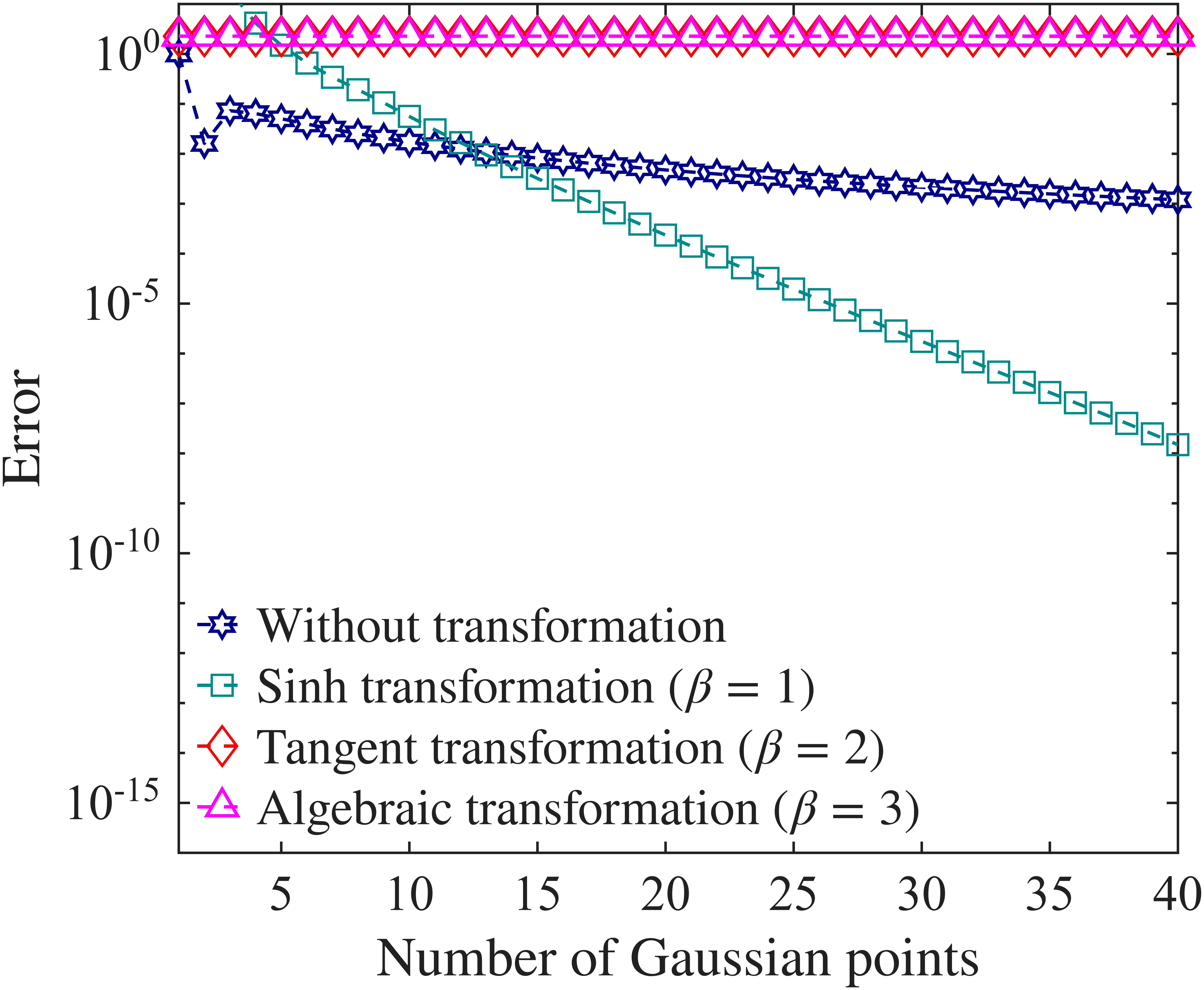}}
    \caption{Errors for different hypergeometric transformations, $n=3,4,5$.
      Rows from top to bottom: $n=3$, $n=4$, $n=5$. The sinh transformation
      ($\beta=1$) consistently outperforms direct Gauss--Legendre and the
      $\beta=2,3$ alternatives.}
    \label{fig:Higher_integrals}
\end{figure}

\paragraph{Triple, quadruple, and quintuple integrals ($n=3,4,5$)} 
To systematically evaluate the proposed framework in higher dimensions, we consider a sequence of $n$-dimensional volume integrals featuring the near-singular exponent $\alpha = n - 1$. This specific exponent corresponds to the magnitude of the gradient of the fundamental solution to the $n$-dimensional Laplace equation (i.e., a generalized inverse-power force field). The triple integral ($n=3$, $\alpha=2$)
\begin{equation}
	\iiint_{[-1,1]^3} \frac{1}{(x-e_1)^2 + (y-e_2)^2 + (z-e_3)^2 + c^2} \,\mathrm{d}x\,\mathrm{d}y\,\mathrm{d}z
\end{equation}
features a regularised $\mathcal{O}(r^{-2})$ kernel, representing an inverse-square force field prevalent in 3D gravitational and electrostatic interactions~\cite{kellogg1953foundations,jackson1999electrodynamics}. The quadruple integral ($n=4$, $\alpha=3$)
\begin{equation}
	\iiiint_{[-1,1]^4} \frac{1}{\bigl((x-e_1)^2 + (y-e_2)^2 + (z-e_3)^2 + (w-e_4)^2 + c^2\bigr)^{3/2}} \,\mathrm{d}x\,\mathrm{d}y\,\mathrm{d}z\,\mathrm{d}w
\end{equation}
extends this hierarchy to four dimensions with an $\mathcal{O}(r^{-3})$ kernel. Such high-dimensional potentials arise frequently in 4D potential theory~\cite{latypov2021evaluation} and the evaluation of inverse multiquadric radial basis functions in meshless approximation methods. The quintuple integral ($n=5$, $\alpha=4$)
\begin{equation}
	\iiint\!\!\!\iint_{[-1,1]^5} \frac{1}{\bigl((x-e_1)^2 + \cdots + (v-e_5)^2 + c^2\bigr)^2} \,\mathrm{d}x\,\mathrm{d}y\,\mathrm{d}z\,\mathrm{d}w\,\mathrm{d}v
\end{equation}
further generalizes the force field to five dimensions with an $\mathcal{O}(r^{-4})$ kernel. Such integrals are natively encountered when spatial models are augmented with multidimensional parameter spaces, such as in high-dimensional uncertainty quantification~\cite{ganapathysubramanian2008sparse}. Figure~\ref{fig:Higher_integrals} compares the sinh ($\beta=1$), tangent ($\beta=2$), and algebraic ($\beta=3$) transformations against direct Gauss--Legendre quadrature. In all cases, the sinh transformation yields exponential convergence; the tangent and algebraic transformations perform worse than direct Gauss--Legendre, corroborating the theoretical prediction of Section~\ref{sec:sinh}. Robust exponential convergence persists up to $n=5$.

% ======================================================================
\section{Significance for variational discretisations}
\label{sec:discussion}
% ======================================================================

The preceding sections analysed the quadrature error of an individual integral.
In this section, we show the accurate evaluation of singular and nearly singular
integrals is critical for the overall accuracy of numerical methods based on
variational formulations. The framework presented here applies to any
discretisation method that relies on numerical integration of weak forms,
including finite element, boundary element, and enriched Galerkin methods.

\subsection{Strang's first lemma and the consistency error}

Consider an abstract variational problem: find $u \in V$ such that
\begin{equation}
	a(u, w) = l(w), \quad \forall w \in V,
\end{equation}
where $V$ is an appropriate Hilbert space, $a(\cdot,\cdot): V \times V \to
\mathbb{R}$ is a bounded and coercive bilinear form, and $l(\cdot): V \to
\mathbb{R}$ is a bounded linear functional. When the integrals defining $a$
and $l$ involve singular or nearly singular kernels, they cannot be evaluated
exactly and must be approximated by numerical quadrature. Let $\mathcal{T}_h$
denote a regular partition of the domain into elements $K$, and let $Q_K$ be a
quadrature rule on $K$ with weights $\omega_{q,K}$ and abscissae
$\boldsymbol{x}_{q,K}$. The exact bilinear and linear forms,
\begin{equation}
	a(v_h, w_h) = \sum_{K \in \mathcal{T}_h} \int_K \mathcal{K}_a(v_h,w_h;\boldsymbol{x})\,\mathrm{d}\boldsymbol{x},
	\qquad
	l(w_h) = \sum_{K \in \mathcal{T}_h} \int_K \mathcal{K}_l(w_h;\boldsymbol{x})\,\mathrm{d}\boldsymbol{x},
\end{equation}
are replaced by their approximate counterparts
\begin{equation}
	a_h(v_h, w_h) = \sum_{K \in \mathcal{T}_h} \sum_{q=1}^{N_q} \omega_{q,K}\,
	\mathcal{K}_a(v_h,w_h;\boldsymbol{x}_{q,K}),
	l_h(w_h) = \sum_{K \in \mathcal{T}_h} \sum_{q=1}^{N_q} \omega_{q,K}\,
	\mathcal{K}_l(w_h;\boldsymbol{x}_{q,K}).
\end{equation}
The fully discrete problem is then: find $\tilde{u}_h \in V_h$ such that
$a_h(\tilde{u}_h, w_h) = l_h(w_h)$ for all $w_h \in V_h$. Strang's first lemma~\cite{strang1974analysis} provides an upper bound for the
total error between $\tilde{u}_h$ and the exact solution $u$:
\begin{equation}
	\|u - \tilde{u}_h\|_V \le C_0 \inf_{v_h \in V_h} \Bigl(
	\|u - v_h\|_V + \sup_{w_h \in V_h} \frac{|a(v_h,w_h) - a_h(v_h,w_h)|}{\|w_h\|_V}
	\Bigr) + C_0 \sup_{w_h \in V_h} \frac{|l(w_h) - l_h(w_h)|}{\|w_h\|_V}, \label{eq:strang}
\end{equation}
where $C_0 = \max(1 + M/\tilde{\alpha},\, 1/\tilde{\alpha})$ depends only on
the continuity constant $M$ and the coercivity constant $\tilde{\alpha}$ of
$a(\cdot,\cdot)$. The first term in the infimum is the best approximation error,
governed solely by the approximation power of $V_h$. The remaining terms are
consistency errors induced purely by numerical quadrature.

Let $E_K(\phi) = \int_K \phi\,\mathrm{d}\boldsymbol{x} - Q_K(\phi)$ denote the
local quadrature error functional on element $K$. The global consistency errors
are bounded by the sum of local errors:
\begin{equation}
	\begin{split}
		|a(v_h, w_h) - a_h(v_h, w_h)| &\le \sum_{K \in \mathcal{T}_h} |E_K(\mathcal{K}_a(v_h,w_h))|,\\
		|l(w_h) - l_h(w_h)| &\le \sum_{K \in \mathcal{T}_h} |E_K(\mathcal{K}_l(w_h))|.
	\end{split}
\end{equation}

\subsection{Effect of quadrature quality on the consistency error}

When the kernel in $\mathcal{K}_a$ or $\mathcal{K}_l$ is singular, standard Gauss--Legendre quadrature applied without transformation converges only algebraically. For an $n$-dimensional integral, the local quadrature error satisfies
\begin{equation}
	|E_K(\phi)| \le \sum_{j=1}^{n} C_j m_j^{-q_j},
\end{equation}
where $q_j > 0$ are dimension-dependent algebraic rates and $m_j$ is the number of quadrature points in the $j$-th direction. In the case of nearly singular integrals, the situation is more subtle: exponential convergence is formally present, but the Bernstein ellipse parameter $\rho_j$ approaches unity as $r_0 \to 0$, so that the exponential rate $\ln \rho_j$ tends to zero. In both cases, the consistency error may dominate the best approximation error, preventing the fully discrete solution from attaining the optimal convergence rate predicted by C\'{e}a's lemma.

The variable transformations developed in Section~\ref{sec:problem} change the character of the quadrature error. For singular integrals, the Duffy transformation cancels the $1/r^\alpha$ singularity, converting the integrand to a form that is regular (or mildly singular) in the radial direction. For nearly singular integrals, the sinh transformation regularises the near-singularity so that the transformed integrand is substantially smoother. 

Bounding the consistency error in Equation~\eqref{eq:strang} requires evaluating the quadrature error of the product integrands $\mathcal{K}_a$ and $\mathcal{K}_l$. To isolate the test function norm $\|w_h\|_V$ from the supremum, the maximum modulus of the transformed integrands on the complex Bernstein ellipse $E_\rho$ is evaluated. By invoking a stable decomposition of the enriched space and applying Bernstein's inequality for polynomials on ellipses to the standard components, the test function norm factors out (see Appendix~\ref{sec:appendix_b}).

Substituting the resulting local bounds into the Strang estimate (Equation~\eqref{eq:strang}) yields the total error bound after transformation (see Appendix~\ref{sec:appendix_b} for the complete derivation):
\begin{equation}\label{eq:convproof}
	\|u - \tilde{u}_h\|_V \le C_0 \inf_{v_h \in V_h} \|u - v_h\|_V
	+ \tilde{C} \sum_{K \in \mathcal{T}_h} \sum_{j=1}^{n} \bigl( \rho_j^{p-1} + M_{\psi}(\rho_j) \bigr) \rho_j^{-2m_j},
\end{equation}
where $p$ is the polynomial degree of the standard finite element space, and $M_{\psi}(\rho_j)$ bounds the analytic continuation of the regularised enrichment function on $E_{\rho_j}$.

Because the variable transformations guarantee $\rho_j > 1$, the consistency terms decay exponentially at an asymptotic rate of $\mathcal{O}(\rho_j^{-(2m_j - p + 1)})$. Provided the quadrature order $m_j$ is chosen sufficiently large to overcome the polynomial growth factor $\rho_j^{p-1}$, the consistency error vanishes at a rate faster than the approximation error $\inf_{v_h \in V_h} \|u - v_h\|_V$. The fully discrete solution therefore recovers the optimal asymptotic convergence rate dictated solely by the approximation properties of the enriched space $V_h^{\text{enr}}$.

\subsection{Numerical illustration}

To illustrate the effect of quadrature quality on the total discretisation error, we consider a one-dimensional problem whose exact solution possesses a severe near-singularity:
\begin{equation}\label{eq:1d_ode}
	-u''(x) = f(x), \quad x \in (0,1), \qquad u(0) = 0,\;\; u(1) =\arctan\left(\frac{1}{\epsilon}\right),
\end{equation}
with the source term $f(x) = \frac{2\epsilon x}{\left(x^2 + \epsilon^2\right)^2}$ and $\epsilon \ll 1$ acting as the near-singularity parameter. The exact solution is $u_{\text{ex}}(x) =\arctan\left(\frac{x}{\epsilon}\right)$. 

The gradient $u_{\text{ex}}'(x) = \frac{\epsilon }{x^2 + \epsilon^2}$ exhibits a sharp peak at $x = 0$, causing standard piecewise-linear finite elements to converge at suboptimal rates. Motivated by the known structure of the near-singularity, the trial space is augmented by the global enrichment function
\begin{equation}\label{eq:enrich}
	\psi(x) =\arctan\left(\frac{x}{\epsilon}\right) - \arctan\left(\frac{1}{\epsilon}\right)x , \qquad \psi(0) = 0,\;\; \psi(1) = 0,
\end{equation}
which exactly reproduces the singular behaviour of $u_{\text{ex}}$ while satisfying homogeneous boundary conditions. The discrete solution is sought in the enriched space $V_h^{\text{enr}} = V_h \oplus \operatorname{span}\{\psi\}$, where $V_h$ is the standard piecewise-linear space. The enrichment introduces the derivative
\begin{equation}\label{eq:dpsi}
	\psi'(x) = \frac{\epsilon}{\epsilon^2 + x^2} - \arctan\left(\frac{1}{\epsilon}\right).
\end{equation}

The integrand exhibits a near-singularity at $x = 0$. To regularise this integral, we introduce the sinh transformation $x = \epsilon \sinh(t)$. To investigate the quadrature-induced consistency error independently of the approximation error, we define computable consistency error indicators. Following the stable decomposition and inverse estimate bounds derived in Appendix~\ref{sec:appendix_b}, the global suprema over the test space in Strang's first lemma are tracked by separating the local quadrature errors of the bilinear and linear forms. We define the bilinear consistency indicator $\eta_{c,a}$ and the linear consistency indicator $\eta_{c,l}$ as:
\begin{align}
	\eta_{c,a} &= \sum_{K \in \mathcal{T}_h} h_K^{1/2} \Bigl( |E_K(u_{\text{ex}}')| + |E_K(u_{\text{ex}}'\psi')| \Bigr), \label{eq:ind_a} \\
	\eta_{c,l} &= \sum_{K \in \mathcal{T}_h} h_K^{1/2} \Bigl( |E_K(f)| + |E_K(f\psi)| \Bigr), \label{eq:ind_l}
\end{align}
where $E_K(\cdot)$ denotes the local quadrature error functional, and $h_K^{1/2}$ is the geometric scaling factor arising from the 1D inverse estimate.

\begin{figure}[!htbp]
	\centering
	\captionsetup[subfigure]{justification=centering}
	
	% Row 1: epsilon = 10^-4 (Total & Bilinear)
	\subcaptionbox{Total error, no trans.\ ($\epsilon=10^{-4}$)}{%
		\includegraphics[width=3.2cm,height=2.6cm]{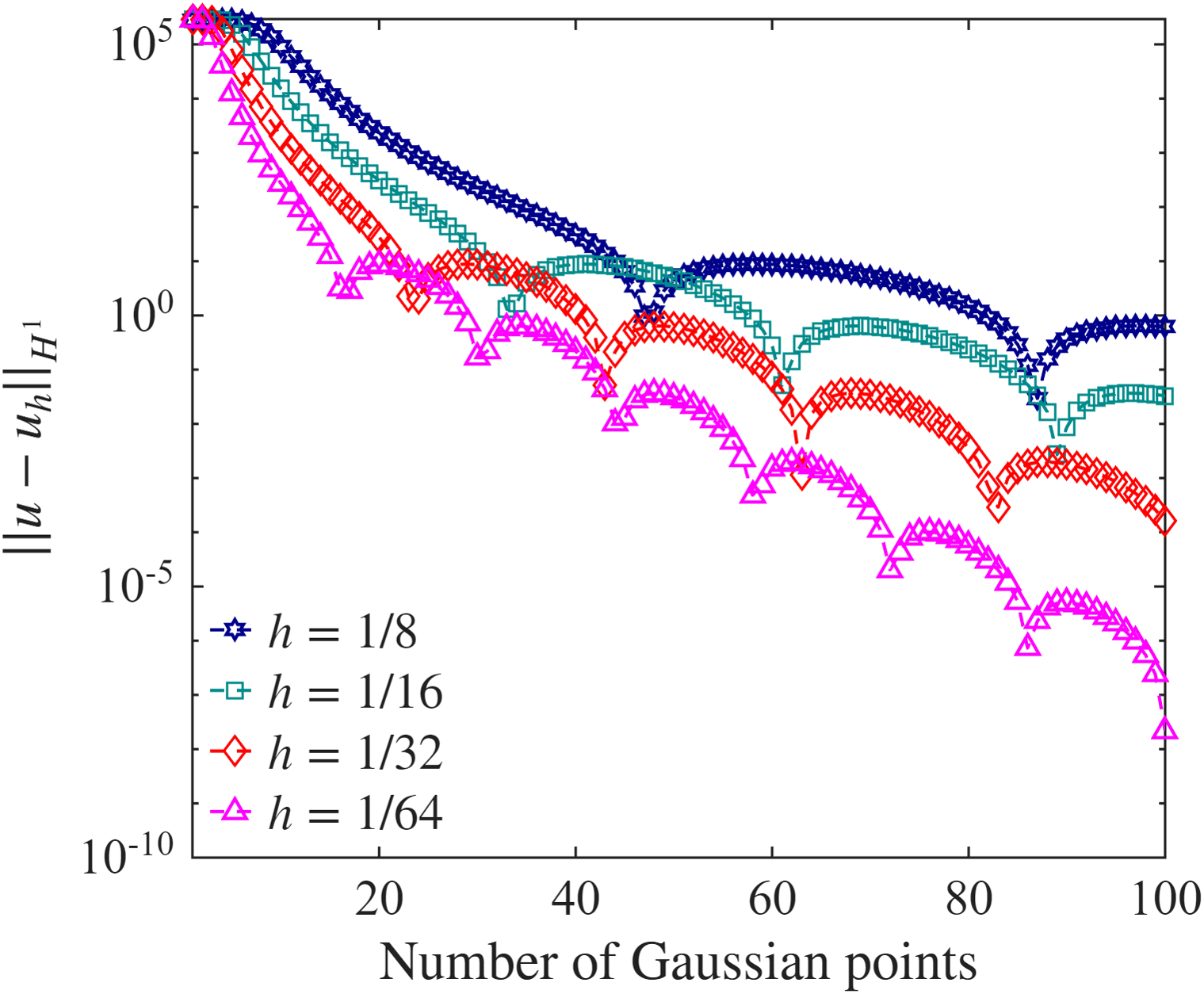}}\hfill
	\subcaptionbox{Total error, with trans.\ ($\epsilon=10^{-4}$)}{%
		\includegraphics[width=3.2cm,height=2.6cm]{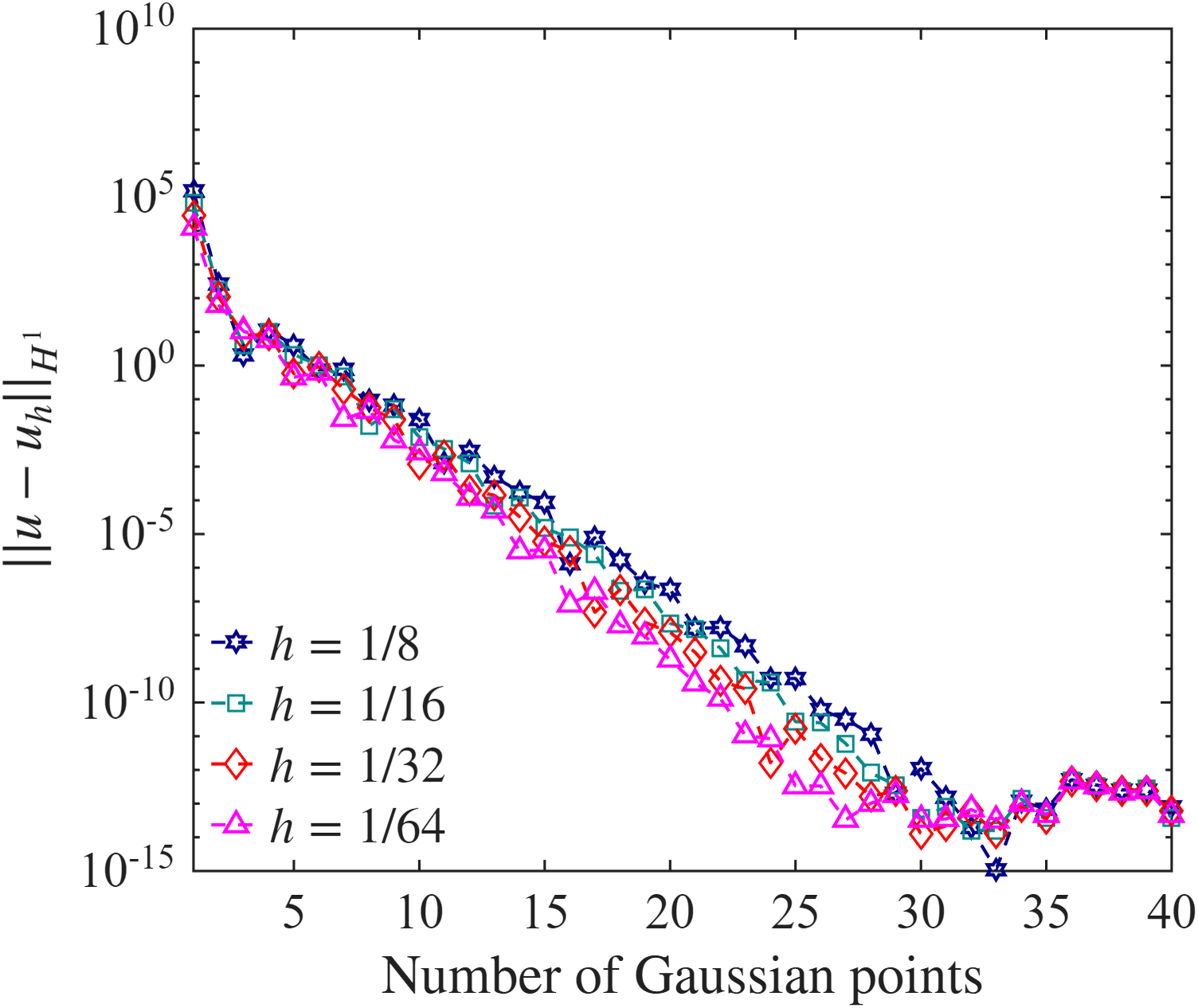}}\hfill
	\subcaptionbox{Bilinear cons., no trans.\ ($\epsilon=10^{-4}$)}{%
		\includegraphics[width=3.2cm,height=2.6cm]{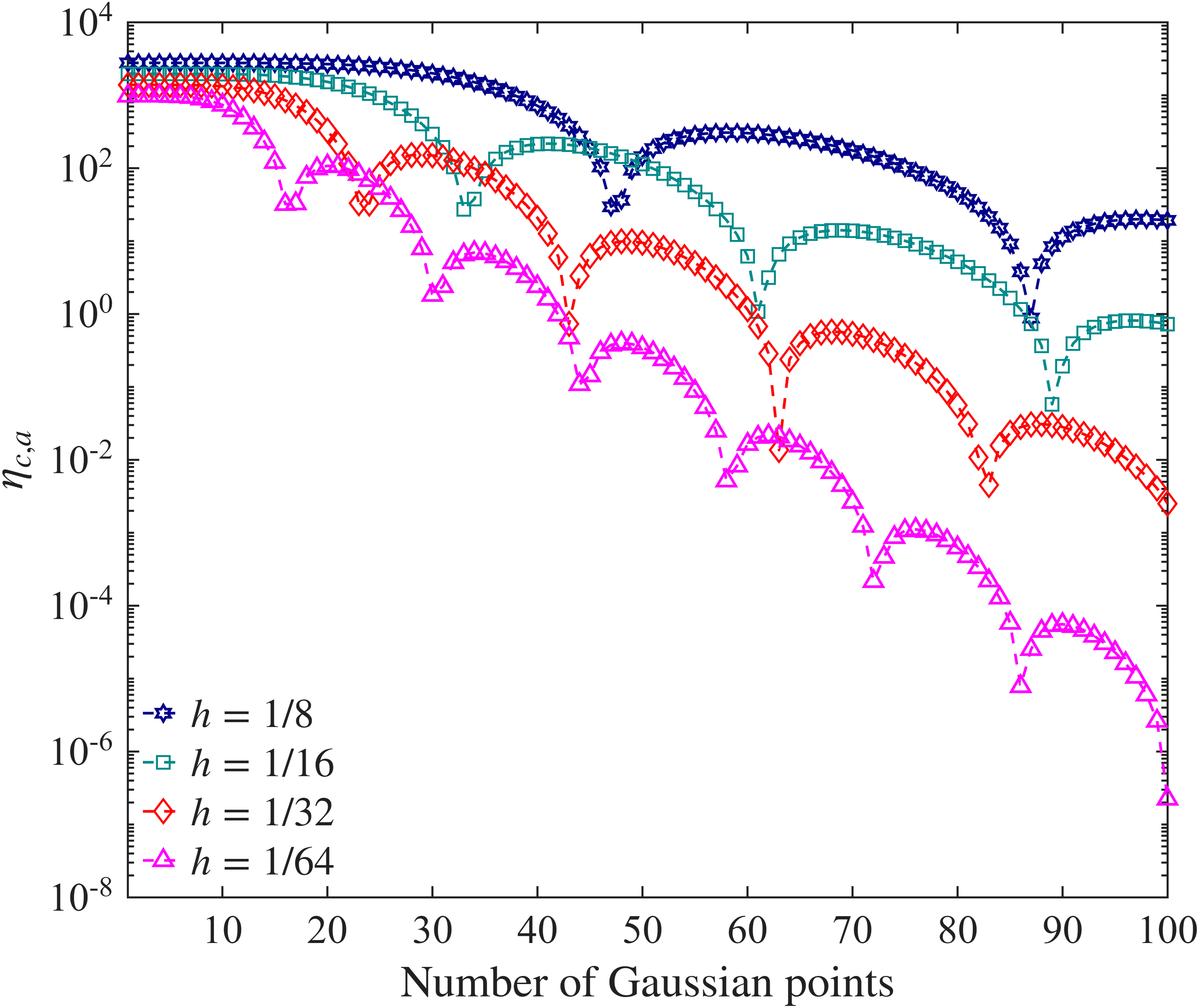}}\hfill
	\subcaptionbox{Bilinear cons., with trans.\ ($\epsilon=10^{-4}$)}{%
		\includegraphics[width=3.2cm,height=2.6cm]{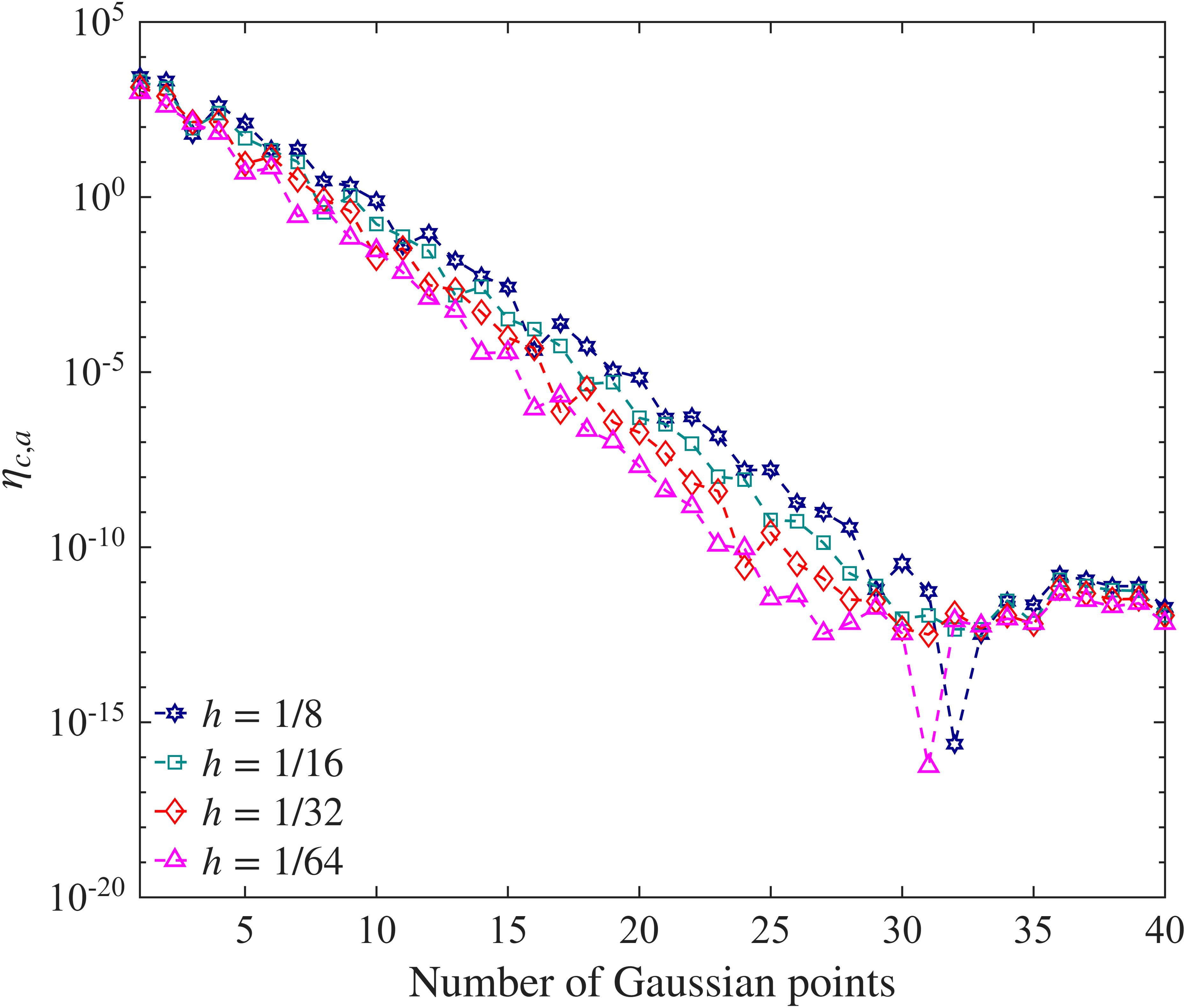}}\\
	
	% Row 2: epsilon = 10^-4 (Linear) & epsilon = 10^-5 (Total)
	\subcaptionbox{Linear cons., no trans.\ ($\epsilon=10^{-4}$)}{%
		\includegraphics[width=3.2cm,height=2.6cm]{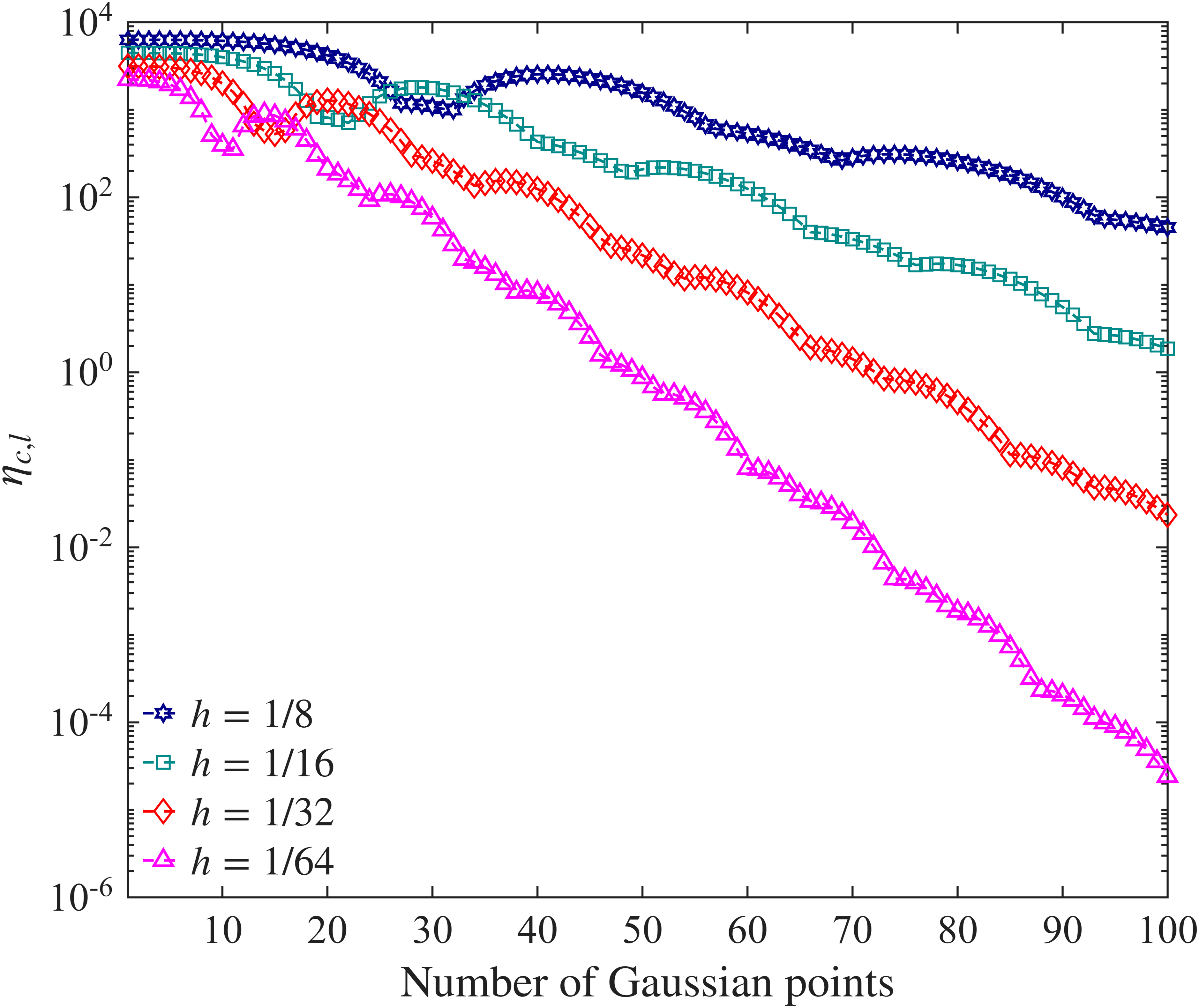}}\hfill
	\subcaptionbox{Linear cons., with trans.\ ($\epsilon=10^{-4}$)}{%
		\includegraphics[width=3.2cm,height=2.6cm]{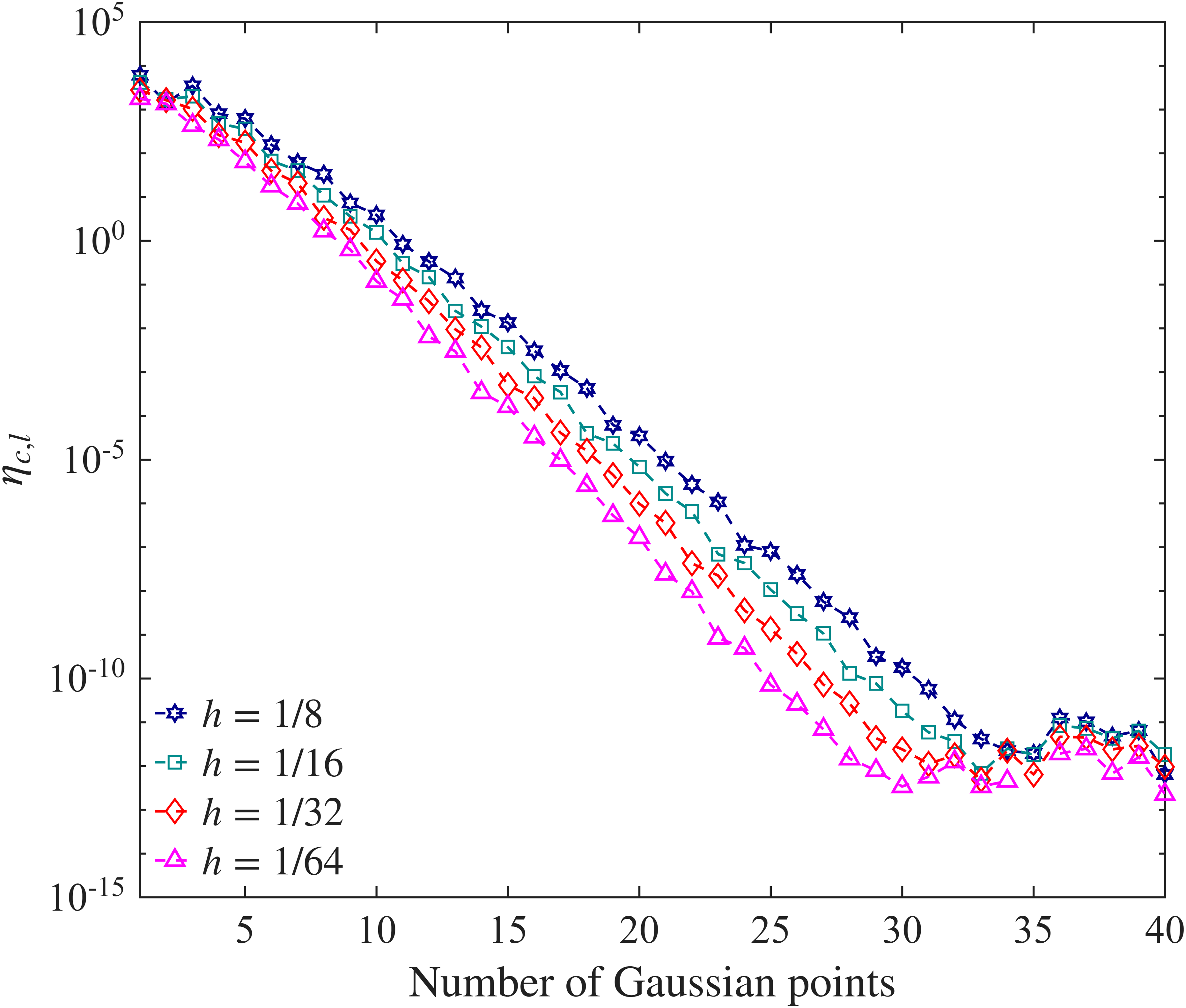}}\hfill
	\subcaptionbox{Total error, no trans.\ ($\epsilon=10^{-5}$)}{%
		\includegraphics[width=3.2cm,height=2.6cm]{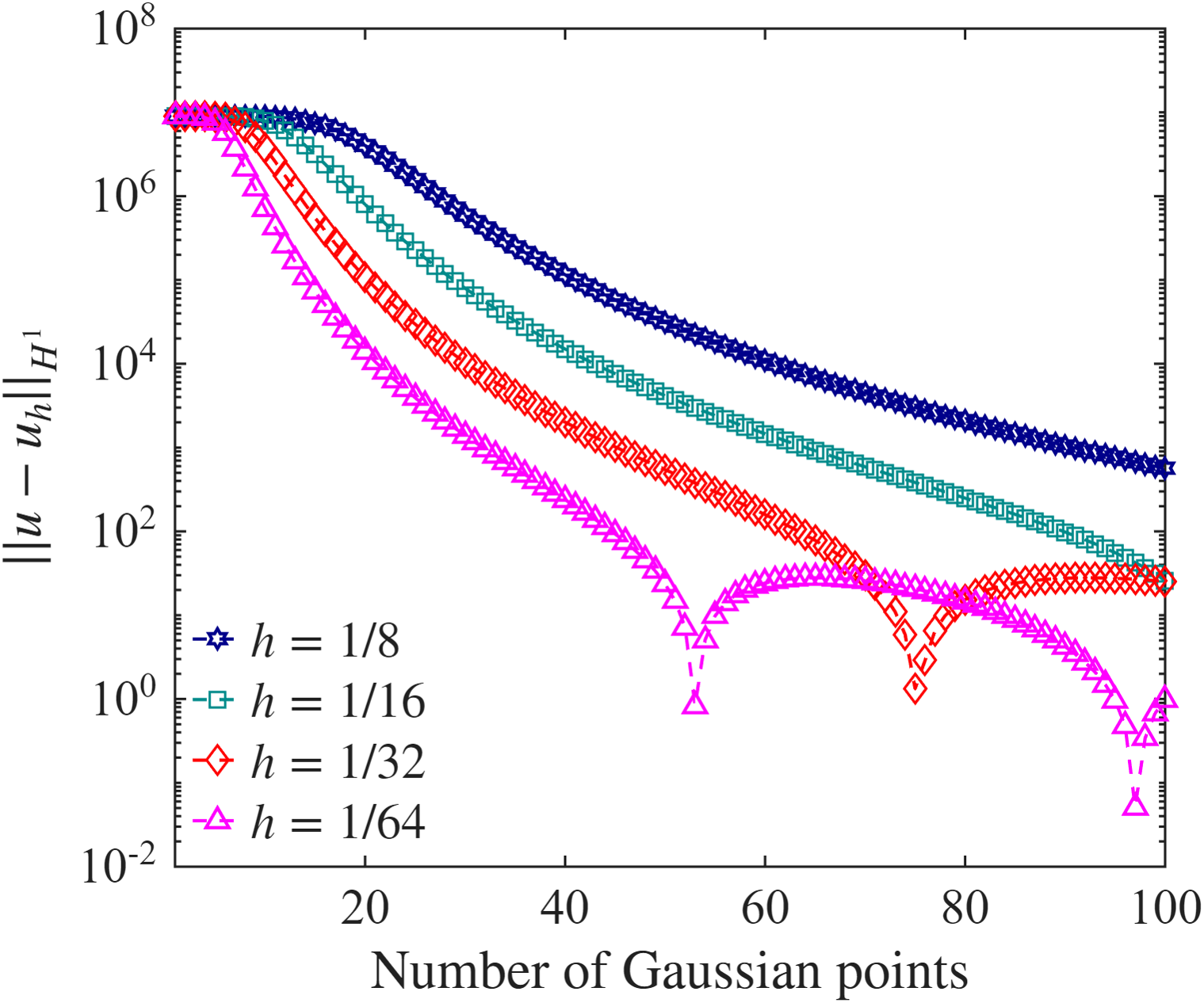}}\hfill
	\subcaptionbox{Total error, with trans.\ ($\epsilon=10^{-5}$)}{%
		\includegraphics[width=3.2cm,height=2.6cm]{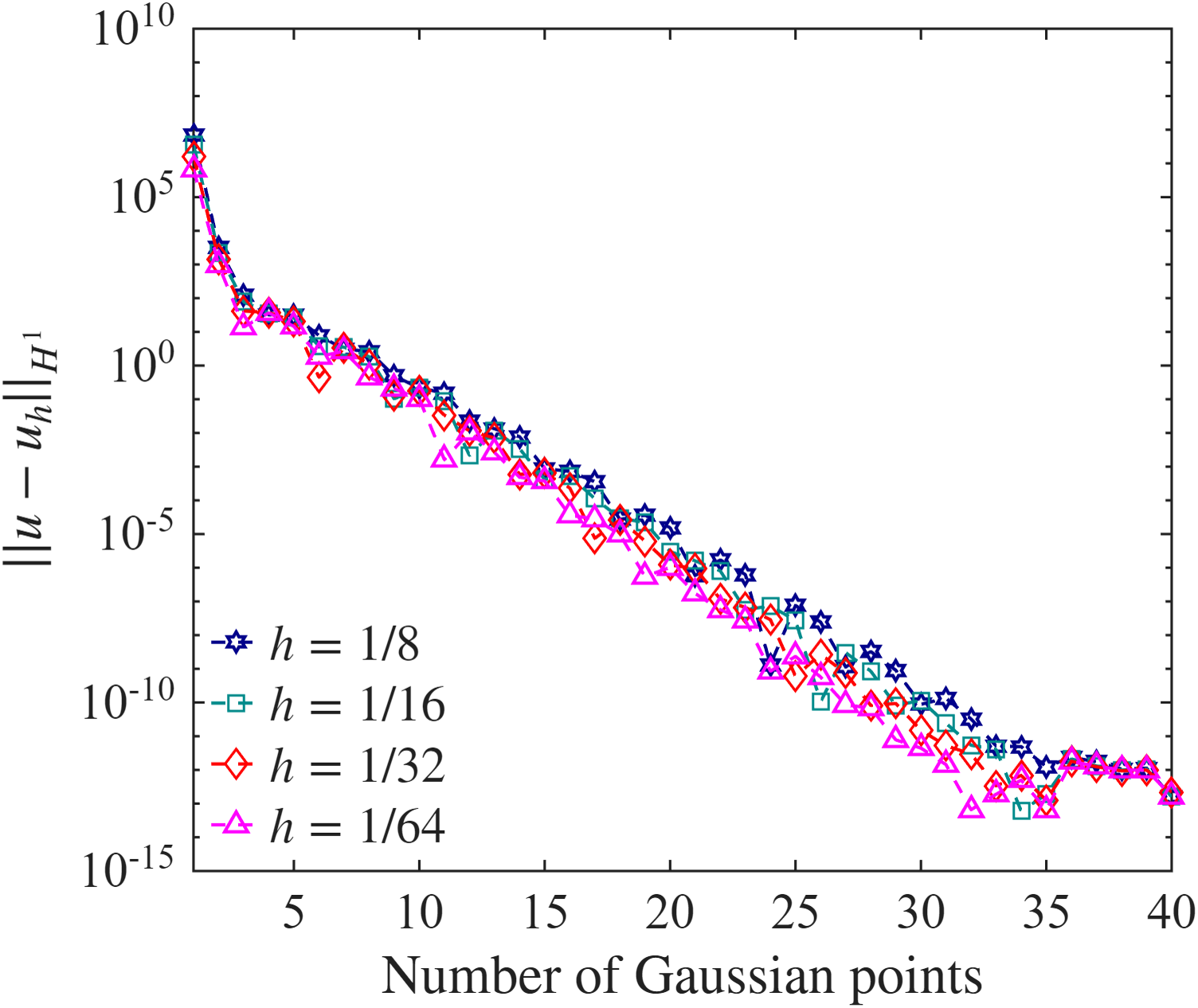}}\\
	
	% Row 3: epsilon = 10^-5 (Bilinear & Linear)
	\subcaptionbox{Bilinear cons., no trans.\ ($\epsilon=10^{-5}$)}{%
		\includegraphics[width=3.2cm,height=2.6cm]{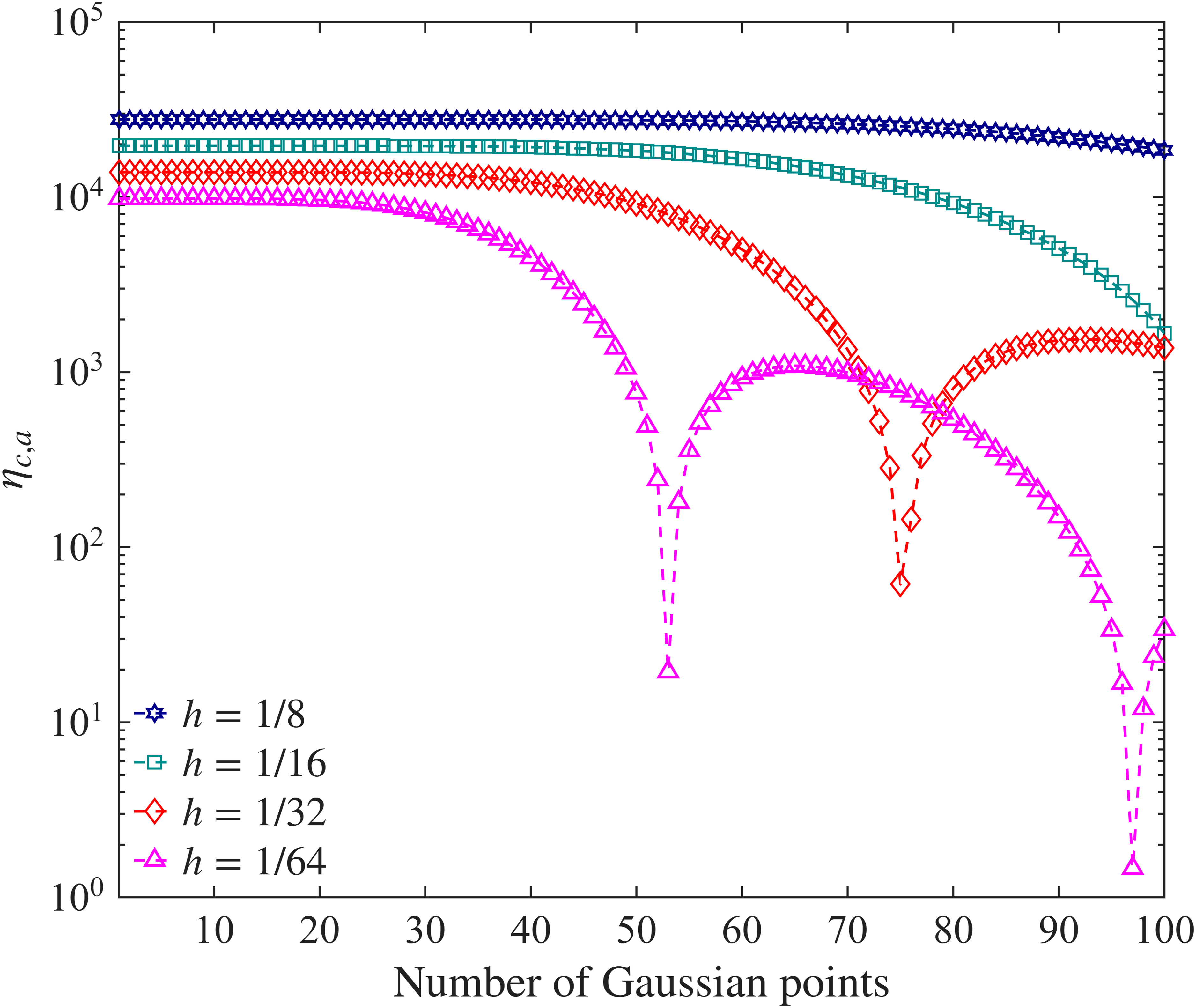}}\hfill
	\subcaptionbox{Bilinear cons., with trans.\ ($\epsilon=10^{-5}$)}{%
		\includegraphics[width=3.2cm,height=2.6cm]{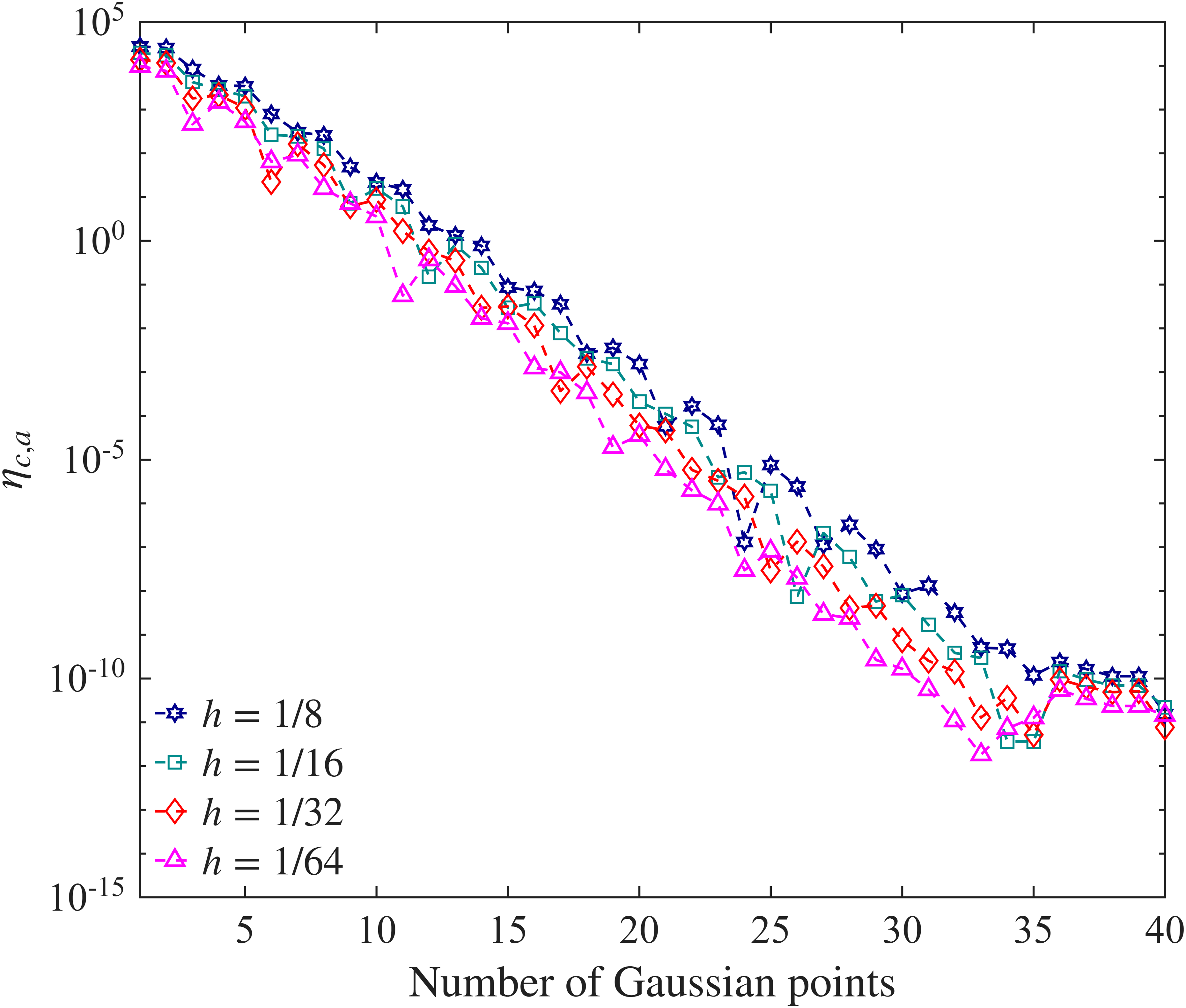}}\hfill
	\subcaptionbox{Linear cons., no trans.\ ($\epsilon=10^{-5}$)}{%
		\includegraphics[width=3.2cm,height=2.6cm]{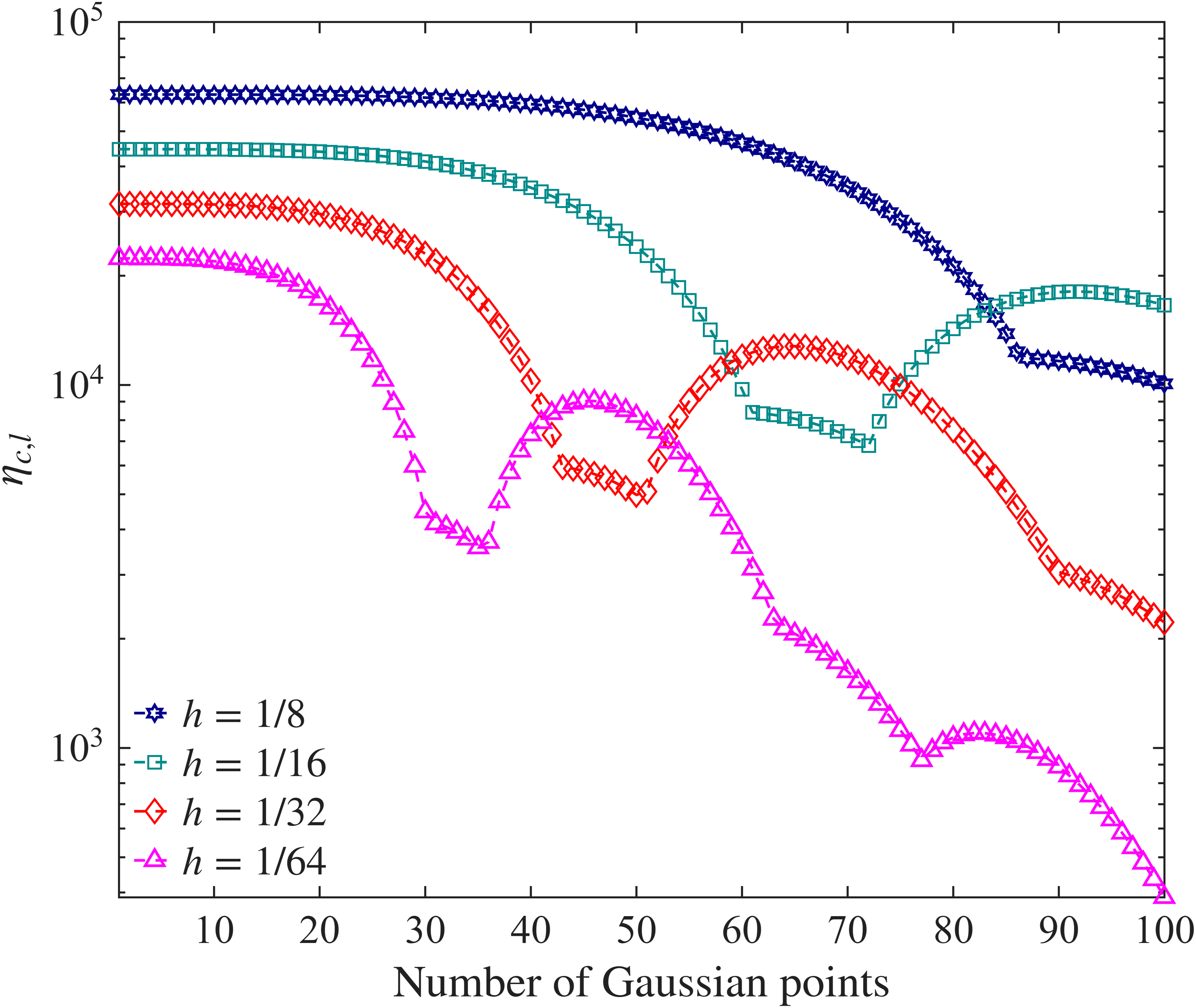}}\hfill
	\subcaptionbox{Linear cons., with trans.\ ($\epsilon=10^{-5}$)}{%
		\includegraphics[width=3.2cm,height=2.6cm]{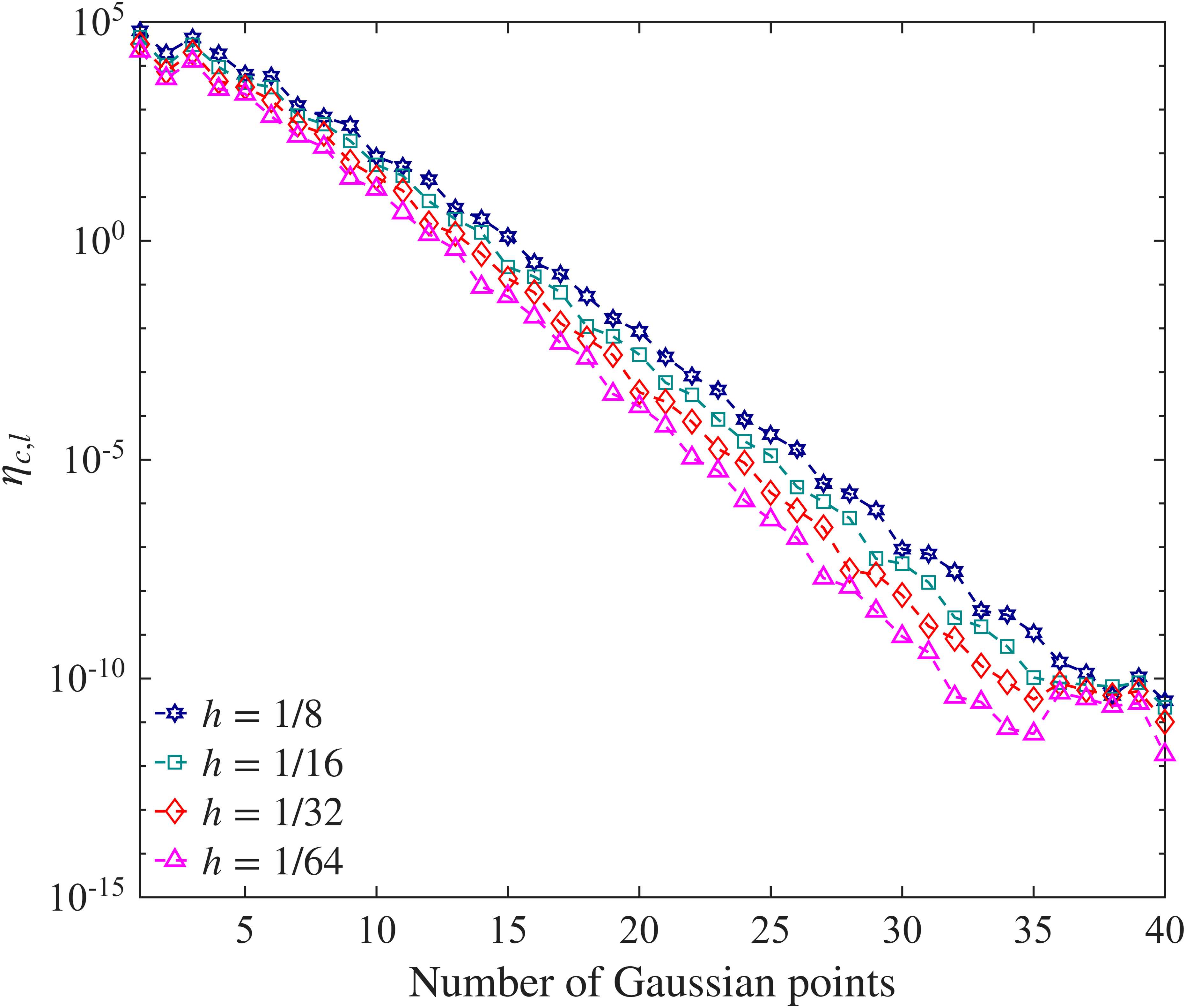}}\\
	
	% Row 4: epsilon = 10^-6 (Total & Bilinear)
	\subcaptionbox{Total error, no trans.\ ($\epsilon=10^{-6}$)}{%
		\includegraphics[width=3.2cm,height=2.6cm]{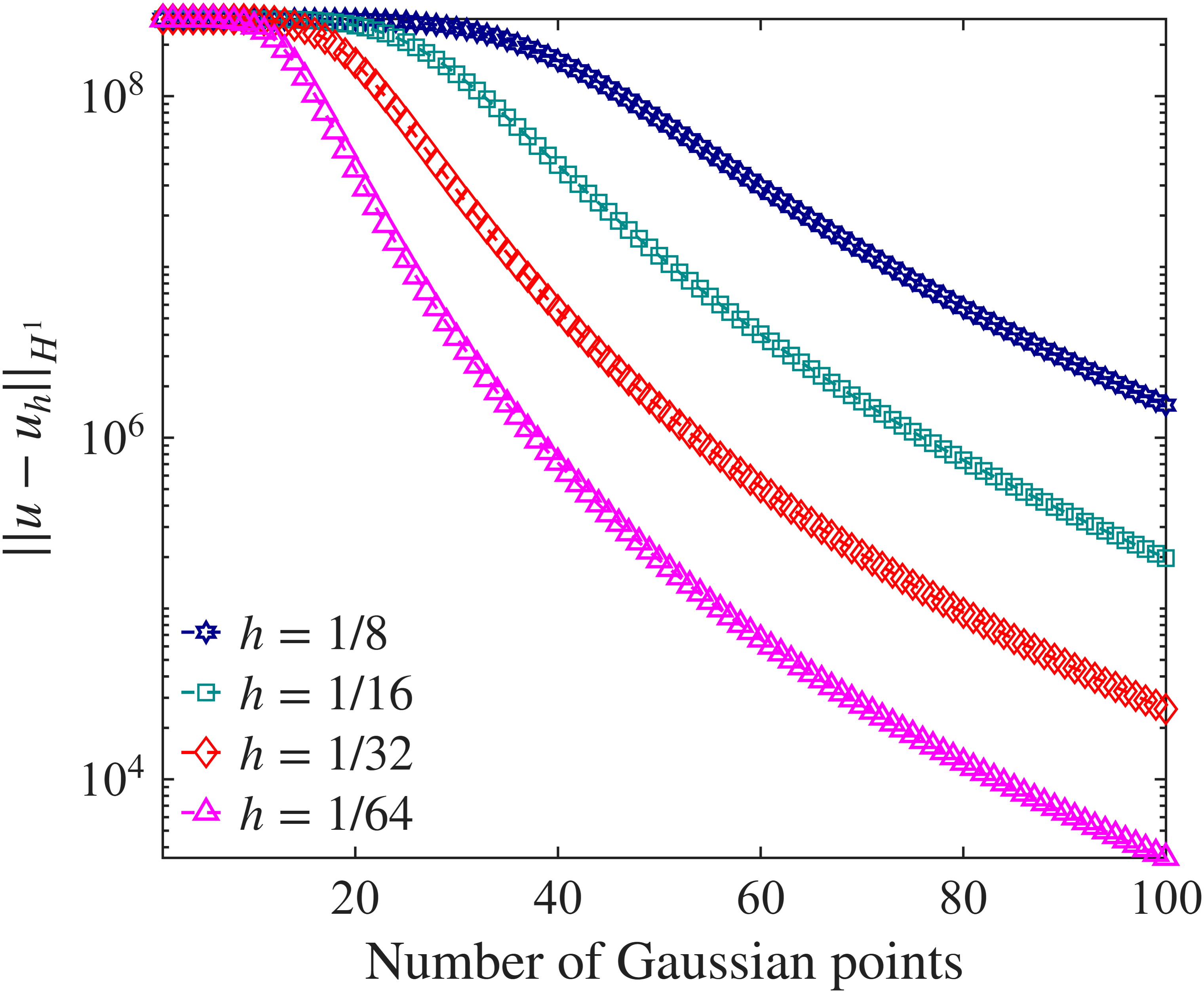}}\hfill
	\subcaptionbox{Total error, with trans.\ ($\epsilon=10^{-6}$)}{%
		\includegraphics[width=3.2cm,height=2.6cm]{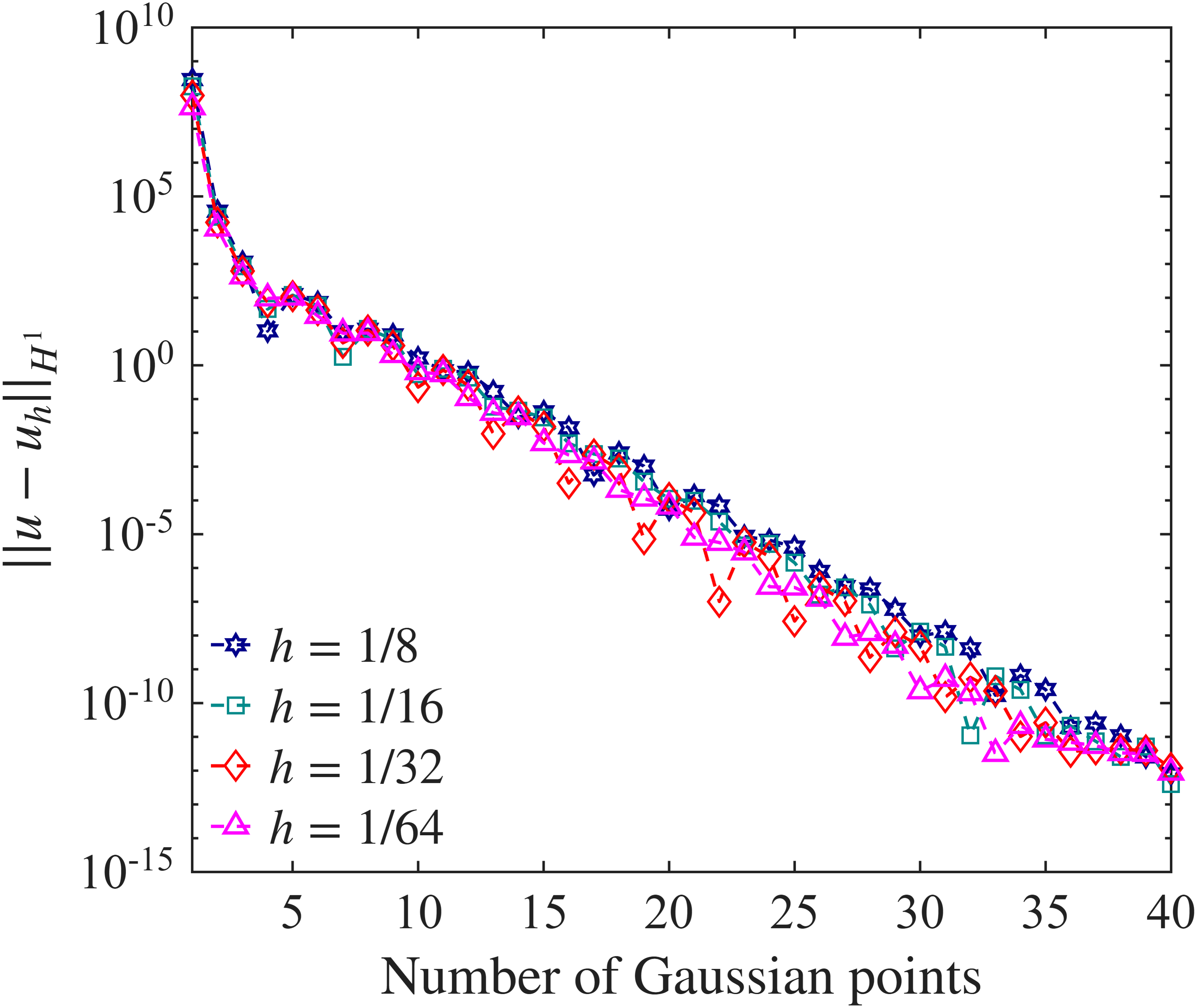}}\hfill
	\subcaptionbox{Bilinear cons., no trans.\ ($\epsilon=10^{-6}$)}{%
		\includegraphics[width=3.2cm,height=2.6cm]{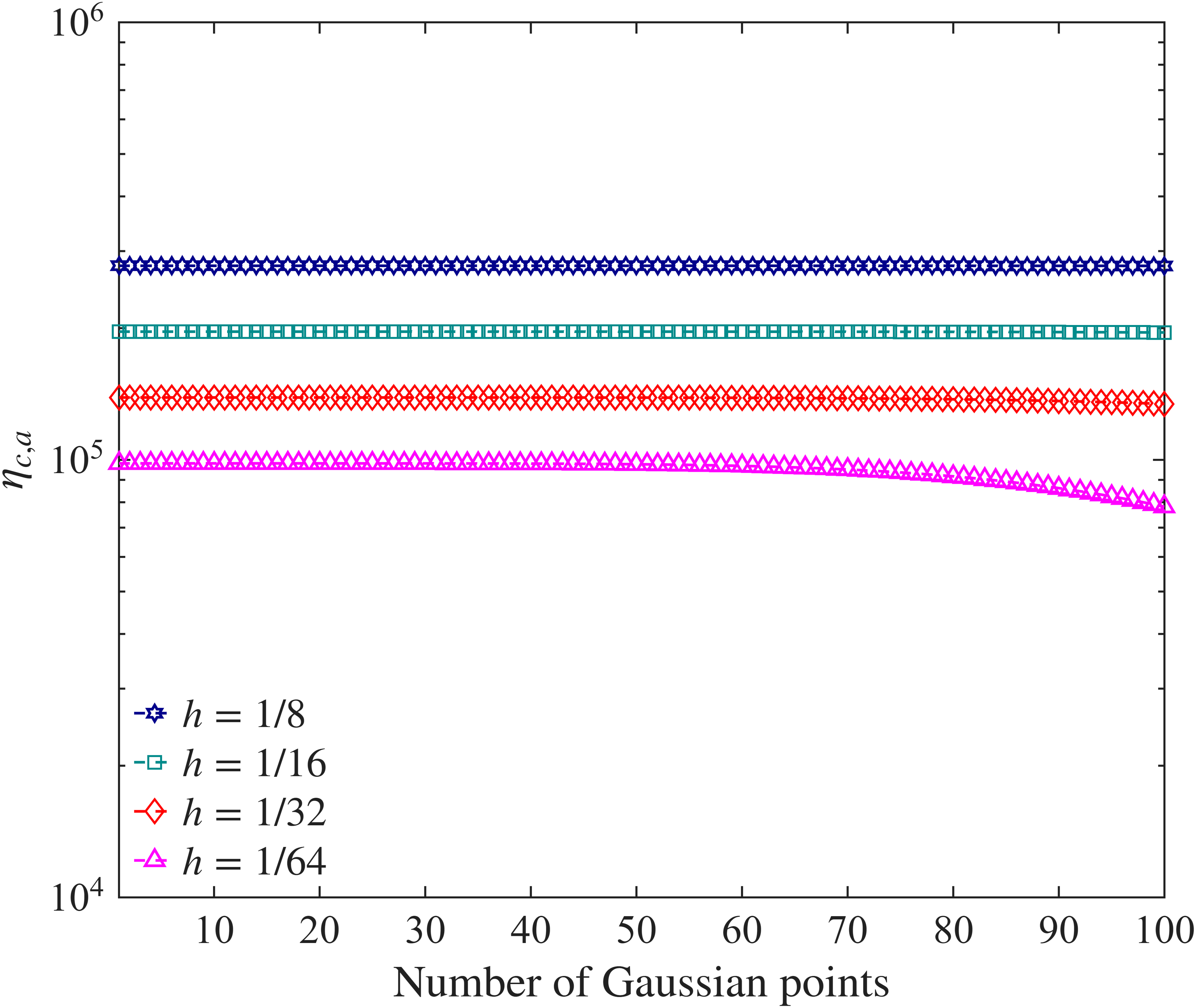}}\hfill
	\subcaptionbox{Bilinear cons., with trans.\ ($\epsilon=10^{-6}$)}{%
		\includegraphics[width=3.2cm,height=2.6cm]{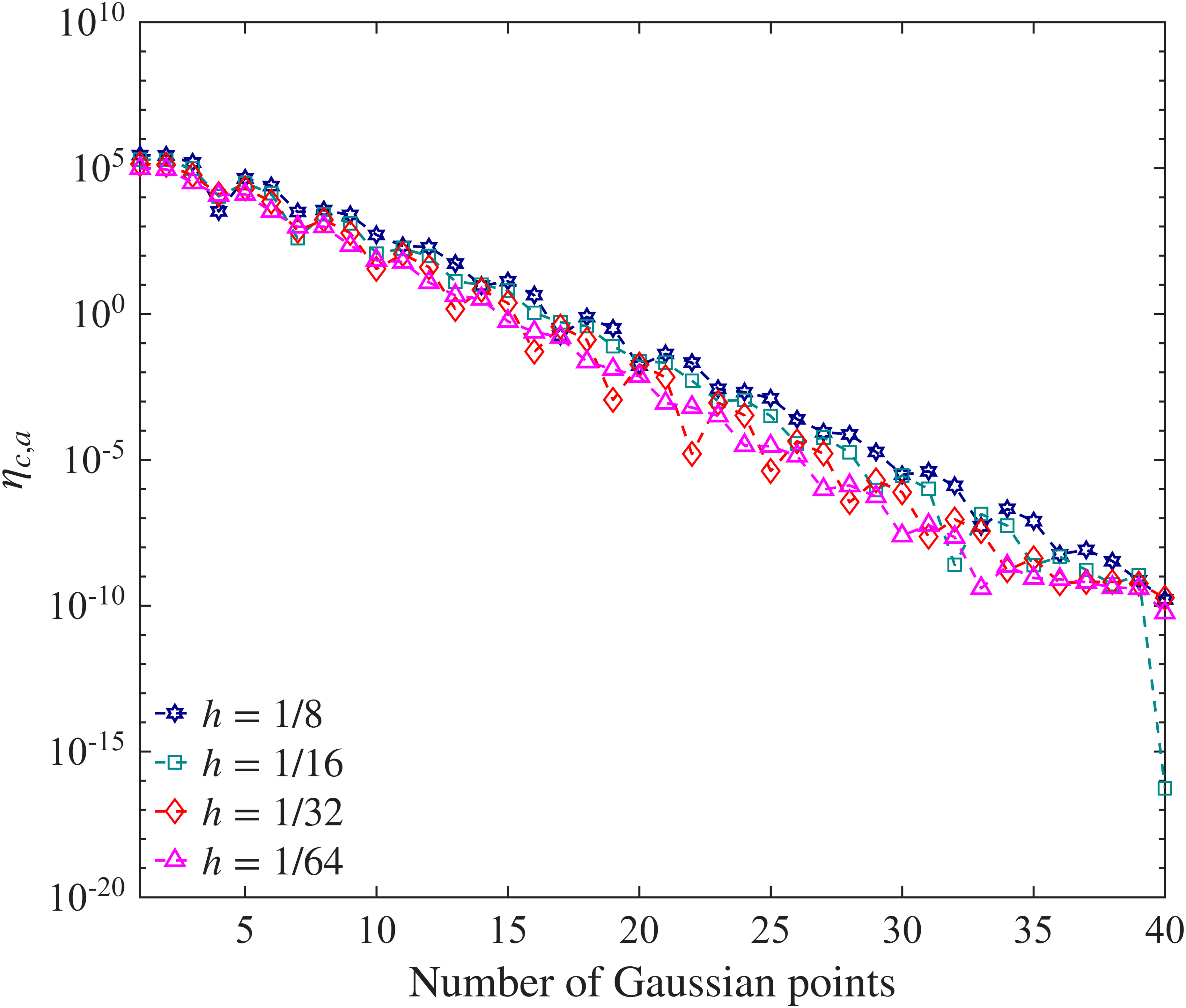}}\\
	
	% Row 5: epsilon = 10^-6 (Linear)
	\subcaptionbox{Linear cons., no trans.\ ($\epsilon=10^{-6}$)}{%
		\includegraphics[width=3.2cm,height=2.6cm]{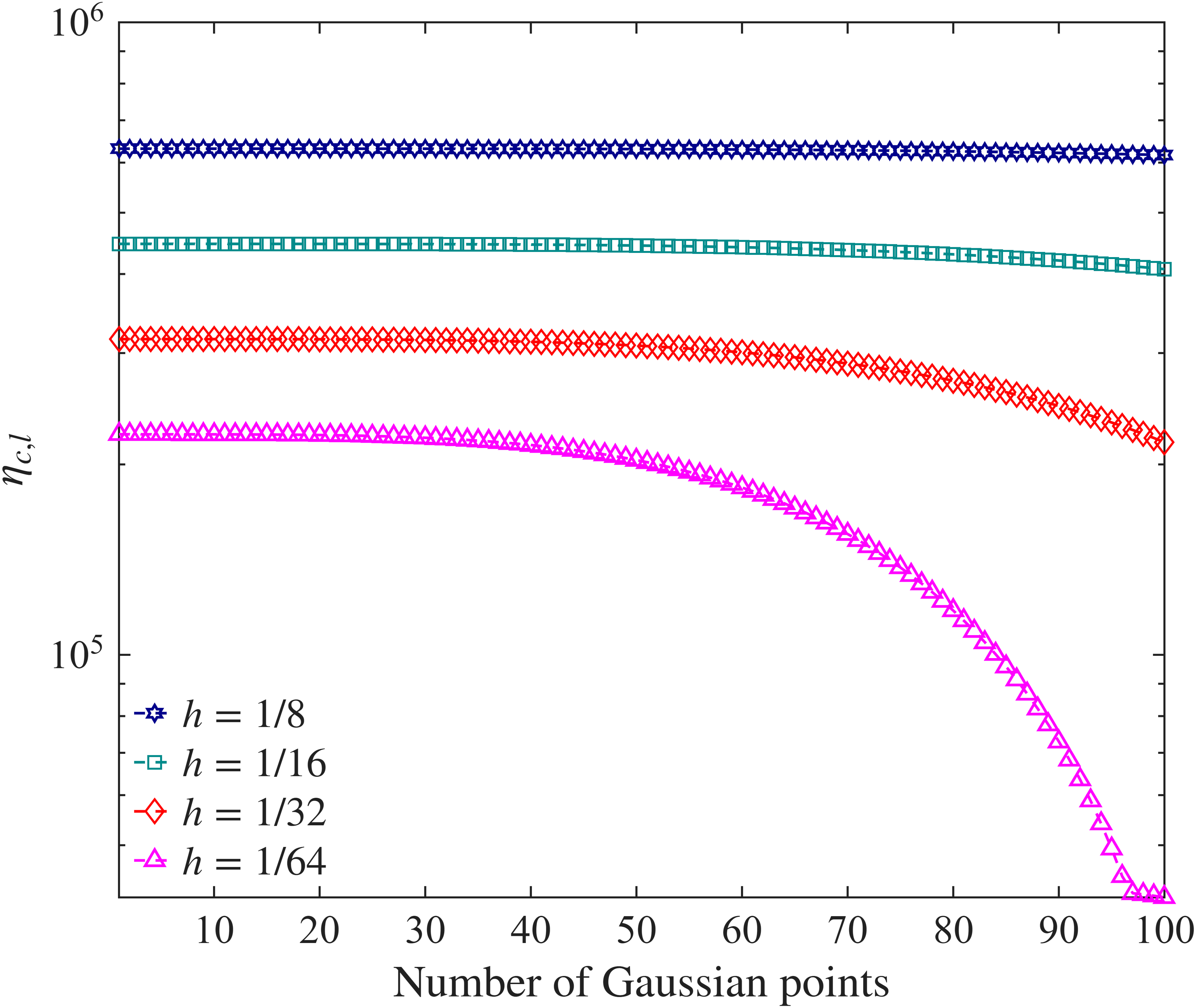}}\hfill
	\subcaptionbox{Linear cons., with trans.\ ($\epsilon=10^{-6}$)}{%
		\includegraphics[width=3.2cm,height=2.6cm]{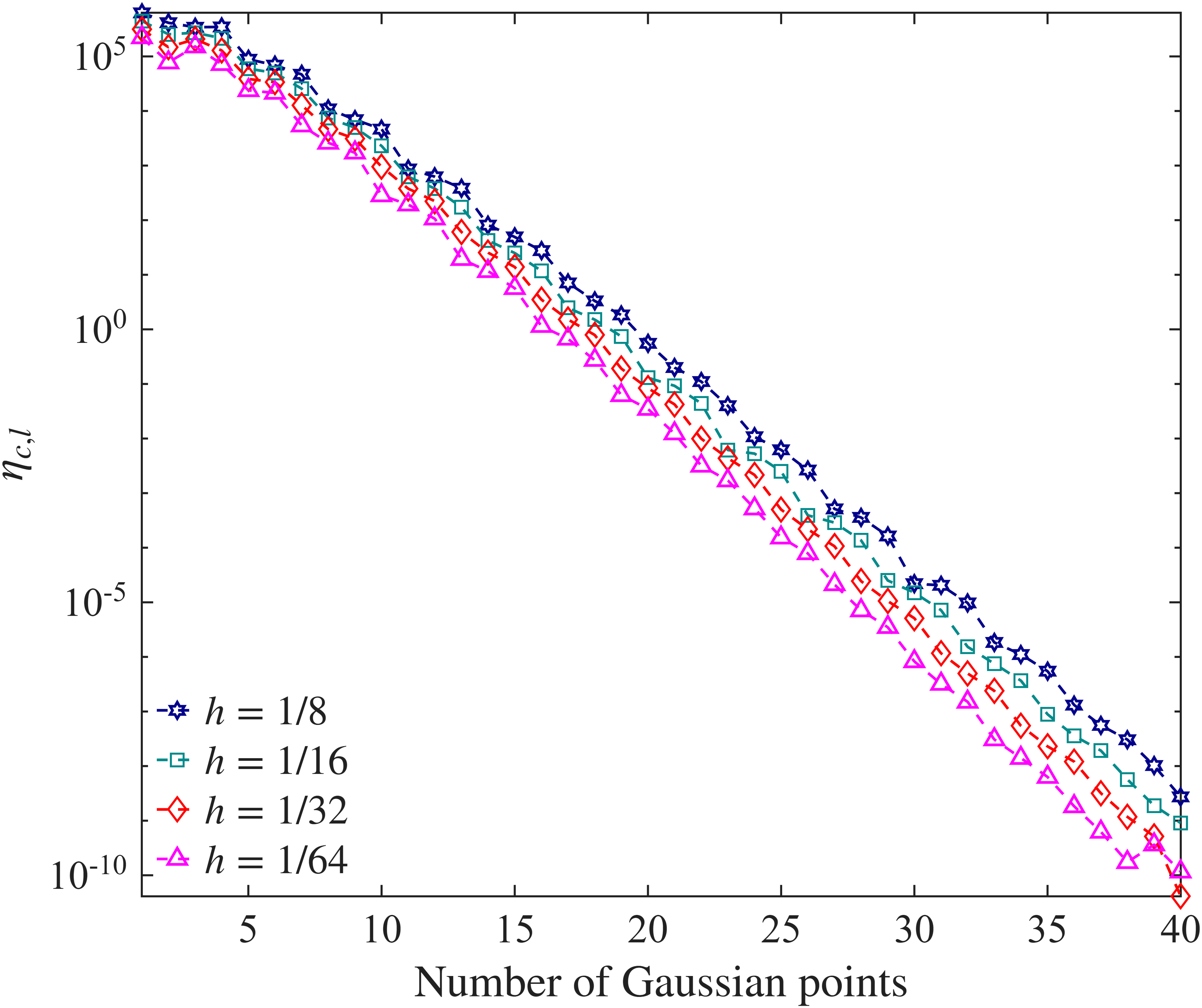}}\hfill
	\makebox[3.2cm]{}\hfill
	\makebox[3.2cm]{}
	
	\caption{Total error $\|u_{\text{ex}}-\tilde{u}_h\|_{H^1}$ and consistency error $\eta_c$ versus the number of Gauss points per element, evaluated for increasingly severe nearly singular parameters $\epsilon \in \{10^{-4}, 10^{-5}, 10^{-6}\}$.}
	\label{fig:tot_con_error}
\end{figure}

Figure~\ref{fig:tot_con_error} displays the total error $\|u_{\text{ex}} - \tilde{u}_h\|_{H^1}$ and the total consistency error $\eta_c$ (as well as its components) as functions of the number of Gauss points per element, for mesh sizes $h \in \{1/8, 1/16, 1/32, 1/64\}$ and near-singularity strengths $\epsilon \in \{10^{-4}, 10^{-5}, 10^{-6}\}$. As depicted, when direct Gauss--Legendre quadrature is applied without transformation, the consistency error decays at an arbitrarily slow algebraic rate. Consequently, the total discretisation error plateaus prematurely, completely dominated by the quadrature error rather than the mesh size $h$. 

Upon applying the sinh transformation, the consistency error transitions to an exponential convergence regime. The observed convergence rates associated with the bilinear and linear forms are comparable, as demonstrated by the analysis in Appendix~\ref*{sec:appendix_b}. Because the consistency error decreases rapidly, the total error strictly follows the theoretical approximation error down to the machine precision limits. Furthermore, as the near-singularity severity increases (i.e., smaller $\epsilon$), the absolute magnitude of the untransformed consistency error grows substantially, confirming the absolute necessity of the transformation to maintain robustness in the numerical scheme.

% ======================================================================
\section{Computational cost and implementation}
\label{sec:complexity}
% ======================================================================

\begin{figure}[!htbp]
    \centering
    \captionsetup[subfigure]{justification=centering}
    \subcaptionbox{$n=2$}{%
        \includegraphics[height=4.cm]{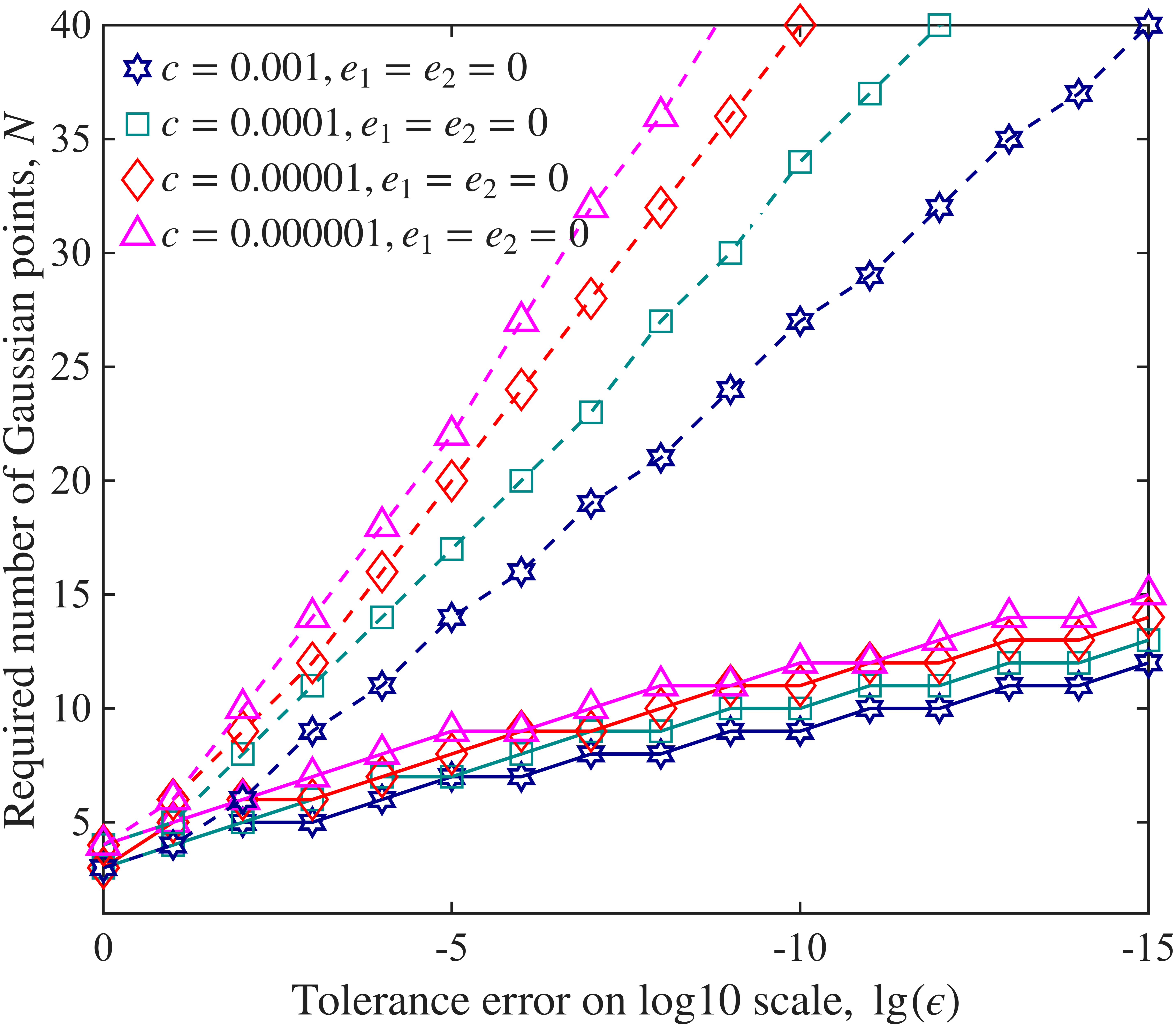}}\qquad
    \subcaptionbox{$n=3$}{%
        \includegraphics[height=4.cm]{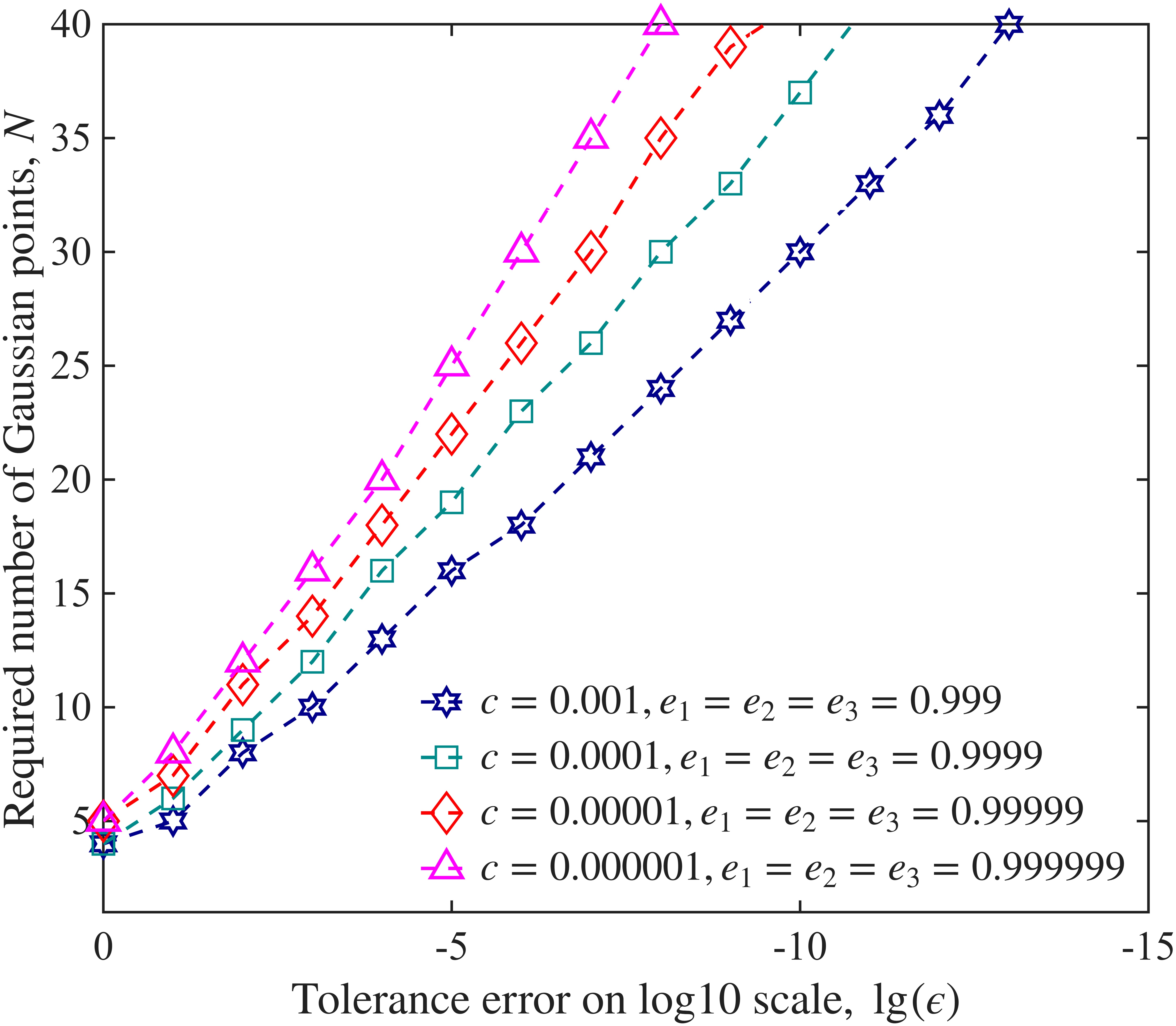}}\\
    \subcaptionbox{$n=4$}{%
        \includegraphics[height=4.cm]{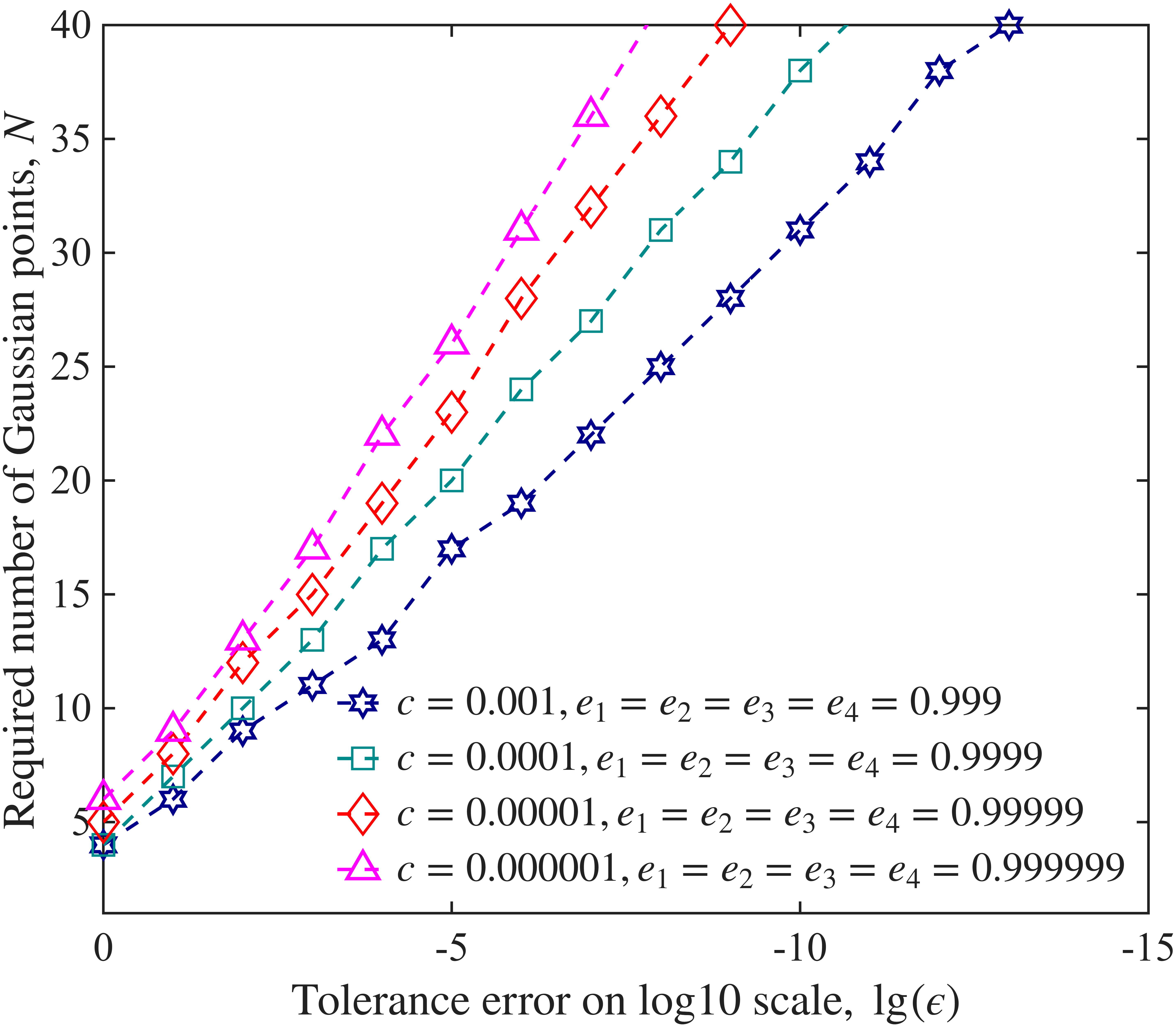}}\qquad
    \subcaptionbox{$n=5$}{%
        \includegraphics[height=4.cm]{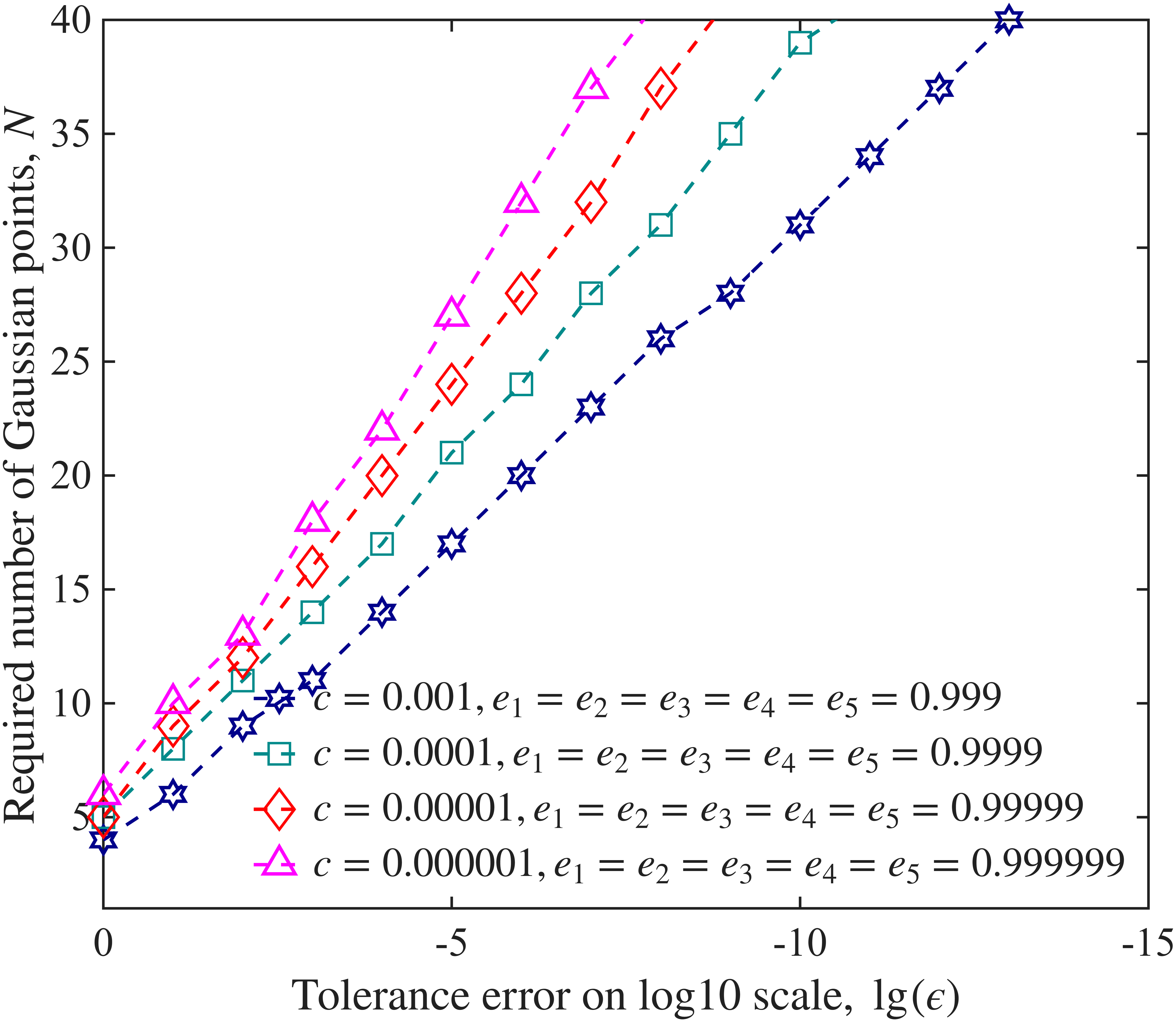}}
    \caption{Required number of Gaussian points to achieve a target error tolerance for $n=2,\dots,5$. In the 2D case $n=2$, the solid and dashed lines correspond to results computed in the Duffy and physical domains, respectively.}
    \label{fig:time_complx}
\end{figure}

\paragraph{Time complexity}
The sinh transformation is employed in both the Duffy domain
and the physical Cartesian domain for arbitrary dimensions. The only difference is the number
of sub-integrals: $2n!$ in the Duffy domain compared to \ $2^n$ in the physical domain.
After applying the sinh transformation to each dimension, the total quadrature
error is
\begin{equation}
    |E_{\text{total}}| = \sum_{j=1}^{n} C_j \rho_j^{-2m_j},
\end{equation}
where $\rho_j \gg 1$ is the Bernstein parameter for the $j$-th dimension.
To obtain a simple closed-form estimate, we distribute the total error budget
$\epsilon$ uniformly across all $n$ dimensions, i.e.\ $C_j \rho_j^{-2m_j}
\le \epsilon/n$ for each $j$. This uniform allocation assumes that the
dimension-dependent constants $C_j$ and the Bernstein parameters $\rho_j$ are of
comparable magnitude across directions; when some directions carry substantially
weaker near-singularities (i.e. larger $\rho_j$), the uniform distribution can be conservative. Solving for $m_j$ under the uniform
assumption yields
\begin{equation}
    m_j \approx \frac{\ln(n C_j / \epsilon)}{2\ln(\rho_j)}.
\end{equation}
The total number of tensor-product nodes is
\begin{equation}
    N_{\text{total}} = \prod_{j=1}^{n} \frac{\ln(n C_j / \epsilon)}{2\ln(\rho_j)}. \label{eq:time_complexity}
\end{equation}
Complexity grows exponentially with $n$ but is anisotropic: dimensions with
weaker near-singularities require fewer nodes.
Figure~\ref{fig:time_complx} illustrates the time complexity for $n=2,\dots,5$.

\paragraph{Space complexity}

The storage requirement scales linearly with the total number of quadrature
points:
\begin{equation}
    S_{\text{total}} \sim \mathcal{O}\Bigl( \prod_{j=1}^{n} m_j \Bigr)
    = \mathcal{O}\Bigl( \prod_{j=1}^{n} \frac{\ln(n C_j / \epsilon)}{2\ln(\rho_j)} \Bigr).
\end{equation}

\paragraph{General algorithm}
Based on the preceding analysis, Algorithm~\ref{alg:snsic_algo} summarises the
complete procedure for evaluating singular integrals with $n \le 2$ and nearly
singular integrals in arbitrary dimensions. For the singular case, the Duffy
transformation is applied to cancel the $1/r^\alpha$ singularity ($\alpha < n$
required), followed by a sinh transformation on $\theta_2$ to handle any angular
near-singularity. The nearly singular case branches on dimension: for $n \le 2$,
the Duffy domain is used with independent sinh transformations on $\theta_1$ and
$\theta_2$; for $n > 2$, the physical domain is used with independent sinh
transformations on each coordinate direction.

\begin{spacing}{1}
    \begin{algorithm}[!htbp]
        \caption{Evaluation of singular and nearly singular integrals.}
        \label{alg:snsic_algo}
        \begin{algorithmic}[1]
            \renewcommand{\algorithmicrequire}{\textbf{Input:}}
            \renewcommand{\algorithmicensure}{\textbf{Output:}}
            \REQUIRE Source $\boldsymbol{x}_s$, element $\mathbf{D}$, integrand $f(\boldsymbol{x})/\|\boldsymbol{x}-\boldsymbol{x}_s\|^\alpha$, quadrature order $m$.
            \ENSURE Integral value $I$.
            \IF{$\boldsymbol{x}_s$ is on the element $\mathbf{D}$ (singular case)}
                \STATE Let $\boldsymbol{x}_0 = \boldsymbol{x}_s$. Find local coordinates $\bar{\boldsymbol{x}}_0$.
                \STATE Subdivide $\mathbf{D}$ into $2\cdot n!$ sub-simplices $\mathbf{D}_i$ sharing $\bar{\boldsymbol{x}}_0$.
                \STATE Initialise $I \leftarrow 0$.
                \FOR{each sub-simplex $\mathbf{D}_i$}
                    \STATE Apply the $n$-dimensional Duffy transformation, mapping $\mathbf{D}_i \to [0,1]^n$.
                    \STATE Compute geometric coefficients $t_1, t_2, t_3$ from the simplex vertices.
                    \STATE Apply sinh transformation $\theta_2 = \theta_2(\Theta_2)$ to regularise the angular variable.
                    \STATE Evaluate $I_i$ using $m$-point Gauss--Legendre quadrature in $(\theta_1, \Theta_2)$.
                    \STATE $I \leftarrow I + I_i$.
                \ENDFOR
            \ELSE[Nearly singular case]
                \STATE Find projection $\boldsymbol{x}_0 \in \mathbf{D}$ nearest to $\boldsymbol{x}_s$; compute $r_0 = \|\boldsymbol{x}_0 - \boldsymbol{x}_s\|$.
                \STATE Find local coordinates $\bar{\boldsymbol{x}}_0$ of $\boldsymbol{x}_0$.
                \IF{$n \le 2$}
                    \STATE Subdivide $\mathbf{D}$ into sub-regions $\mathbf{D}_i$ in the Duffy domain.
                    \FOR{each sub-region $\mathbf{D}_i$}
                        \STATE Apply Duffy transformation. Compute geometric coefficients.
                        \STATE Apply sinh transformation to $\theta_1$ and $\theta_2$.
                        \STATE Evaluate $I_i$ with Gauss--Legendre in $(\Theta_1, \Theta_2)$.
                    \ENDFOR
                \ELSE
                    \STATE Subdivide $\mathbf{D}$ into $2^n$ hyper-rectangles in the physical domain.
                    \FOR{each hyper-rectangle}
                        \STATE Compute $c_k = \mathbf{S}^{(k)}\!\cdot\!\mathbf{S}^{(k)}$ for each direction $k=1,\dots,n$.
                        \STATE For each direction $k$, compute sinh transformation parameters $\Theta_{k}(a_k)$, $\Theta_{k}(b_k)$.
                        \STATE Apply sinh transformation $\xi^{(k)} = \xi^{(k)}(\Theta_k)$ to each direction.
                        \STATE Evaluate $I_i$ with $m$-point Gauss--Legendre in $(\Theta_1,\dots,\Theta_n)$.
                    \ENDFOR
                \ENDIF
                \STATE $I \leftarrow \sum_i I_i$.
            \ENDIF
            \RETURN $I$.
        \end{algorithmic}
    \end{algorithm}
\end{spacing}

% ======================================================================
\section{Concluding remarks}
\label{sec:conclusion}
% ======================================================================

We have developed a unified framework for the numerical evaluation of singular
and nearly singular integrals based on variable transformations. The principal
findings are:

\begin{enumerate}
    \item Singular integrals. The Duffy transformation provides a valid
      method in any dimension $n$: the Jacobian cancels the radial singularity
      when $\alpha < n$, yielding exponential convergence in the radial direction
      for $\alpha \le n-1$ and algebraic convergence of order $2(n-\alpha)$ for
      $n-1 < \alpha < n$. For well-shaped simplices, angular convergence is
      exponential regardless of $n$. For distorted simplices when $n>2$, the
      angular variables couple through $c_2$, creating an $(n-2)$-dimensional
      near-singular manifold that degrades the convergence rate. This is a
      robustness limitation of the Duffy method.

    \item Nearly singular integrals. The Duffy domain admits a complete
      separable treatment only for $n=2$, where independent sinh transformations
      on $\theta_1$ and $\theta_2$ suffice. For $n>2$, the same geometric
      obstruction that affects distorted singular simplices becomes fatal: the
      near-singular manifold has dimension $\ge 1$ and cannot be resolved by
      a tensor-product rule. Switching to the physical domain resolves this
      obstruction: the near-singularities align with the coordinate axes, and
      the sinh transformation applied independently to each dimension yields
      exponential convergence for any $n$.

    \item Transformation selection. A Bernstein-ellipse convergence
      analysis proves that among the three closed-form members of the Gauss
      hypergeometric family, only the sinh transformation ($\beta=1$) consistently
      improves convergence. The tangent ($\beta=2$) and algebraic ($\beta=3$)
      transformations introduce secondary near-singularities through their
      Jacobians and are ineffective.

    \item Variational discretisations. A Strang-lemma analysis
      demonstrates that the improved quadrature accuracy directly translates to
      optimal asymptotic convergence rates. The consistency error decays
      exponentially, eliminating the quadrature bottleneck and allowing the
      discrete solution to achieve the rate dictated by approximation theory.
\end{enumerate}

Future work may address the extension to curved elements with rigorous treatment
of cross-terms in the physical domain, adaptive transformation strategies, and
the coupling with sparse-grid quadrature to mitigate the exponential growth of
tensor-product cost in very high dimensions.

\appendix

\section{Convergence of the tangent and algebraic transformations}
\label{apdx_beta23}

\subsection{Tangent transformation ($\beta=2$)}

The upper integration limit after the tangent transformation is
\begin{equation}
    \Theta_1 = \frac{\arctan(\sqrt{c}/r_0)}{\sqrt{c}\,r_0}
    \;\xrightarrow{r_0\to 0}\; \frac{\pi}{2\sqrt{c}\,r_0}.
\end{equation}
The Jacobian $\sec^2(r_0\sqrt{c}\,\Theta)$ has singularities at
$r_0\sqrt{c}\,\Theta_s = \pi/2 + k\pi$. Mapping to the $p$-domain:
\begin{equation}
    p_s^{(\beta=2)} = \frac{2\Theta_s}{\Theta_1} - 1
    = \frac{\pi}{\arctan(\sqrt{c}/r_0)} - 1
    \;\xrightarrow{r_0\to 0}\; \frac{\pi}{\pi/2} - 1 = 1.
\end{equation}
The Bernstein parameter tends to unity and the convergence is arbitrarily slow.

\subsection{Algebraic transformation ($\beta=3$)}

The upper limit is
\begin{equation}
    \Theta_1 = \frac{1}{r_0^2\sqrt{r_0^2 + c}}
    \;\xrightarrow{r_0\to 0}\; \frac{1}{r_0^2\sqrt{c}}.
\end{equation}
The Jacobian $(1 - c r_0^4\Theta^2)^{-3/2}$ has singularities at
$\Theta_s = \pm 1/(r_0^2\sqrt{c})$. Mapping to the $p$-domain gives
$p_s^{(\beta=3)} \approx 1$, again collapsing the Bernstein parameter to unity.

\subsection{Behaviour in the physical domain}

The preceding analysis was conducted for a single variable in the Duffy domain.
We now verify that the same qualitative conclusions hold in the physical
Cartesian domain, where the sinh transformation is applied independently to
each coordinate direction. The analysis is presented for the first coordinate
$\xi^{(1)} \equiv x_1$; the treatment of the remaining coordinates is identical.

\subsubsection{Sinh transformation ($\beta=1$) in the physical domain}

After applying the sinh transformation to $x_1$ in Eq.~\eqref{eq:cart_denom}
with $\alpha = 1$, the sub-integral becomes
\begin{equation}
    I_l = \int_{-1}^{1}\int_{a_2}^{b_2}\!\cdots\!\int_{a_n}^{b_n}
    \frac{f(\mathbf{x})}
    {\bigl(r_0^2 + \sum_{j=1}^{n}\sum_{k=1}^{n} \mathbf{S}^{(j)}\!\cdot\!\mathbf{S}^{(k)}\,
      \bm{\xi}^{(j)}\bm{\xi}^{(k)}\bigr)^{1/2}}
    \,\frac{\mathrm{d}x_1}{\mathrm{d}\theta_1}\,\mathrm{d}\theta_1\,\mathrm{d}x_2\cdots\mathrm{d}x_n.
\end{equation}
The Jacobian of the sinh transformation in the physical domain takes the form
\begin{equation}
    \frac{\mathrm{d}x_1}{\mathrm{d}\theta_1} = \frac{r_0 k_1}{\sqrt{c_1}} \cosh(k_1\theta_1 + k_2),
\end{equation}
with the coefficients
\begin{equation}
	\begin{split}
    k_1 = \frac{1}{2}\arcsinh\!\Bigl(\frac{\sqrt{c_1}(b_1 - x_{0,1})}{r_0}\Bigr)
        - \frac{1}{2}\arcsinh\!\Bigl(\frac{\sqrt{c_1}(a_1 - x_{0,1})}{r_0}\Bigr),\\
    k_2 = \frac{1}{2}\arcsinh\!\Bigl(\frac{\sqrt{c_1}(b_1 - x_{0,1})}{r_0}\Bigr)
        + \frac{1}{2}\arcsinh\!\Bigl(\frac{\sqrt{c_1}(a_1 - x_{0,1})}{r_0}\Bigr),
   	\end{split}
\end{equation}
where $x_{0,i}$ denotes the $i$-th component of $\mathbf{x}_0$, $c_1=\mathbf{S}^{(1)}\!\cdot\!\mathbf{S}^{(1)}$. The $\cosh$
function has its nearest singularity in the complex $\theta_1$-plane at
$\theta_1 = i\pi/(2k_1) - k_2/k_1$. Because $k_1$ and $k_2$ involve
$\arcsinh(\cdot/r_0)$, they grow only logarithmically as $r_0 \to 0$,
pushing the singularity away from the real axis. The transformed integrand
therefore exhibits no near-singular behaviour, and Gauss--Legendre quadrature
converges rapidly.

\subsubsection{Tangent transformation ($\beta=2$) in the physical domain}

For $\alpha = 2$, the tangent transformation is applied. After transformation,
the Jacobian becomes
\begin{equation}
    \frac{\mathrm{d}x_1}{\mathrm{d}\Theta_1} =\frac{ r_0 k_1}{\sqrt{c_1}} \sec^2(k_1\Theta_1 + k_2),
    \label{eq:jacobian_beta2_cart}
\end{equation}
where now
\begin{equation}
	\begin{split}
    k_1 = \frac{1}{2}\arctan\!\Bigl(\frac{\sqrt{c_1}(b_1 - x_{0,1})}{r_0}\Bigr)
        - \frac{1}{2}\arctan\!\Bigl(\frac{\sqrt{c_1}(a_1 - x_{0,1})}{r_0}\Bigr),\\
    k_2 = \frac{1}{2}\arctan\!\Bigl(\frac{\sqrt{c_1}(b_1 - x_{0,1})}{r_0}\Bigr)
        + \frac{1}{2}\arctan\!\Bigl(\frac{\sqrt{c_1}(a_1 - x_{0,1})}{r_0}\Bigr).
    \end{split}
\end{equation}
The Jacobian~\eqref{eq:jacobian_beta2_cart} has singularities at
$\Theta_1 = (\pm\pi - 2k_2)/(2k_1)$. Mapping these to the standard interval
$[-1,1]$ via $p = 2(\Theta_1 - \Theta_1(a_1))/(\Theta_1(b_1) - \Theta_1(a_1)) - 1$
yields
\begin{equation}
    p_s = \frac{\pm\pi - \arctan\!\bigl(\frac{\sqrt{c_1}(b_1-x_{0,1})}{r_0}\bigr)-\arctan\!\bigl(\frac{\sqrt{c_1}(a_1-x_{0,1})}{r_0}\bigr)}        {\arctan\!\bigl(\frac{\sqrt{c_1}(b_1-x_{0,1})}{r_0}\bigr)     - \arctan\!\bigl(\frac{\sqrt{c_1}(a_1-x_{0,1})}{r_0}\bigr)}.
\end{equation}
When the projection point lies on the isoparametric domain boundary, i.e. $b_1 = x_{0,1}$ or
$a_1 = x_{0,1}$, the corresponding $\arctan$ terms vanish, and $p_s$
approaches $\pm 1$. The Bernstein ellipse parameter therefore collapses to
unity when the source projects onto the element boundary, which is the
most critical case in practice.

\subsubsection{Algebraic transformation ($\beta=3$) in the physical domain}

For $\alpha = 3$, the algebraic transformation is applied. The Jacobian is
\begin{equation}
    \frac{\mathrm{d}x_1}{\mathrm{d}\Theta_1} = \frac{r_0^3}{(1 - c_1 r_0^4\Theta_1^2)^{3/2}},
    \label{eq:jacobian_beta3_cart}
\end{equation}
with singularities at $\Theta_1 = \pm 1/(r_0^2\sqrt{c_1})$. The integration limits are
\begin{equation}
    \Theta_1(a_1) = \frac{a_1 - x_{0,1}}{r_0^2\sqrt{r_0^2 + c_1(a_1 - x_{0,1})^2}},
    \qquad
    \Theta_1(b_1) = \frac{b_1 - x_{0,1}}{r_0^2\sqrt{r_0^2 + c_1(b_1 - x_{0,1})^2}}.
\end{equation}
Mapping the singularities to $[-1,1]$ gives
\begin{equation}
    p_s = \frac{\pm 2 - \frac{\sqrt{c_1}(a_1-x_{0,1})}{\sqrt{r_0^2+c_1(a_1-x_{0,1})^2}}-\frac{\sqrt{c_1}(b_1-x_{0,1})}{\sqrt{r_0^2+c_1(b_1-x_{0,1})^2}}}{\frac{\sqrt{c_1}(a_1-x_{0,1})}{\sqrt{r_0^2+c_1(a_1-x_{0,1})^2}}-\frac{\sqrt{c_1}(b_1-x_{0,1})}{\sqrt{r_0^2+c_1(b_1-x_{0,1})^2}}}.
\end{equation}
As $r_0 \to 0$, the terms $\frac{\sqrt{c_1}(a_1-x_{0,1})}{\sqrt{r_0^2+c_1(a_1-x_{0,1})^2}}$ or 
$\frac{\sqrt{c_1}(b_1-x_{0,1})}{\sqrt{r_0^2+c_1(b_1-x_{0,1})^2}}$ approach $\pm 1$. Consequently, the numerator and
denominator tend to values that drive $p_s$ to $\pm 1$, collapsing the Bernstein
ellipse parameter to unity.

\smallskip\noindent
In summary, the physical-domain analysis confirms the conclusions drawn in the
Duffy domain. For $\beta=1$, the logarithmic growth of $\arcsinh$ pushes complex
singularities away from the real axis, preserving exponential convergence. For
$\beta=2$ and $\beta=3$, the Jacobian singularities map to $p_s = \pm 1$ as
$r_0 \to 0$,
collapsing the Bernstein parameter to unity. The sinh transformation is
therefore uniquely effective in the physical domain as well.

% ======================================================================
\section{Derivation of the post-transformation error bound}
\label{sec:appendix_b}

We derive the bound for the total discretisation error presented in Equation \ref{eq:convproof}. This rigorous derivation demonstrates how the normalisation $||w_h||_V$ in Strang's first lemma is absorbed into the convergence constants by analysing the quadrature error in the complex plane, avoiding the algebraic fallacies associated with factoring quadrature errors of polynomial products.

\subsection*{B.1. Strang's Lemma and Stable Decomposition}

Let $I_h u \in V_h^{enr}$ be the interpolant of the exact solution $u$. Strang's first lemma bounds the total error by the approximation error plus the consistency errors arising from both the bilinear form $a(\cdot,\cdot)$ and the linear form $l(\cdot)$:
\begin{equation}
	\begin{split}
		||u - \tilde{u}_h||_V \le C_0 \inf_{v_h \in V_h} ||u - v_h||_V + C_0 \sup_{w_h \in V_h} \frac{|a(I_h u, w_h) - a_h(I_h u, w_h)|}{||w_h||_V} \\ +C_0 \sup_{w_h \in V_h} \frac{|l(w_h) - l_h(w_h)|}{||w_h||_V}.
		\label{eq:strang_full}
	\end{split}
\end{equation}

Given a generic element $K \in \mathcal{T}_h$, the exact local integrands for the bilinear and linear forms are denoted as $\mathcal{K}_a(I_h u, w_h; x) = \Phi_{I_h u}(x) \Psi_{a, w_h}(x)$ and $\mathcal{K}_l(w_h; x) = \Phi_l(x) \Psi_{l, w_h}(x)$, respectively. Here, $\Phi_{I_h u}$ and $\Phi_l$ capture the (nearly) singular state and source variables. The terms $\Psi_{a, w_h}$ and $\Psi_{l, w_h}$ represent the test function evaluated through the appropriate linear differential operators (e.g., the spatial gradient $\nabla w_h$ for the bilinear form, and $w_h$ for the linear form). Let $E_K(\cdot)$ denote the local quadrature error functional such that the global consistency errors are bounded by $\sum_{K} |E_K(\mathcal{K}_a)|$ and $\sum_{K} |E_K(\mathcal{K}_l)|$.

Using the standard stable decomposition condition for enriched finite element spaces, any discrete test function $w_h \in V_h^{enr}$ can be uniquely decomposed into a standard polynomial part and an enriched part, $w_h = w_{std} + c_\psi \psi$, satisfying:
\begin{equation}
	||w_{std}||_V \le C_{stab}||w_h||_V, \quad |c_\psi| \le \frac{C_{stab}}{||\psi||_V}||w_h||_V.
	\label{eq:stable_decomp}
\end{equation}
By the linearity of the quadrature error operator, the local errors split accordingly:
\begin{align}
	E_K(\mathcal{K}_a) &= E_K(\Phi_{I_h u} \Psi_{a, w_{std}}) + c_\psi E_K(\Phi_{I_h u} \Psi_{a, \psi}), \label{eq:split_a} \\
	E_K(\mathcal{K}_l) &= E_K(\Phi_l \Psi_{l, w_{std}}) + c_\psi E_K(\Phi_l \Psi_{l, \psi}). \label{eq:split_l}
\end{align}

\subsection*{B.2. Bounding the Bilinear Form Consistency Error}

After applying the appropriate variable transformation (e.g. the sinh transformation), the integration over $K$ is mapped to the reference domain $\hat{K} = [-1, 1]$. The transformed integrand is analytic in a region bounded by the Bernstein ellipse $E_\rho$ with parameter $\rho > 1$. The quadrature error for an $m$-point Gauss-Legendre rule is bounded by $|E_{\hat{K}}(F)| \le C \frac{M(\rho)}{\rho - 1} \rho^{-2m}$, where $M(\rho) = \max_{\xi \in E_\rho} |F(\xi)|$.

\textbf{Standard Component:} Given the mapped test function $\Psi_{w_{std}}$ is a polynomial of degree $p-1$, we apply the Bernstein's inequality for polynomials on ellipses, which bounds its maximum on $E_\rho$ by its real maximum:
\begin{equation}
	\max_{\xi \in E_\rho} |\Psi_{w_{std}}(\xi)| \le \rho^{p-1} ||\Psi_{w_{std}}||_{L^\infty(\hat{K})}.
\end{equation}
Mapping back to the physical element and utilising the standard finite element inverse estimate $||\Psi||_{L^\infty(K)} \le C_{inv} h_K^{-d/2} ||\Psi||_{L^2(K)}$, we obtain:
\begin{equation}
	\max_{\xi \in E_\rho} |\Psi_{w_{std}}(\xi)| \le \rho^{p-1} C_{inv} h_K^{-d/2} ||w_{std}||_{V,K} \le \rho^{p-1} C_{inv} C_{stab} h_K^{-d/2} ||w_h||_V.
\end{equation}
Incorporating the geometric Jacobian $J_K \propto h_K^d$, the quadrature bound for the standard part of the bilinear form isolates $||w_h||_V$:
\begin{equation}
	|E_K(\Phi_{I_h u} \Psi_{w_{std}})| \le \left[ C_{inv} C_{stab} h_K^{d/2} \left( \max_{E_\rho} |\Phi_{I_h u}| \right) \right] ||w_h||_V \frac{\rho^{p-1-2m}}{\rho - 1}.
\end{equation}

\textbf{Enriched Component:} The enrichment function $\Psi_\psi$ is not a polynomial, and Bernstein's inequality does not apply. However, consider that the sinh transformation regularises the nearly singular behaviour, its analytic continuation is bounded on $E_\rho$ by a finite constant $M_{\psi}(\rho)$. Utilising the stable decomposition bound for $c_\psi$ from \eqref{eq:stable_decomp}:
\begin{equation}
	|c_\psi E_K(\Phi_{I_h u} \Psi_\psi)| \le \left[ \frac{C_{stab}}{||\psi||_V} h_K^d \left( \max_{E_\rho} |\Phi_{I_h u}| \right) M_{\psi}(\rho) \right] ||w_h||_V \frac{\rho^{-2m}}{\rho - 1}.
\end{equation}

\subsection*{B.3. Bounding the Linear Form Consistency Error}

The linear form consistency error follows a similar analytic structure. Applying the Bernstein inequality to the standard component $\Psi_{l, w_{std}}$ bounds its maximum by the local $L^2(K)$ norm. To relate this back to the global energy norm, we invoke the global Poincar\'e inequality $||w_{std}||_{L^2(\Omega)} \le C_P ||w_{std}||_V$, where $C_P$ is the Poincar\'e constant associated with the domain diameter. This yields:
\begin{equation}
	|E_K(\Phi_l \Psi_{l, w_{std}})| \le \left[ C_{inv} C_{stab} C_P h_K^{d/2} \left( \max_{E_\rho} |\Phi_l| \right) \right] ||w_h||_V \frac{\rho^{p-1-2m}}{\rho - 1}.
\end{equation}
Because $C_P$ carries the physical dimension of length, this inequality remains dimensionally consistent with the physical integral.

Similarly, the enriched component of the linear form relies on the bounded analytic continuation of the transformed source term product on $E_\rho$, denoted $M_{l,\psi}(\rho)$. Utilising the stable decomposition directly bypasses the Poincar\'e inequality:
\begin{equation}
	|c_\psi E_K(\Phi_l \Psi_{l, \psi})| \le \left[ \frac{C_{stab}}{||\psi||_V} h_K^d \left( \max_{E_\rho} |\Phi_l| \right) M_{l,\psi}(\rho) \right] ||w_h||_V \frac{\rho^{-2m}}{\rho - 1}.
\end{equation}

\subsection*{B.4. Global Assembly and Optimal Convergence}

Summing the local standard and enriched bounds for both forms, the global test function norm $||w_h||_V$ can now be factored out of the local quadrature errors. Substituting into the normalised consistency terms in Strang's lemma \eqref{eq:strang_full} cancels $||w_h||_V$ entirely:
\begin{align}
	\sup_{w_h \in V_h} \frac{|a(I_h u, w_h) - a_h(I_h u, w_h)|}{||w_h||_V} &\le \sum_{K \in \mathcal{T}_h} \tilde{C}_{a,K} \left( \rho^{p-1} + M_\psi(\rho) \right) \rho^{-2m}, \\
	\sup_{w_h \in V_h} \frac{|l(w_h) - l_h(w_h)|}{||w_h||_V} &\le \sum_{K \in \mathcal{T}_h} \tilde{C}_{l,K} \left( \rho^{p-1} + M_{l,\psi}(\rho) \right) \rho^{-2m},
\end{align}
where $\tilde{C}_{a,K}$ and $\tilde{C}_{l,K}$ absorb the mesh-dependent geometric constants and the maximum moduli of $\Phi_{I_h u}$ and $\Phi_l$, respectively.

As the variable transformations guarantee $\rho > 1$, both consistency errors decay exponentially at an asymptotic rate of $\mathcal{O}(\rho^{-(2m - p + 1)})$. For a sufficiently large quadrature order $m$, these consistency errors become subdominant to the best approximation error $C_0 \inf_{v_h \in V_h} ||u - v_h||_V$, recovering the optimal asymptotic convergence rate dictated by the enriched finite element space.

% ======================================================================

\section*{Acknowledgments}
This work was funded by Shanghai Pujiang Program (22PJD074) and the National Natural Science Foundation of China (Grant No. 12571466).

\bibliographystyle{siamplain}
\bibliography{references.bib}

\end{document}

%% file: ex_shared.tex
\usepackage{lipsum}
\usepackage{amsfonts}
\usepackage{graphicx}
\usepackage{epstopdf}
\usepackage{algorithmic}
\ifpdf
\DeclareGraphicsExtensions{.eps,.pdf,.png,.jpg}
\else
\DeclareGraphicsExtensions{.eps}
\fi

\newsiamremark{remark}{Remark}
\newsiamremark{hypothesis}{Hypothesis}
\crefname{hypothesis}{Hypothesis}{Hypotheses}
\newsiamthm{claim}{Claim}
\newsiamremark{fact}{Fact}
\crefname{fact}{Fact}{Facts}

\headers{High-dimensional singular and nearly singular quadrature}{Jing Hu, Luqiang He and Shuyu Sun}

\title{Singular and nearly singular quadrature in arbitrary dimensions\thanks{Submitted to the editors DATE.}}
\author{
	Jing Hu\thanks{College of Civil Engineering, Tongji University, Shanghai, China (\email{jhu@tongji.edu.cn}, \email{luqianghe@tongji.edu.cn}). Corresponding author: Jing Hu.}
	\and
	Luqiang He\footnotemark[1]
	\and
	Shuyu Sun\thanks{School of Mathematical Science,Tongji University, Shanghai, China (\email{suns@tongji.edu.cn}).}
}

\usepackage{amsopn}